\documentclass[12pt,a4paper]{article}

\usepackage[UKenglish]{isodate}% http://ctan.org/pkg/isodate

\usepackage[a4paper,margin=2cm, marginparwidth=2cm]{geometry}

\usepackage{amsmath,amssymb,amsthm}

\usepackage{hyperref}

\usepackage{graphicx}
\graphicspath{{./figures/}}

\usepackage[table,dvipsnames]{xcolor}
\usepackage{multirow}

\makeatletter
\renewcommand{\@biblabel}[1]{\quad#1.}
\makeatother

\usepackage{caption}
\usepackage{subcaption}

\newcommand{\bea}{\begin{eqnarray}}
\newcommand{\eea}{\end{eqnarray}}
\newcommand{\beq}{\begin{equation}}
\newcommand{\eeq}{\end{equation}}

\providecommand{\eqref}[1]{(\ref{#1})}

\newcommand{\figref}[1]{Fig.\ \ref{#1}}
\newcommand{\Figref}[1]{Fig.\ \ref{#1}} % Form to use at start of sentence
\newcommand{\tabref}[1]{Table~\ref{#1}}

\newcommand{\appref}[1]{Appendix~\ref{#1}}

\newcommand{\tsedef}[1]{\textsc{#1}}
\newcommand{\urlname}[1]{\href{http://#1}{\texttt{#1}}}

\begin{document}
%  \begin{flushright}
%  \textbf{Confidential}. \\ Not to be redistributed without permission of the author.\\
%  \today \\
% %%% (\texttt{\jobname.tex}  LaTeX-ed on \today ) \\
%  \end{flushright}
%  \vspace*{0.5cm}

\begin{center}
	{\Large\textbf{Morphology of Vaccine RD\&D Translation }	}
	\\[0.5\baselineskip]
	{\large Martin Ho\textsuperscript{1,2\S*},
	Henry CW Price\textsuperscript{3,4\S},
	Tim S Evans\textsuperscript{3,4\ddag},
	Eoin O'Sullivan\textsuperscript{1,2\ddag}
	}
	\\[0.5\baselineskip]
	\textbf{1} Centre for Science Technology \& Innovation Policy, University of Cambridge, Cambridge CB3 0HU, United Kingdom
	\\
	\textbf{2} Department of Engineering,  University of Cambridge, Cambridge CB3 0HU, United Kingdom
	\\
	\textbf{3} Centre for Complexity Science, Imperial College London, London SW7 2AZ, United Kingdom
	\\
	\textbf{4} Theoretical Physics Group, Department of Physics, Imperial College London, London SW7 2AZ, United Kingdom
	\\[0.5\baselineskip]
	
	% Insert additional author notes using the symbols described below. Insert symbol callouts after author names as necessary.
	%
	% Remove or comment out the author notes below if they aren't used.
	%
	% Primary Equal Contribution Note
	\S These authors contributed equally to this work.
	
	% Additional Equal Contribution Note
	% Also use this double-dagger symbol for special authorship notes, such as senior authorship.
	\ddag These authors also contributed equally to this work.
	
	% \textcurrency Current Address: Dept/Program/Center, Institution Name, City, State, Country % change symbol to "\textcurrency a" if more than one current address note
	% \textcurrency b Insert second current address
	% \textcurrency c Insert third current address
	
	% \dag Deceased
	
	% Group/Consortium Author Note
	% \textpilcrow Membership list can be found in the Acknowledgments section.
	
	% Use the asterisk to denote corresponding authorship and provide email address in note below.
	* Corresponding author: \texttt{wtmh3@eng.cam.ac.uk}
	
\end{center}

% ****************************************************************************************************
\begin{abstract}

\noindent
Translation as a concept coordinates participation in innovation but remains a qualitative construct. We provide multivariate accounting of linkages between market entries of vaccines, clinical trials, patents, publications, funders, and grants to quantify biomedical translation. We found that the most prevalent types of biomedical translation are those between basic and applied research (52\%) followed by those between research and product development (36\%). Although many biomedical stakeholders assume knowledge flows one-way from upstream research to downstream application, knowledge feedbacks that mediate translation are prevalent. We also cluster biomedical funders based on the types of translations they fund. Large-scale funding agencies such as NIH are similarly involved in early-stage translation, whereas pharmaceuticals and ``mission-oriented’’ agencies such as DARPA involve diverse translation types, and each leaves different translation footprints.

\end{abstract}

Translation, particularly in biomedical contexts, is the process of turning knowledge developed from laboratory observations into interventions and applications \cite{RN1365}. Despite the rising awareness of translational research, the magnitude and longitudinal distribution of translation required to advance scientific knowledge into (e.g.\ therapeutic) applications is unclear. Given knowledge flows exist between multiple innovation phases \cite{B45} and are multidirectional and non-sequential \cite{RN1365,RN285,RN1703}, translation would benefit from being studied as multiple subtypes rather than a singular process \cite{RN1918}.

The US agency responsible for the approval of medical advances, the Food and Drug Administration (FDA), has a ``Critical Path Initiative'' \cite{FDA04} which provides a mental framework for advances in a medical setting. In this picture, knowledge is described to proceed along a ``critical path'' passing through five phases: basic research, prototype design or delivery, preclinical development, clinical development, and FDA filing \cite{RN84}. One aspect not often considered is whether the same amount of effort is required to translate between these five phases and whether the majority of translational effort proceeds in the sequence stated by the Critical Path. In science policy, technological investments as a function of phases along the Critical Path is often portrayed as multi-modal, with the chasm between investment peaks reflecting deficiencies in translational research as a ``valley of death'' \cite{RN1993,RN245}. Biomedical communities attribute insufficient translational research to its intrinsic higher risk, greater difficulty, lack of awareness, and institutional silos among actors along the Critical Path \cite{RN1910}. The under-supply of translational research is, in turn, hypothesized to contribute to declining biomedical RD\&D (research, development and demonstration) productivity \cite{RN1366}.

Just as understanding the structure of a protein is vital to deducing its function and applications, this study deduces the ``morphology'' of translation processes to inform translational efforts for future drugs. Prior statistical studies of translation used nearly entire datasets from the US Patent and Trademark Office, Web of Science, or National Institutes of Health (NIH) Research Portfolio Online Reporting Tools to investigate (i) who funded translational research \cite{RN1919,RN1937,RN212,RN211} and (ii) what proportion of basic research became translated \cite{RN1918,RN212,RN1920,RN1626}. For instance, it is found that 29\% of publications associated with FDA new molecular entities were NIH-funded, justifying spendings on basic science on a country-level \cite{RN211}. However, such aggregates hide a wider range of patterns.

To understand the translational effort required to advance new biomedical technologies, we instead analyze translation itself for specific biomedical advances.
We begin by proposing an analytical framework built from the phases in the ``linear'' and ``chain-linked'' models of science \cite{B45,RN285}, which also resemble the steps on the ``critical path'' used by the FDA \cite{FDA04,RN84}. 
In our framework, we define six innovation stages in technological development: 
\tsedef{basic research}, 
\tsedef{applied research,} 
\tsedef{product development}, 
\tsedef{process development}, 
\tsedef{demonstration}, 
and \tsedef{application} (see Appendix \ref{app:network}). \tsedef{Dissemination} is defined as a link within an innovation stage whereas \tsedef{Translation} is represented by a link between two different innovation stages, as illustrated in \figref{fig:framework}.
This framework is concordant with the assertion that translation is ``multistep and recursive'' \cite{RN1939}. 

To illustrate our approach we use data underlying eight FDA-authorized vaccines \cite{RePEc:arx:papers:2302.13076}. 
\appref{app:data} provides a method to generate a citation network containing more exhaustive knowledge points for any given regulator-approved therapeutics and it allows us to decompose this citation data into different types of dissemination and translation across multiple innovation phases over a seventy-year period.

Our first results concern the frequency of biomedical translation. Our networks contain 2.6 million links (representing knowledge diffusion) among 310,871 publications, 128,366 patents, 180 clinical trials, and 121,919 grants that gave rise to the marketing authorization of eight vaccines. Among the 2.6 million diffusion events, 1,911,119 (75\%) are knowledge dissemination events within the same innovation phase, and 653,361 (25\%) are knowledge translation across innovation phases. Among the eight vaccines studied, the most prevalent types of translation in descending order are:
basic research-to-applied research (median: 26.4\%, interquartile range - IQR:11.7\%), %%%(median: 26.42\%, interquartile range - IQR:11.65\%),  
applied research-to-basic research (median:26.3\%, IQR:16.0\%), %%%(median:26.35\%, IQR:16.02\%), 
basic research-to-product development (median:26.3\%, IQR:14.5\%), %%%(median:26.33\%, IQR:14.54\%), 
applied research-to-product development (median:10.5\%, IQR:6.4\%), %%%(median:10.46\%, IQR:6.41\%), 
process development-to-product development (median:4.1\%, IQR:2.7\%), %%%(median:4.14\%, IQR:2.65\%), 
product development-to-process development (median:1.8\%, IQR:2.4\%). %%%(median:1.80\%, IQR:2.44\%). 
These prevalences are partly due to research publications receiving the largest observed amount of funding in support of the vaccines. With over half of observed translational effort resting at the early research phase, translation at later development and demonstration phases either commanded or received less resources or is less visible in our data.

Using the same results, we also estimate the recursiveness and direction of biomedical translation with reference to the ``linear model'' of science shown in \figref{fig:framework}a. In this picture, dissemination events (between documents from the same phase) have no direction. Most translation events symbolize knowledge advancing from upstream to downstream in the linear model (left-to-right in \figref{fig:framework}a); citations from applied research documents to product development documents are one example. Other translation events represent knowledge moving from downstream to upstream, for example citations from product development documents to applied research documents. 
What our data shows is that: 
(i) 74.5\% of knowledge is disseminated at similar innovation phases --- approximating the amount of knowledge required for translation and indicating one-in-four knowledge flow is translational,
(ii) 17.9\% of translation travel upstream to downstream towards application, 
and 
(iii) 7.6\% of translation travel downstream to upstream --- acting as knowledge feedbacks to nurture early innovation phases. 
In summation, a net 10.3\% of knowledge flows from research to development, demonstration, and application. 
This can be understood as a net rate of biomedical translation.

Our second set of results elucidates the growth pattern of biomedical translation over time. \Figref{fig:simple_cdf} shows the accumulation of translation and dissemination for the eight vaccines from 1950 to 2020. The curves in \figref{fig:simple_cdf} are increasing and are broadly sigmoidal, as fitted to the logistic function in \figref{fig:scurve}, indicating that the volume of documents useful for the development of any one vaccine increases until we begin to find solutions for key problems at which point the growth slows, stopping once the product is developed. 
The growth stage in translation events is almost always ahead of dissemination (Imvanex is the only exception by our measures, see \tabref{tab:scurve}) but in many cases the difference is small, one year or so. 
The volume of dissemination is greater than that of translation at all times. These patterns may mean a large volume of knowledge dissemination is initiated after translation. Translation in this case may be amassing awareness, interest, and demand from participants from other innovation phases and application domains.

\begin{figure}
  \centering
    \includegraphics[width=0.9\textwidth]{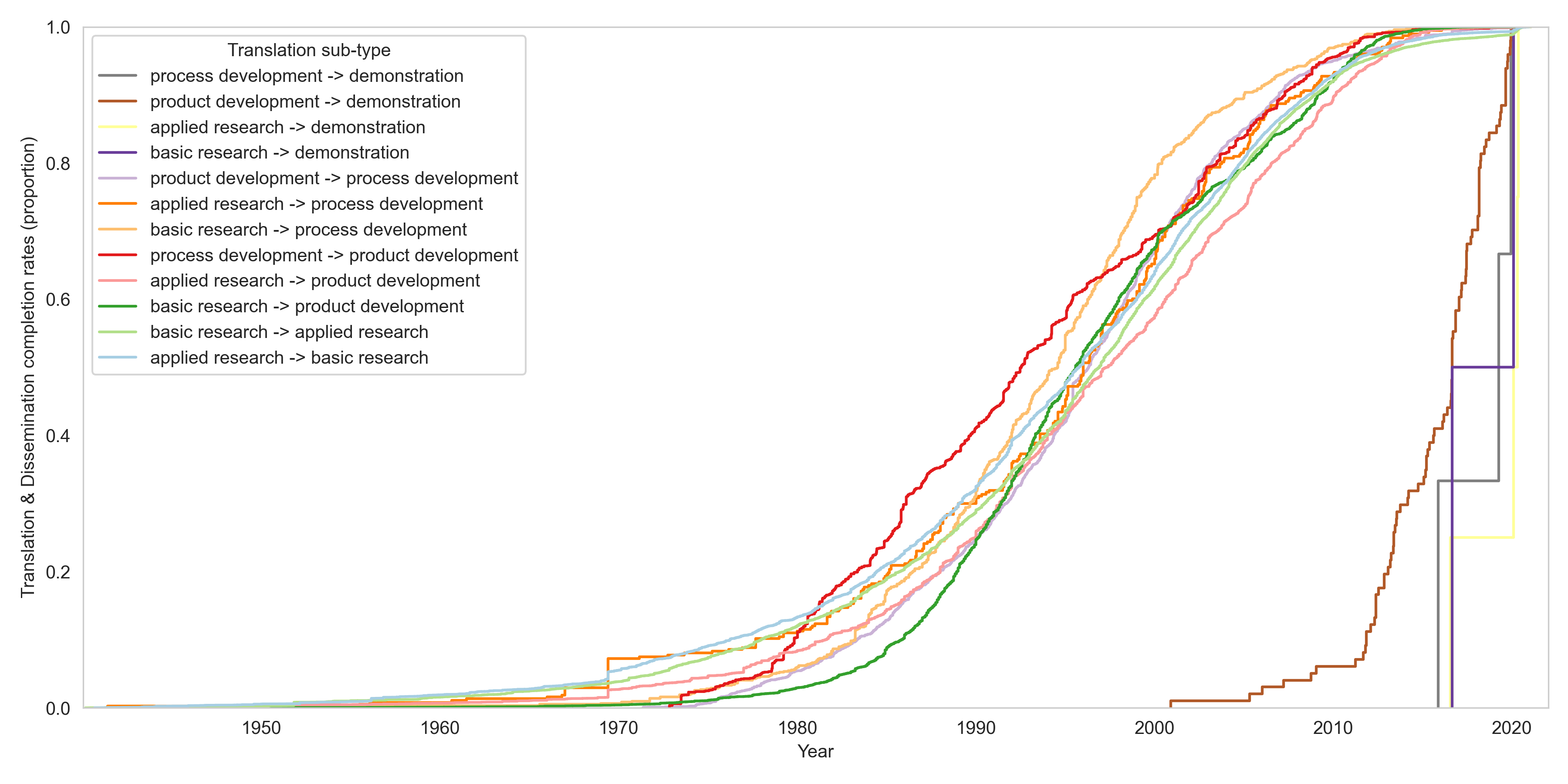}
  \caption{Translation subtypes completion rates as a function of time.
  	 The fraction of translation events for different subtypes that occurred up to time $t$, as a function of the time of each event (the date of the newer of the two documents defining an event) for our total data. Illustrative data from AstraZeneca Vaxzeria.
  	 See also Appendix \ref{app:scurve} for other views of this data.  
  	 For some translational subtypes there are too few events to be shown clearly.
          }  % end of caption
  \label{fig:inter_cdf}
\end{figure}

The classification of translation events into subtypes (Appendices \ref{app:pub} and \ref{app:pat}) gives further insights into the temporal pattern of knowledge diffusion and this is shown in \figref{fig:inter_cdf}. For vaccines approved between 2020-22, applied research to basic research and basic research to applied research translations both emerged in the 1960s and reached 10\% completion by 1980. This is followed by the emergence of the following translation types in chronological order: 
from applied research-to-process development, 
from process development-to-product development, 
from applied research-to-product development, 
from basic research-to-process development, 
from product development-to-process development, 
and 
from basic research-to-product development
--- all reaching 10\% completion between 1980 and 1990. 
Although translations involving patents always emerge later than translations purely between publications, the rates of the former always surpass the latter. The exact timespans of translation types are available in \figref{fig:scurvefit_subtypes}.

\begin{figure}
     \centering
     \begin{subfigure}[b]{0.9\textwidth}
         \centering
         \includegraphics[width=\textwidth]{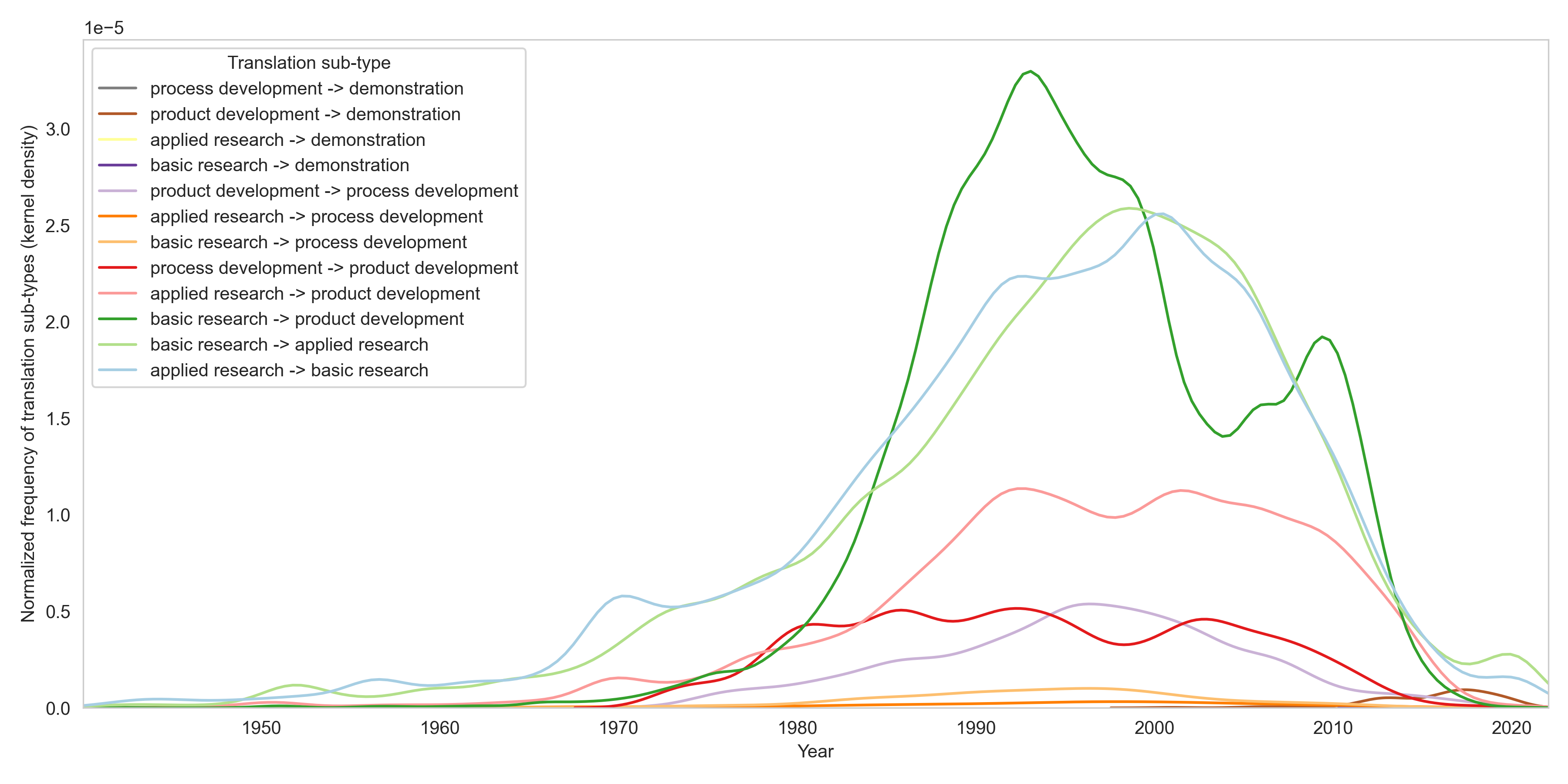}
         \caption{Translation subtypes frequencies as a function of time after kernel density estimation used to produce smooth curves.
     For some translational subtypes there are too few events to be shown clearly
     }

     \end{subfigure}
     \hfill
     \begin{subfigure}[b]{0.9\textwidth}
         \centering
         \includegraphics[width=\textwidth]{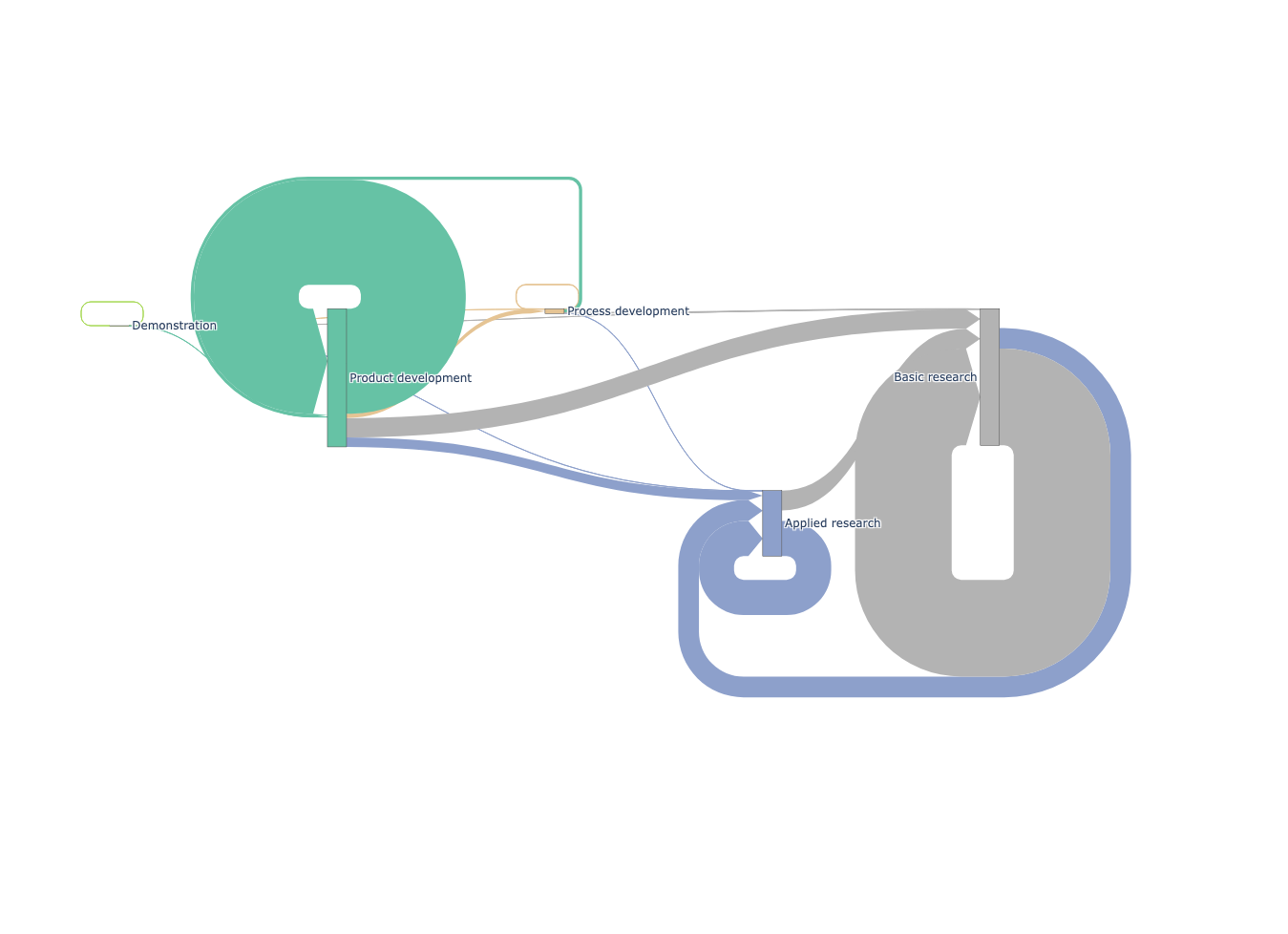}
         \caption{Translation subtypes frequencies represented as knowledge flow between five of the document types. There are too few events related to application, i.e.\ vaccine approval, to be included here. The width of each arrow is proportional to the number of the corresponding events. The height of the nodes is set by the larger of the total in- and out-flows. The flow rates are not normalized to retain information about absolute volume of the various translation subtypes.}
     \end{subfigure}
     \hfill
 		\caption{The number of events against time for different translation subtypes. Illustrative data from AstraZeneca Vaxzeria. See Appendix \ref{app:figures} for full data.
 	}
		\label{fig:inter_kde}    
\end{figure}

A possible explanation for this order is that a threshold amount of basic research translation is the necessary risk reduction for translation from the research to development phase, which involves more private sector entities. The rates of development-related translations might accelerate due to their proprietary nature and the monopoly incentives of patents or because the biomedical community has, at this point, agreed on the problem and solution approaches \cite{RN2211}. \Figref{fig:inter_kde} shows the volume of these translation types are often multi-modal with peaks approximately 20 years apart --- the typical patent expiration --- showing the propensity to translate might be affected by patenting. Translations involving clinical trials are, unsurprisingly, the last to occur because, in our dataset, clinical trials are the last phase before marketing authorizations. Notably, at any given completion rate in \figref{fig:inter_cdf}, the time lag among translations of basic research, applied research, produce development, and process developments are within 15 years. In other words, translations among early innovation phases are cooperative. On the other hand, translations to the demonstration phase and beyond are intrinsically conditional on pharmaceuticals’ decision to conduct clinical trials, typically only emerging 20-30 years after early-stage translations in our data. These differing distributions lead us to ask whether funders have differing translation footprints. Similar longitudinal translation patterns are observed across all eight vaccines (\appref{app:figures}).

\begin{figure}
  \centering
    \includegraphics[width=0.9\textwidth]{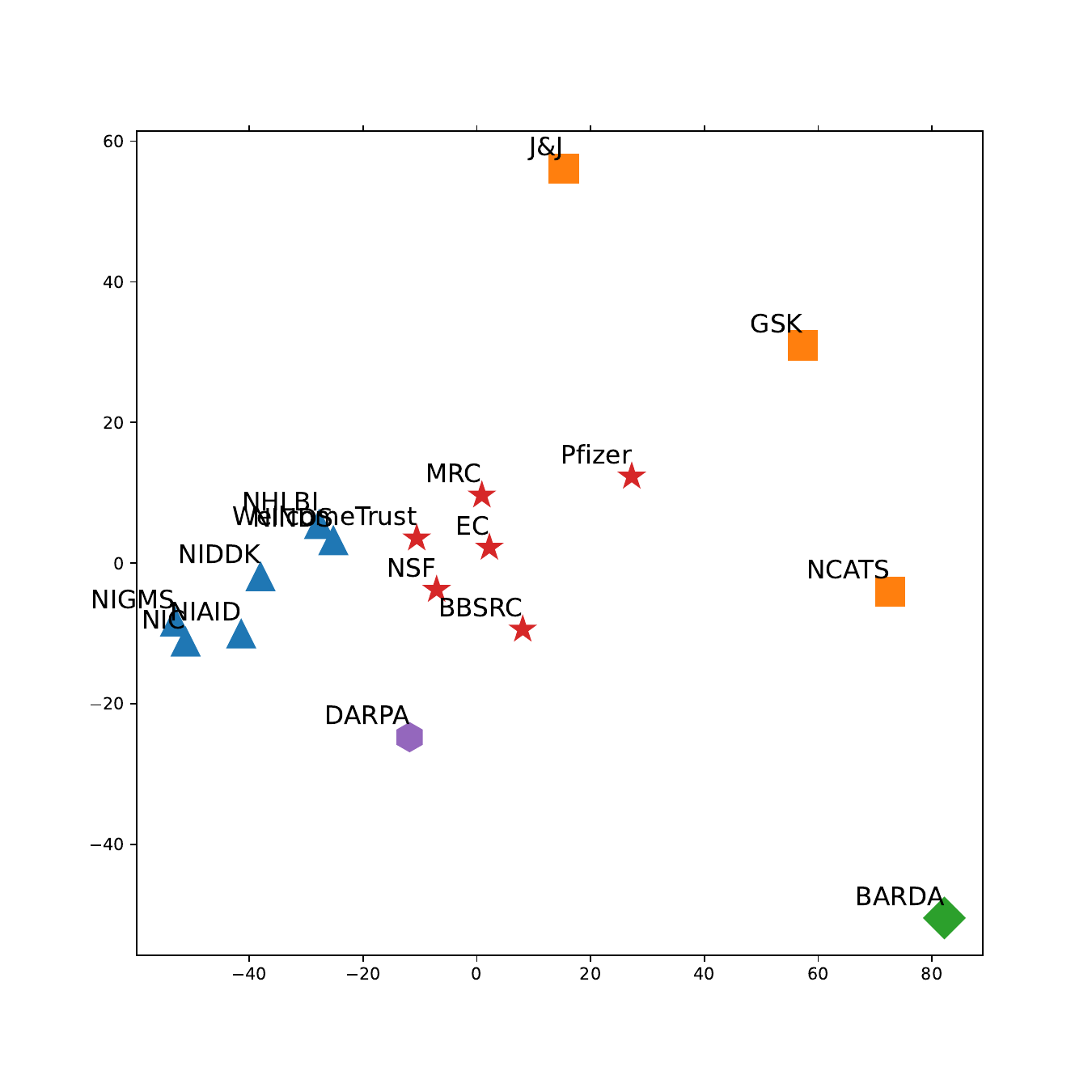}
  \caption{Analysis of similarity between funders based on a cosine distance matrix derived from the fraction of translation and dissemination subtypes per vaccine. For each funder, the fraction of edge subtypes for each vaccine is compared to other funders using cosine similarity to produce a distance matrix. This is used for principal component analysis where coordinates come from the two most significant eigenvalues. The colours and shapes of points correspond to distinct clusters produced using k-means clustering methods with 5 clusters requested.}
  \label{fig:pca}
\end{figure}

Our third result concerns funders. The funders included in our analysis are mostly US funders and a few UK and EU funders, as they most frequently occur in our dataset. Through a principal component analysis (PCA) and k-means clustering of funder-associated translation subtypes (\figref{fig:pca}), 
we found that daughter institutes of the \href{https://www.nih.gov/}{National Institutes of Health} (NIH, a U.S.\ government agency for medical research)
--- 
National Institute of General Medical Sciences (NIGMS), 
National Cancer Institute (NIC), 
National Institute of Allergy and Infectious Diseases (NIAID), 
National Heart, Lung, and Blood Institute (NHLBL), 
National Institute of Neurological Disorders and Stroke (NINDS) 
--- are clustered together. 
This clustering suggests these funders are similar to each other with respect to their translation and dissemination patterns. Another PCA cluster that is close to the NIH cluster is composed of the European Commission (EC), U.K.\ Medical Research Council (MRC), Wellcome Trust, National Science Foundation (NSF), and Biotechnology and Biological Sciences Research Council (BBSRC). There are also funders who are far away from the clusters: Defense Advanced Research Projects Agency (DARPA), Pfizer, GSK, Johnson\&Johnson (J\&J), National Center for Advancing Translational Sciences (NCATS), and Biomedical Advanced Research and Development Authority (BARDA) in ascending distance to any clusters. While being distant from the main NIH cluster, these funders are also dissimilar among themselves in terms of knowledge diffusion. The results from the PCA is also supported by other clustering methods (detailed in Appendix \ref{app:fundersim}).

To our knowledge, this is the first time that the translation behaviors of biomedical, or any type of, RD\&D funders are quantified. The PCA, combined with the first set of results, explains in high-context the: (i) Similarities of basic research funders. It is widely accepted that NIH, NSF, and UKRI (parent institutions of MRC and BBSRC) occupy earlier innovation phases because their remits are to ``seek fundamental knowledge'', ``promote the progress of science'', and support ``research and innovation system''. This belief is now backed up with statistics: when it comes to translation, basic research funders are most involved in basic research-to-product development translation; (ii) Differences between these early innovation phase funders with pharmaceuticals at later innovation phases. Compared with basic research funders, pharmaceuticals are also heavily involved in basic research-to-product development but also in applied research-to-production development translation and applied research-to-basic research feedback; (iii) Differences between early innovation phase funders with mission-oriented funders. It was hypothesized that some biomedical funders, such as NCATS, BARDA and DARPA, are deliberately organized around a wider range of innovation phases to improve the valorization of early-stage discoveries \cite{RN1141,RN448}. We are able to show that these, often called ``mission-oriented'' agencies, are responsible for a large proportion of product development-to-clinical demonstration translation along with a strong involvement in a mix of basic research-to-product development, basic research-to-applied research, and applied research-to-basic research translations. Mission-oriented agencies distinguish themselves by being involved in a wider range of translation types; finally, (iv) Although funders at early innovation phases translate similarly, funders at late innovation phases- the pharmaceuticals and mission-oriented agencies - all translate differently. This firstly suggests biomedical translation requires a high degree of specialization, and secondly implies pharmaceuticals and mission-oriented agencies uniquely allocate their resources (and place their bets) for translations.

Prior statistical studies of research translation called for more comprehensive definitions of knowledge links \cite{RN1918,RN1626}; they provided partial pictures of the translation journey: first-degree citations between patent and research \cite{RN1919,RN212,RN1626}, drug and research \cite{RN211}, or drug and patent \cite{RN1937}. This study builds on this request by accounting for the permutations between innovation phases to cover biomedical translation journeys more completely. In addition, the linking of publication, patent, clinical trial, grant, funder, and therapeutic data from multiple sources enabled analysis of biomedical funders beyond the NIH. The results show statistically what it means to be a ``mission-oriented'' funder and the unique ways in which pharmaceuticals valorize, successfully or unsuccessfully, biomedical knowledge through translation.

Although our analysis is a more complete articulation of biomedical translation, there are important limitations and assumptions. In terms of data, we assume publications represent research, patents represent development, and clinical trials represent demonstration. Furthermore, documents vary by citation practices and constraints. For instance, patents need to strike a balance between minimizing citations to demonstrate novelty and citing enough to not infringe \cite{RN1710}; translational science, such as bioinformatics and clinical trial designs, are embedded in multiple innovation journeys but not often acknowledged \cite{RN1365}. Despite variations in citation practices, we do not ``normalize'' or ``correct'' the numbers of translation types because, firstly, regardless of document type, each citation factually represents a minimum viable increment of novelty. Secondly, taking away the absolute frequency of translation subtypes would conceal the deficiency in some of them.
In terms of data classification, the categorization of research into ``basic'' and ``applied'' and patents into ``product'' and ``process'' remains open-ended. This study relied on Australia’s and UK’s national research classification systems (Appendix \ref{app:pub}). We also improved an existing method \cite{RN794} to classify patents into product and process developments. In terms of scope, the present study does not include data on other innovation phases such as manufacturing. In addition, the analysis is based only on backward citations from vaccines. Given a large part of Operation Warp Speed emphasized on manufacturing readiness \cite{RN350} and that therapeutics go beyond vaccines \cite{RN1947}, future analysis of biomedical translation should also consider drugs, medical devices, other biologics and classes of translational science as well as including manufacturing indicators. Future studies may also correlate the percentage distribution of translation subtypes (\figref{fig:inter_kde}) with RD\&D efficiency metrics to inform biomedical resource allocations; and also use the time lag between dissemination and translation (\figref{fig:simple_cdf}) to classify or diagnose emerging technologies.

Despite these limitations, the novelty of this proof-of-concept study is the linking of multiple types of innovation data to measure biomedical translation. The broad implication is that the timing, magnitude, and combination of translation funding may affect the capability to innovate. For countries or firms without the breadth to participate in all phases of innovation, our analysis illuminates the complementary/collaborative innovators to pursue.

Using vaccine datasets, the results in this study represent a first step to realize this vision of quantifying translation. The study stylized several new facts about biomedical translation. First, translation most frequently occurs in the early innovation phases and overall proceeds in the direction from research, to development and demonstration albeit permeated by knowledge feedbacks. Second, translation is sigmoidal over time and that various subtypes of translations are cooperative at early stages of vaccine innovation. Third, basic research funders, such as NIH daughters and NSF, fund similar types of translations whereas mission-oriented funders and pharmaceuticals fund a wider range of translation subtypes differently. These results, when viewed in terms of proportion, are concordant with the linear model of science; when viewed longitudinally, resonate with the chain-linked model of science.

\definecolor{TSEazure}{RGB}{1, 169, 244}
% #1e41e7 Palatinate Blue
\definecolor{TSEPalatinateBlue}{RGB}{30, 65, 231}
% #a9e7f0 Blizzard Blue
\definecolor{TSEBlizzardBlue}{RGB}{169, 231, 240}
% #3c96b4 Glacial Blue Ice
\definecolor{TSEGlacialBlueIce}{RGB}{60, 150, 180}
\begin{figure}
  \centering
    \includegraphics[width=0.9\textwidth]{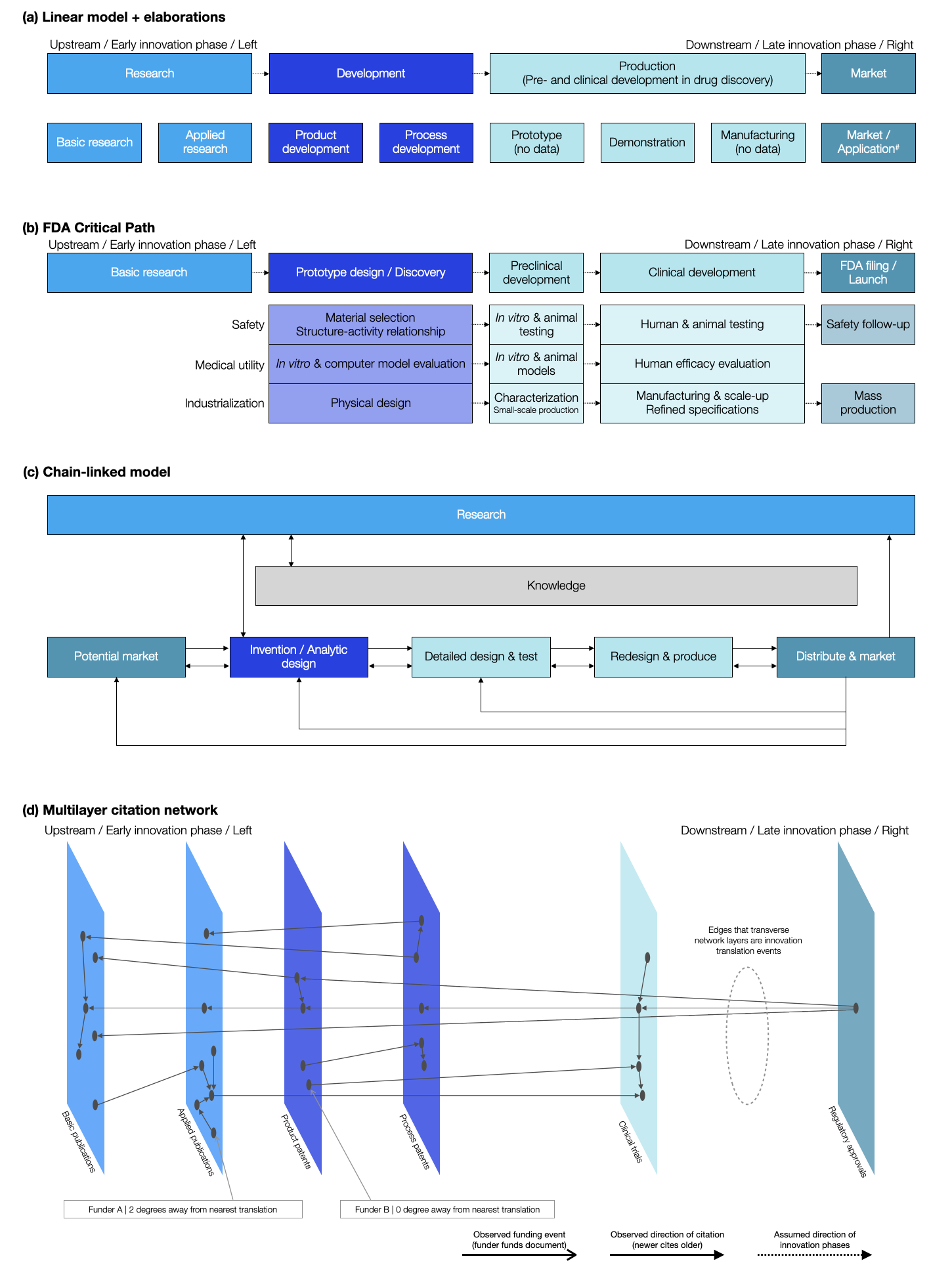}
    \\
    %\framebox[width][pos]
    \raisebox{8cm}[0pt][0pt]{
    	\fbox{\tiny
    		\begin{tabular}{rc}
    			%\textbf{\color{TSEazure} research} 
    			\textbf{research}    &  \textbf{\color{TSEazure} $\bullet$} \\ % (basic then applied), 
    			%\textbf{\color{TSEPalatinateBlue} development} 
    			\textbf{development} & {\color{TSEPalatinateBlue} $\bullet$}  \\ %(product then process), 
    			%\textbf{\color{TSEBlizzardBlue} production} 
    			\textbf{production}  & {\color{TSEBlizzardBlue} $\bullet$} \\
    			%\textbf{\color{TSEGlacialBlueIce} market} 
    			\textbf{market}      & {\color{TSEGlacialBlueIce} $\bullet$} \\
    		\end{tabular}		
    	}% fbox
    }% raisbox
  \caption{Correspondence among Linear innovation model \cite{B45,godin2006linear}, FDA Critical Path \cite{FDA04,RN84}, Chain-linked innovation model \cite{RN285} \& Multi-layer network \cite{RePEc:arx:papers:2302.13076}. 
  	{\small
  	(a) The linear model \cite{B45,godin2006linear} has been used since 1945 and provides a benchmark to study the linearity and nonlinearity of innovation; 
  	(b) The FDA Critical Path \cite{FDA04,RN84}, proposed in 2004, can be considered a tailored version of (a) for drug discovery; 
  	(c) The chain-linked model \cite{RN285} is an elaboration of (a) from 1986. The thickness and chronological order of arrows in (c) were unquantified before this study; 
  	(d) A network model proposed by this study to quantify the arrows (i.e.\ translation) in (a)--(c).   
  	%\cite{B45,RN285,RN84,RN1993}.
  	The colour of each component is consistent across all four diagrams and correspond to different phases in the innovation process. 
  	Likewise, the horizontal position of elements in (a), (b) and (d) is consistent with the six stages in the innovation process.
  	From left to right in (a) (b) and (d) we have:
  	%\textbf{\color{TSEazure} research} 
  	\textbf{research} \textbf{\color{TSEazure} $\bullet$} (basic then applied), 
  	%\textbf{\color{TSEPalatinateBlue} development} 
  	\textbf{development} {\color{TSEPalatinateBlue} $\bullet$} (product then process), 
  	%\textbf{\color{TSEBlizzardBlue} production} 
  	\textbf{production} {\color{TSEBlizzardBlue} $\bullet$}, 
  	%\textbf{\color{TSEGlacialBlueIce} market} 
  	\textbf{market} {\color{TSEGlacialBlueIce} $\bullet$}. 	
   } % end of change in font size
  } % end of caption
\label{fig:framework}
\end{figure}

\clearpage

% BBBBBBBBBBBBBBBBBBBBBBBBBBBBBBBBBBBBBBBBBBBBBBBBBBBBBBBBBBBBBB
%
% BIBLIOGRAPHY
%
%\bibliography{plos_ref}
\bibliography{innovation}

%
% End of bibliography
%
% BBBBBBBBBBBBBBBBBBBBBBBBBBBBBBBBBBBBBBBBBBBBBBBBBBBBBBBBBBBBBB

\section*{Acknowledgements}
M.H. and E.O. are financially supported by the Gatsby Charitable Foundation (grant NMZM/325). All authors contributed to the design of the research. M.H. and H.C.W.P. obtained the data. H.C.W.P., T.S.E., and M.H. analyzed the data. M.H. wrote the original draft, but all authors contributed to subsequent revisions. The authors declare no competing interests. All data are available in the manuscript or the supplementary materials.

\section*{Supplementary materials}
Appendices A to G

% AAAAAAAAAAAAAAAAAAAAAAAAAAAAAAAAAAAAAAAAAAAAAAAAAAAAAAAAAAAAAA

\clearpage

%%% The text of the appendices is in a separate file,
%%% The command below reads in that text here to produce
%%% a single document suitable for preprints (arXix) 
%%% and initial submission to journals
%%% This is the file containing the text of the appendices/supplementary information

%%% To create a single file with main text and appendices just include the line
%%%   \input{innovationTranslationAppendix}
%%% at the end (just before the \end{document}) of the main file.

%%% To get these as a separate pdf file create a new LaTeX file
%%% (e.g. as needed for a supplementary mateiral file for a journal)
%%% that just reads in this file.

% If you want an appendix then I found we need to work harder to get hyperef to work properly
% APPENDIX
\begin{center}
\Large\textbf{Appendices}
\end{center}
%\section*{\Large APPENDIX}
%\section*{\Large Supplementary Information.}
\appendix
\renewcommand{\thesection}{\Alph{section}}
\renewcommand{\theequation}{\thesection\arabic{equation}}
\renewcommand{\thefigure}{\thesection\arabic{figure}}
\renewcommand{\thetable}{\thesection\arabic{table}}
\numberwithin{equation}{section}
\numberwithin{figure}{section}
\numberwithin{table}{section}
\setcounter{section}{0}
% Need to reset Hyperref counters too
% https://tex.stackexchange.com/questions/71162/reset-section-numbering-between-unnumbered-chapters
\renewcommand{\theHsection}{\Alph{section}}
\renewcommand{\theHequation}{\thesection\arabic{equation}}
\renewcommand{\theHfigure}{\thesection\arabic{figure}}
\renewcommand{\theHtable}{\thesection\arabic{table}}

% *************************************************************************
\section{Glossary}
\label{app:glossary}

\begin{tabular}{r c p{0.7\textwidth}}
Abbreviation & Country & Description \\ \hline
       BARDA & USA & \href{https://www.medicalcountermeasures.gov/barda/}{Biomedical Advanced Research and Development Authority} 
                  is a US government agency that develops medical countermeasures that address the public health including emerging infectious diseases and the medical consequences of serious accidents, incidents and attacks.\\
       BBSRC & UK  & \href{https://www.ukri.org/councils/bbsrc/}{Biotechnology and Biological Sciences Research Council}. Part of the UKRI.  \\
       DARPA & USA &  \href{https://www.darpa.mil/}{Defense Advanced Research Projects Agency} of the U.S.A. invests 
	              in breakthrough technologies for U.S. security\\
          EC & EU  & European Commission \\
	     EMA & EU  & \href{https://www.ema.europa.eu/en}{European Medicines Agency}, authorises medicines in the European Union. \\
	     FDA & USA & \href{https://www.fda.gov/}{U.S. Food and Drug Administration}, authorises medicines in the U.S.A. \\
         GSK &     & multinational drug company \\
	     IQR &     & interquartile range, difference of 75\% and 25\% quartile values. \\
        J\&J &     & Johnson\&Johnson, multinational drug company \\	
         MRC & UK  & \href{https://www.ukri.org/councils/mrc/}{U.K. Medical Research Council}. Part of the UKRI. Funds research at the forefront of science to prevent illness,  develop therapies and improve human health.  \\
       NCATS & USA & \href{https://ncats.nih.gov/}{National Center for Advancing Translational Sciences} aims to transform scientific discoveries into new treatments and cures for disease that can be delivered faster to patients. Part of NIH.\\
       NIAID & USA & \href{https://www.niaid.nih.gov/}{National Institute of Allergy and Infectious Diseases}, part of NIH. \\
         NIC & USA & \href{https://www.cancer.gov/}{National Cancer Institute}, part of NIH. The principal US government agency for cancer research and training.\\
       NIGMS & USA & \href{https://www.nigms.nih.gov/}{National Institute of General Medical Sciences}, part of NIH. \\
         NIH & USA & \href{https://www.nih.gov/}{National Institutes of Health}, a U.S.\ government agency for medical research.\\
       NHLBL & USA & \href{https://www.nhlbi.nih.gov/}{National Heart, Lung, and Blood Institute}, a U.S.\ government agency for prevention and treatment of heart, lung, and blood disorders. \\
       NINDS & USA & \href{https://www.ninds.nih.gov/}{National Institute of Neurological Disorders and Stroke}, part of NIH. \\
         NSF & USA & \href{https://www.nsf.gov/}{National Science Foundation}, a U.S.\ government agency that supports science and engineering. \\
      Pfizer &     & \href{https://www.pfizer.com/}{Pfizer} is an American multinational pharmaceutical and biotechnology company.\\
         PCA &     &  principal component analysis, standard machine learning method. \\
     {RD\&D} &     &  research, development and demonstration. \\
   Wellcome  & UK  & \href{https://wellcome.org/}{Wellcome} is a global charitable foundation based in the UK funding curiosity-driven research in climate change, infectious disease and mental health. \\
        UKRI & UK  & \href{https://www.ukri.org/}{UK Research and Innovation}, a UK government agency which aims to create an outstanding research and innovation system in the UK. \\ 
\end{tabular}
% *************************************************************************
\section{Research translation vs translational science}
\label{app:translation}

The phrases  ``translation'' and ``translational science''  are used to describe different processes and it is important to clarify what this study measures. \tsedef{Translation} is the process of rendering discoveries, inventions, technologies, and innovation useful for certain end-users. Translation has been defined in the biomedical community as ``the process of turning observations in the laboratory, clinic, and community into interventions that improve the health of individuals''  \cite{RN1365}; ``A process of knowledge generation and transfer that enables those utilising the developed knowledge to apply it… knowledge flows can be multidirectional and non-sequential''  \cite{RN1703}; and ``The inclination of biomedical researchers to ultimately help patients''  \cite{RN1936}. 
On the other hand, \tsedef{translational science} are the scientific principles and technologies that facilitate translation: ``The field of investigation focused on understanding the scientific and operational principles underlying each step of the translational process'' \cite{RN1364} and the ``methods and tools to facilitate and establish scientific backbone of the translation process'' \cite{RN1936}. Examples of  translational science include biomarkers, clinical trial protocols, bioinformatics, and drug manufacturing standards. 
We specifically investigate translation in this study. To allow consistent and transferable accounting of translation, we consider the arrows between innovation phases (\figref{fig:framework}) as translation in this study.

\section{Innovation as a multi-layer citation network}
\label{app:network}

In the lens of complexity science, the linear model of science (\figref{fig:framework}) \cite{B45,RN1705} that has been in use in the innovation policy community for decades can be considered a directed acyclic graph because it is a network in which connections have directions and cannot form loops. The chain-linked model, on the other hand, is an undirected cyclic graph for opposite reasons. Formally, a network (or graph) is a set of nodes, and pairs of nodes can be connected by an edge (or arrow). In our networks, each node represents a single document which is one of six types: an innovation outcome represented by FDA/EMA regulatory authorisation, a clinical trial, a patent representing either product or process development \cite{RN837}, or an academic publication representing either basic or applied science \cite{B45}. So, our networks are examples of what are called multi-layer networks, as each type of node can be visualised as placed on a different layer. We grow the networks for up to three degrees following the method of \cite{RePEc:arx:papers:2302.13076}. Each arrow within the layers is considered a knowledge dissemination event; an arrow traversing across layers is considered a knowledge translation event.

% ******************************************
\section{Vaccine data set}
\label{app:data}

This study uses data on vaccine development because it contains clear translation linkages among publications, patents, clinical trials, and regulatory authorizations. In addition, the concept of translation is most established in the biomedical community. We have access to the data on three types of document: clinical trials from \urlname{ClinicalTrials.gov}; patents from \urlname{Lens.org}~\cite{RN1697}; and publication data from \urlname{Dimensions.ai}~\cite{RN1696}. All three data sources contain citation data. For example, each US patent contains ``prior publication data'' and ``related U.S. application data'', which allows us to form patent-to-publication and patent-to-patent citations in our network respectively.  The full dataset of this study is described by \tabref{tab:empirical_data} and \tabref{tab:descriptive_stats} and originates from \cite{RePEc:arx:papers:2302.13076} where a full description of the process used to obtain and clean the data can be found.

\begingroup
\renewcommand{\arraystretch}{1.5}
\begin{table}[!ht]
	\centering
	\small
	\caption{\label{tab:empirical_data}Information on the vaccines analysed here. The number of nodes and edges are those present in the multilayer citation network created from a multi-step snowball sample starting from the vaccine approval document.
	}
	
	\begin{tabular}{lllllll}
		\hline
		\textbf{Vaccine}          &\textbf{Technology}  & \textbf{Disease} & \textbf{Developer} & \textbf{Year first} & \textbf{Source} & \textbf{Data}\\
		                          & \textbf{platform}   & \textbf{targeted} & & \textbf{approved} &\textbf{node} & \textbf{source}\\
		\hline
		Spikevax  & \multirow{2}{*}{mRNA} & \multirow{3}{*}{COVID-19}& Moderna &2020 & \cite{RN1723} & \cite{RN1723,RN1857,RN1858}\\
		\cline{1-1}  \cline{4-7}
		Comirnaty & &&BioNTech& 2020& \cite{RN1724} & \cite{RN1724, RN1859, RN1860} \\\cline{1-2}  \cline{4-7}
		
		Vaxzeria   & \multirow{2}{*}{Viral Vector}  && AstraZeneca& 2020&\cite{RN1722} & \cite{RN1722, RN1861}\\ \cline{1-1} \cline{3-7}
		
		Zabdeno  & & Ebola & Janssen &2020& \cite{RN1721} & \cite{RN1721}\\
		\hline
		Dengvaxia  & \multirow{2}{*}{Live Attenuated} & Dengue & Sanofi Pasteur&2019&\cite{RN1725} & \cite{RN1725,RN1862} \\
		\cline{1-1} \cline{3-7}
		Imvanex &&Smallpox&Bavarian Nordic&2013& \cite{RN1726} & \cite{RN1726,RN1863,RN1864}\\
		\hline
		Nuvaxovid & \multirow{2}{*}{Subunit}  &COVID-19&Novavax&2022&\cite{RN1650} & \cite{RN1650, RN1865, RN1866}  \\
		\cline{1-1} \cline{3-7}
		Shingrix  && Shingles &GSK&2017&\cite{RN1728} &\cite{RN1728,RN1867}\\
		\hline
		
	\end{tabular}
	
\end{table}

\endgroup

\begin{table}[!ht]
	\centering
	\small
	\caption{\label{tab:descriptive_stats}Basic network properties of the eight vaccine networks.
	}

	\begin{tabular}{lllllll}
		\hline
		\textbf{Vaccine}        && \multicolumn{2}{c}{\textbf{Nodes}}  &&  & \multirow{2}{*}{\textbf{Edges}}\\
		\textbf{network}           & \textbf{Publication}   & \textbf{Patent} & \textbf{Clinical trials}  & \textbf{Funders} & \textbf{Grants} &\textbf{} \\
		\hline
		Spikevax  & 62,112 & 24,407 &10  & 1,286 & 25,043 & 786,563 \\
		
		Comirnaty & 37,383 & 8,127 & 76  & 1,289 & 18,744 & 340,161  \\
		
		Vaxzeria   &  58,210 & 32,367 & 5  & 1,274 & 21,528 & 648,877 \\
		
		Zabdeno  & 77,359 & 47,145 & 9 & 1,371 & 27,561 & 953,002 \\
		
		Dengvaxia  & 9,986  & 2,681 & 30  & 505 & 2,079 & 81,716 \\
		
		Imvanex & 38,979 & 5,298 & 24  & 922 & 13,129 & 357,320\\
		
		Nuvaxovid & 13,855 &  1,348 &  4  & 924 & 7,547 & 104,182 \\
		
		Shingrix  & 12,987 & 6,993 & 22 & 753 & 6,288 & 174,881 \\
		\hline

	\end{tabular}
	
\end{table}

% --------------------------------------------------------------------------
\subsection{Classifying publications into basic and applied research}
\label{app:pub}

Publication data from \urlname{Dimensions.ai}~\cite{RN1696} are classified by their Field of Research (FoR) in the Australian and New Zealand Standard Research Classification, Broad Research Areas (BRA) in Australia's National Health and Medical Research Council, Research Activity Classifications (RAC) in UK's Health Research Classification System. We classify a publication as \tsedef{basic research} if its FoR classification is ``Biological sciences'', its BRA is ``Basic science'', or RAC group is ``Underpinning research activity''. Other publications, whose FoR categories are chiefly ``Biomedical and clinical sciences'', are classified as \tsedef{applied research}.

% --------------------------------------------------------------------------
\subsection{Classifying patents into product and process development}
\label{app:pat}

\tsedef{Product innovation} is ``a new technology or combination of technologies introduced commercially to meet a user or a market need''; \tsedef{Process innovation} is an innovation in the way an organization conducts its business, often oriented to improve the effectiveness and efficiency of production \cite{RN837}. We adopt and improve the patent claim classification process by Banholzer et al. \cite{RN794} which searches for process and product innovation keywords from a patent's claims, followed by the overall classification of a patent into process or product based on the ratio of process-to-product claims. In the original 2019 study \cite{RN794}, Enlgish keywords for product claims were: 

\begin{quote}
	device, machine, material, tool, apparatus, compound, composition, substance, article, devices, machines, materials, tools, apparatuses, compounds, compositions, substances, articles.
\end{quote}
By way of comparison, the keywords for process claims in the same study are:
\begin{quote}
	 method, process, procedure, use, utilisation, utilization, usage, methods, processes, procedures, uses, utilisations, utilizations, usages.

\end{quote}  
First, we retrieve all the claims of each patent in the network from \urlname{Lens.org}~\cite{RN1697}. 
Second, we cleaned the text and removed stop words from both \cite{RN794} and \texttt{NTLK} from the patent claims.
Third, we compare the patent claims against a list of process development keywords provided above \cite{RN794}: if there is an overlap between the patent claim and the list of keywords, we assign a value \texttt{TRUE} to the claim and \texttt{FALSE} if not. Upon inspection of the patent classification, we identified the need to further stratify process claims into production-related or production-unrelated processes because processes can either be related to a product's production process or the process of using the product by end-users. Hence, we repeated the third step against a list of production process keywords: 
\begin{quote}
	produce, producing, purifying, purify, stabilize, stabilise, stabilizing, stabilising, isolate, isolation, isolating, analyzing, analyze, analyse, analysing, forming, form, remove, removing, prepare, preparing, making, make, link, linking, separate, separating, process, manufacture, manufacturing, coat, coating, solubilize, solubilizing, solubilise, solubilising, synthesizing, synthesize, construct, constructing, optimize, optimise, optimizing, optimising, convert, converting.
\end{quote}
If a patent claim is \texttt{TRUE} for the process keywords and \texttt{TRUE} for production process keywords, it is classified as a production process (PP) claim; if a patent claim is \texttt{FALSE} for process keywords, it is classified as a product (P) claim; if a patent claim is \texttt{TRUE} for process keywords and \texttt{FALSE} for production process keywords, it is classified as a non-production process (NPP) claim. Finally, for each patent, we check for the fraction of claims classified as P, PP, and NPP. The overall patent is classified as product development, if the largest proportion is either P or NPP, and process development if the largest proportion is PP. 

A downside of further classifying patents and publications is that we lose between 2.8\% and 6.6\% of edges in the vaccine networks because some edges are not labelled with classification or claim data.

% **************************************************************

\section{S-Curves}
\label{app:scurve}

\begin{figure}[hbt]
	\centering
		\begin{subfigure}[b]{0.24\textwidth}
		\centering
		\includegraphics[width=\textwidth]{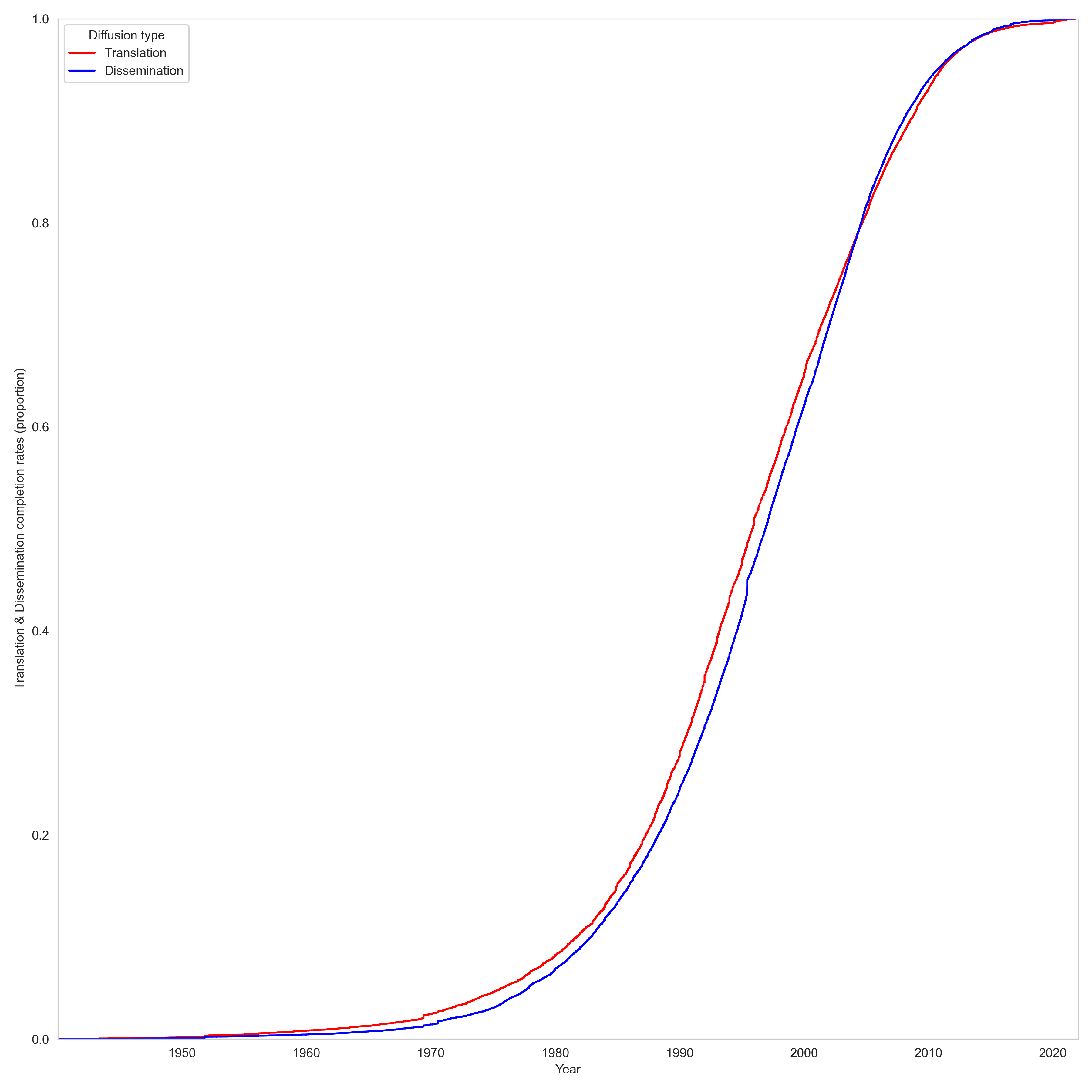}
		\caption{Zabdeno, Ebola, Janssen, 2020,\\AVV}
		%  \label{fig:h_v_d_zabdeno}
	\end{subfigure}
	\hfill
	\begin{subfigure}[b]{0.24\textwidth}
		\centering
		\includegraphics[width=\textwidth]{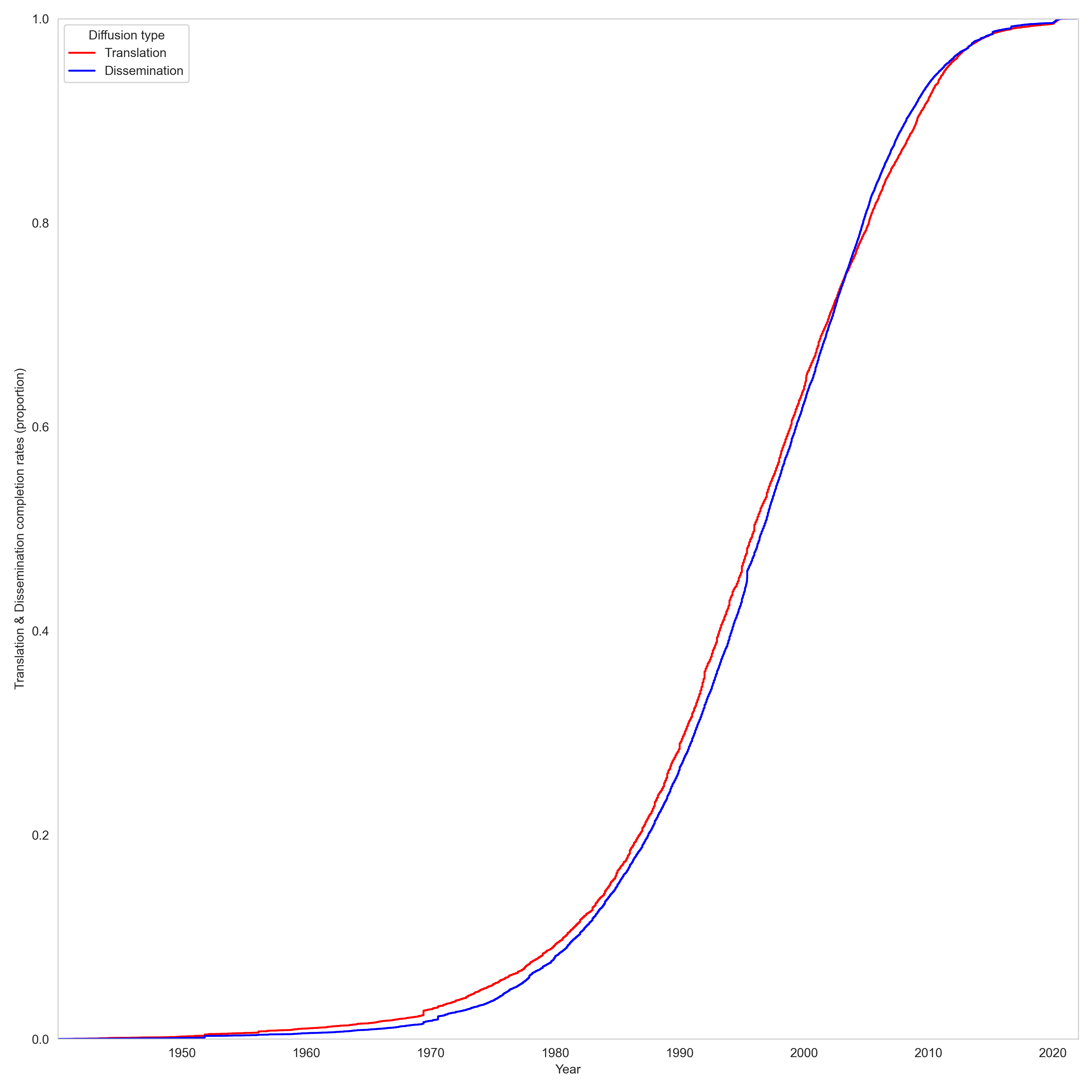}
		\caption{Vaxzevria, COVID-19, AstraZeneca, 2020, AVV}
		%  \label{fig:h_v_d_vaxzevria}
	\end{subfigure}
	\hfill
	\begin{subfigure}[b]{0.24\textwidth}
		\centering
		\includegraphics[width=\textwidth]{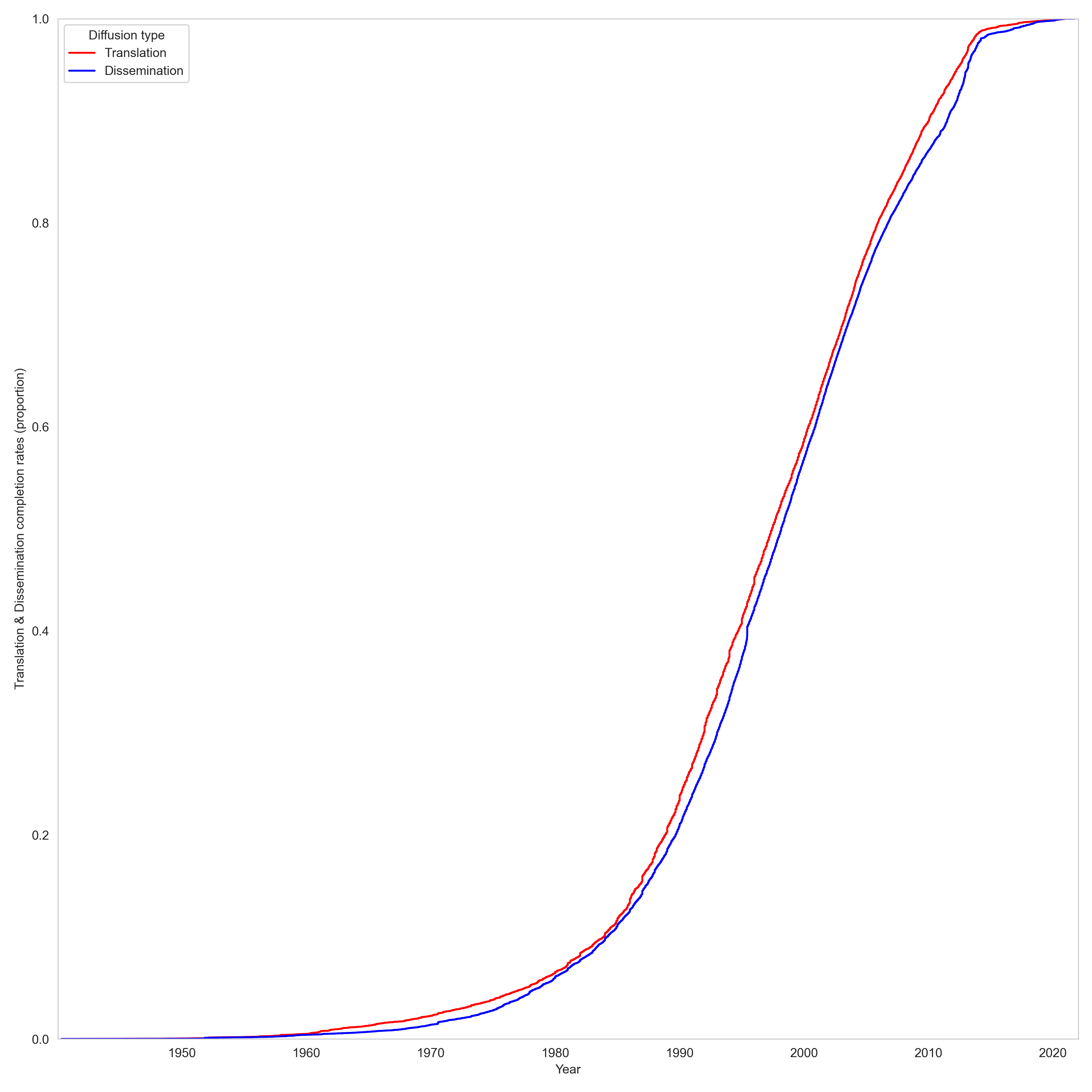}
		\caption{Spikevax, COVID-19, Moderna, 2020, mRNA}
		%  \label{fig:h_v_d_spikevax}
	\end{subfigure}
	\hfill
	\begin{subfigure}[b]{0.24\textwidth}
		\centering
		\includegraphics[width=\textwidth]{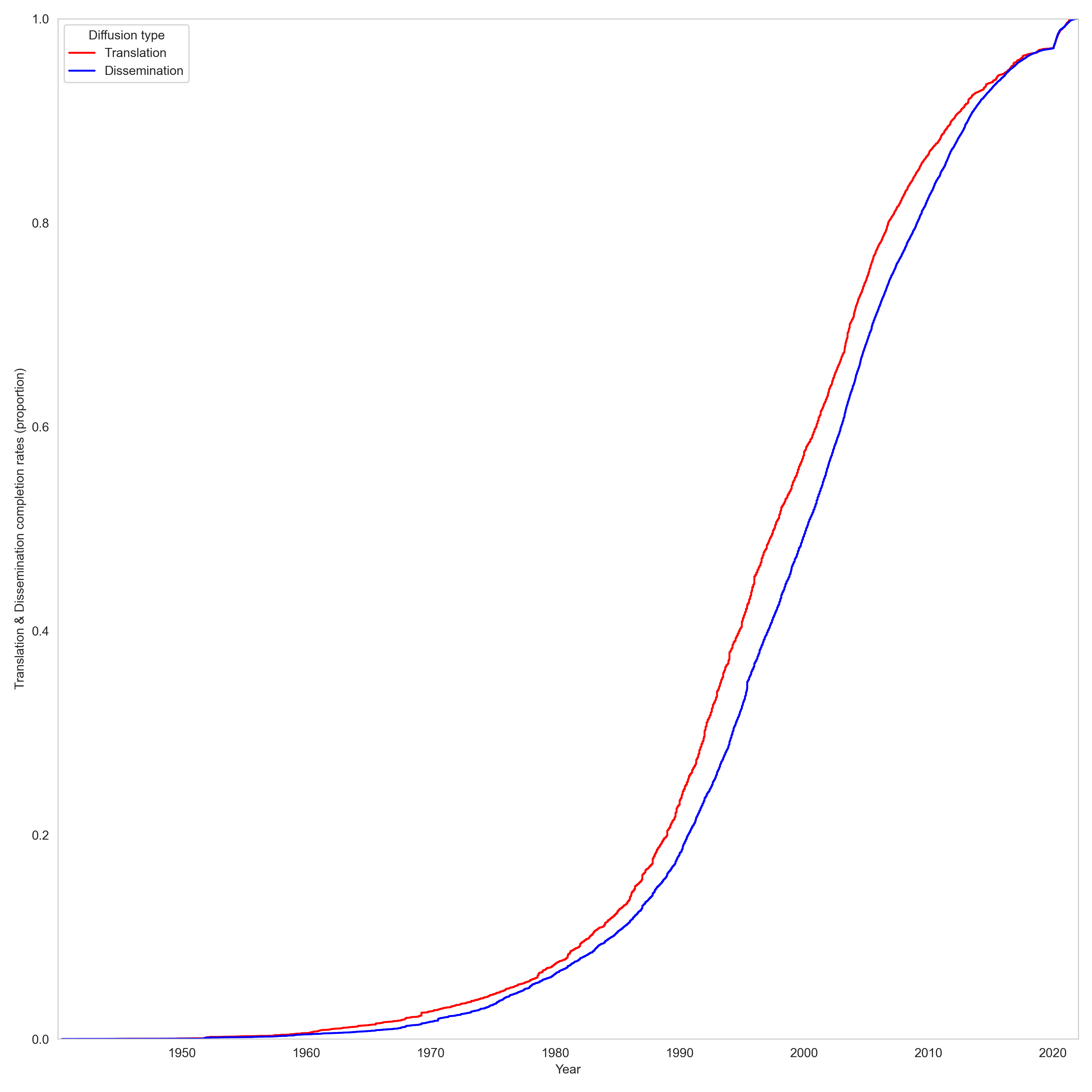}
		\caption{Comirnaty, COVID-19, BioNTech/Pfizer, 2020, mRNA}
		%  \label{fig:h_v_d_comirnaty}
	\end{subfigure}
	\hfill
	\begin{subfigure}[b]{0.24\textwidth}
		\centering
		\includegraphics[width=\textwidth]{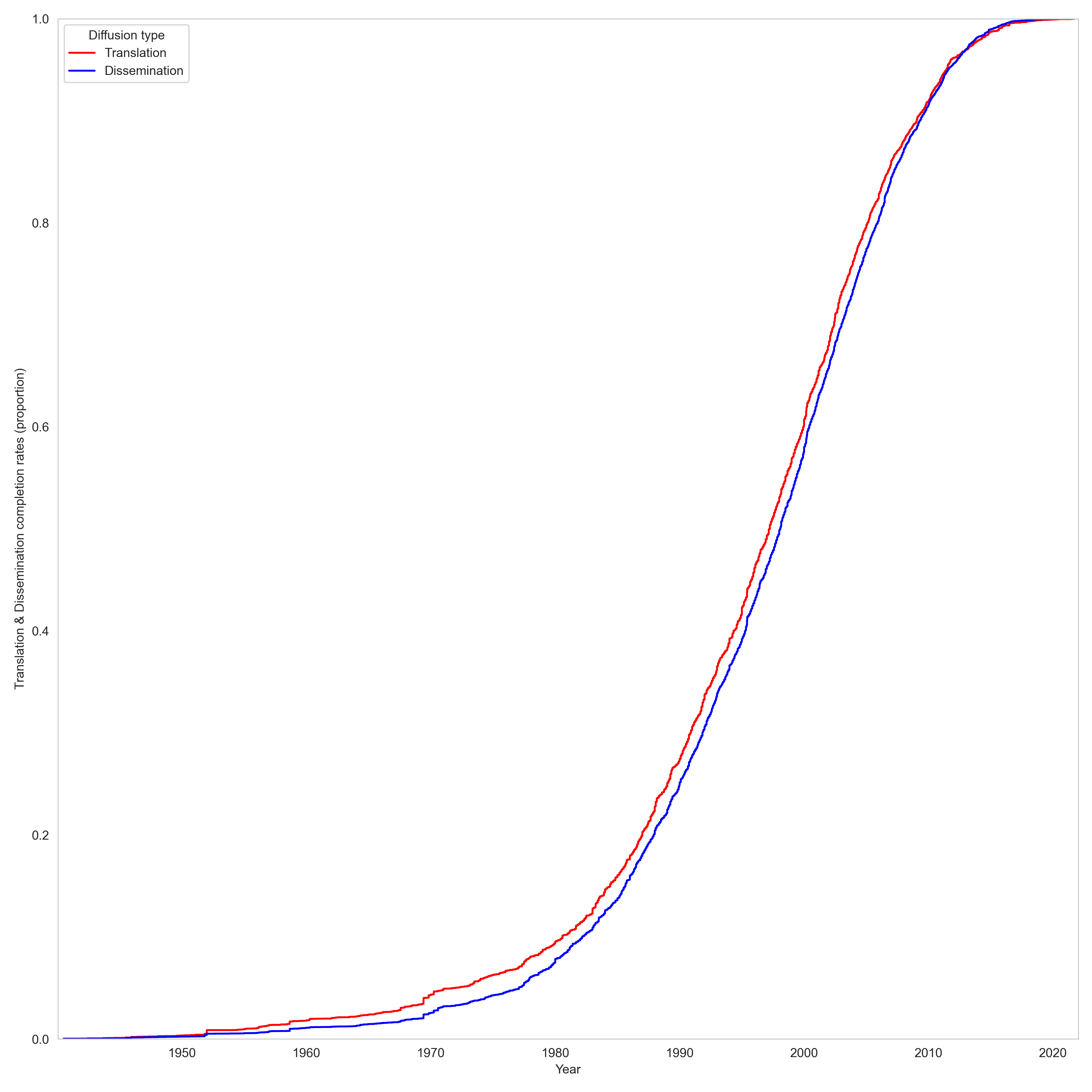}
		\caption{Dengvaxia, Dengue, Sanofi, 2019,\\ WPV}
		%  \label{fig:h_v_d_dengvaxia}
	\end{subfigure}
	\hfill
	\begin{subfigure}[b]{0.24\textwidth}
		\centering
		\includegraphics[width=\textwidth]{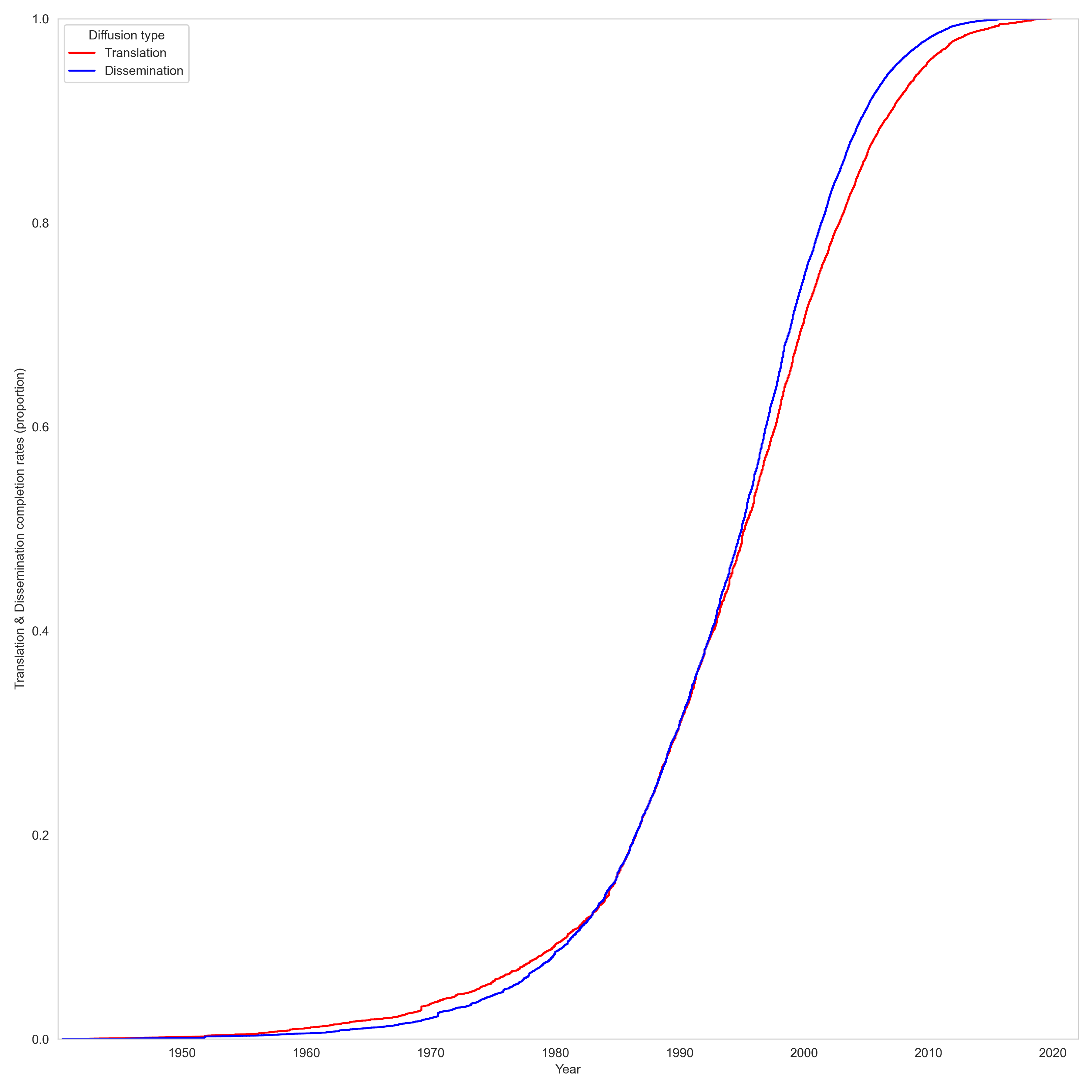}
		\caption{Imvanex, Smallpox, Bavarian Nordic, 2013, WPV}
		%  \label{fig:h_v_d_imvanex}
	\end{subfigure}
	\hfill
	\begin{subfigure}[b]{0.24\textwidth}
		\centering
		\includegraphics[width=\textwidth]{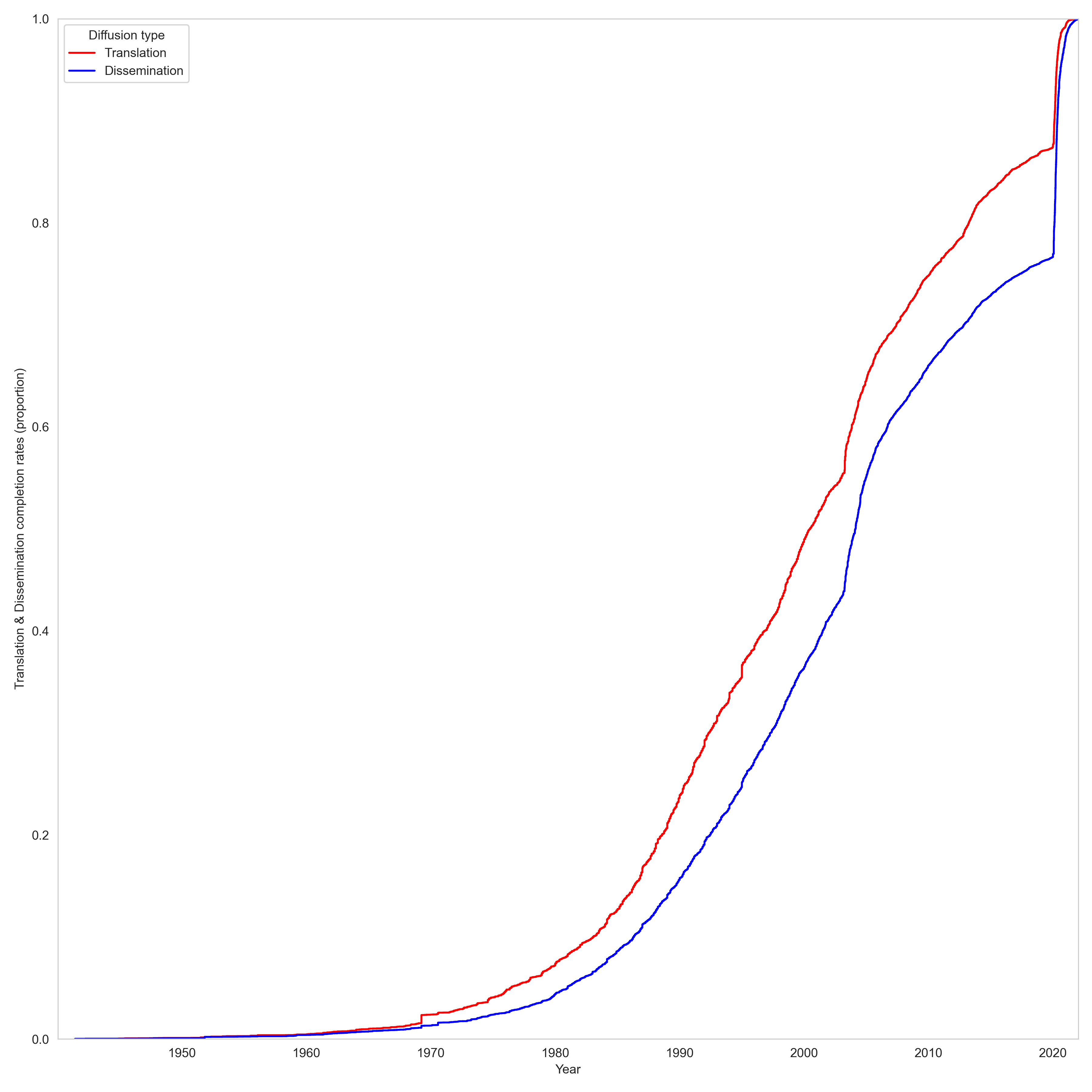}
		\caption{Nuvaxovid, COVID-19, Novavax, 2022, subunits}
		%  \label{fig:h_v_d_nuvaxovid}
	\end{subfigure}
	\hfill
	\begin{subfigure}[b]{0.24\textwidth}
		\centering
		\includegraphics[width=\textwidth]{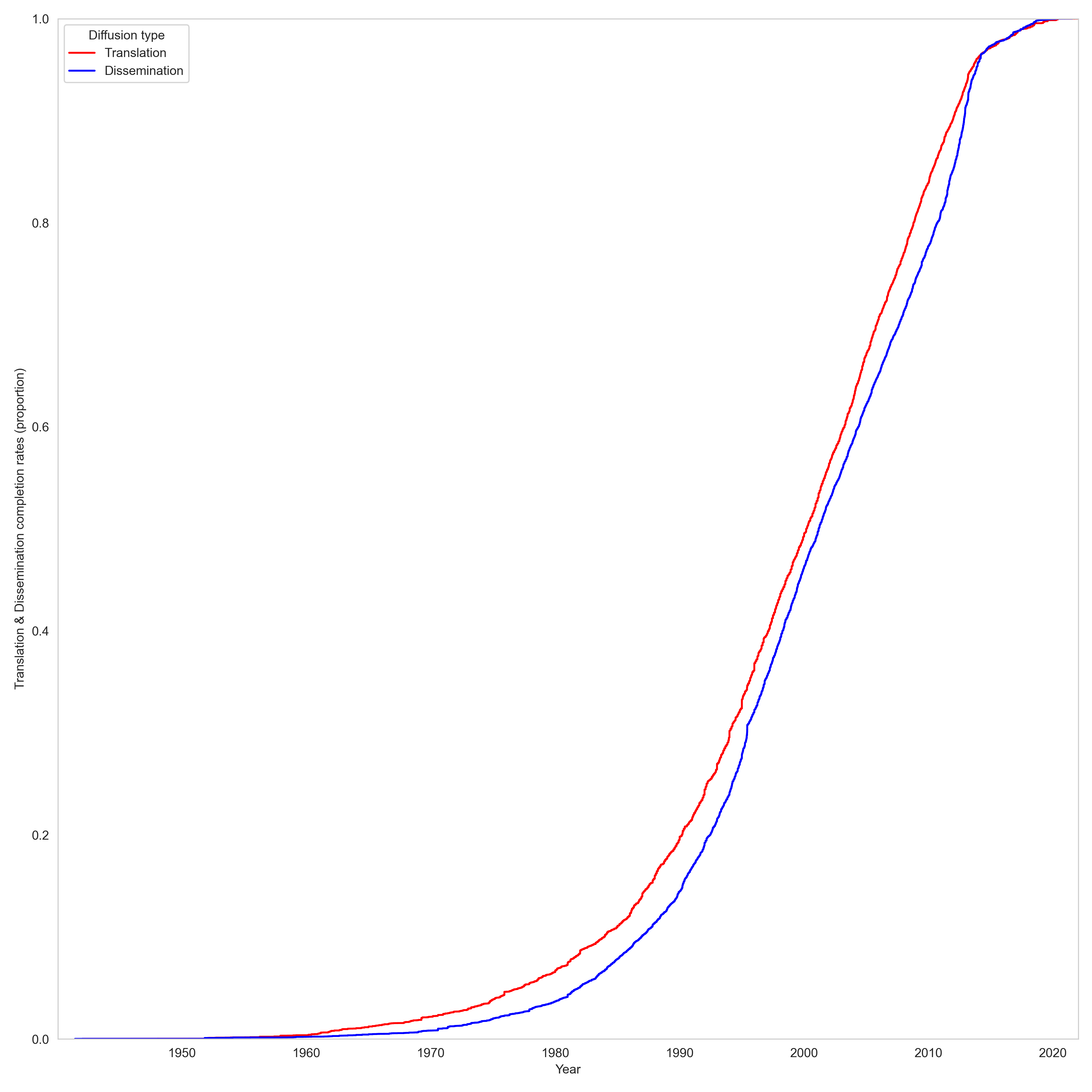}
		\caption{Shringrix, Shingles, GSK, 2017,\\ subunits}
		%  \label{fig:h_v_d_shringrix}
	\end{subfigure}
	\hfill
	\caption{Translation \& dissemination completion rates $f(t)$. We look at the fraction of events that have occurred up to time $t$. Here, an event is the citation of the given type of an older document where the date of the newer document defines the time of such a translation or dissemination event. The events are  (i) a translation event, which is a citation between different layers (categories) of documents, or (ii) a dissemination event, which is a citation within one layer.} 
	\label{fig:simple_cdf}
\end{figure}

\begin{figure}[htb]
	\centering
	\begin{subfigure}[t]{0.48\textwidth}
		\centering
		\includegraphics[width=\textwidth]{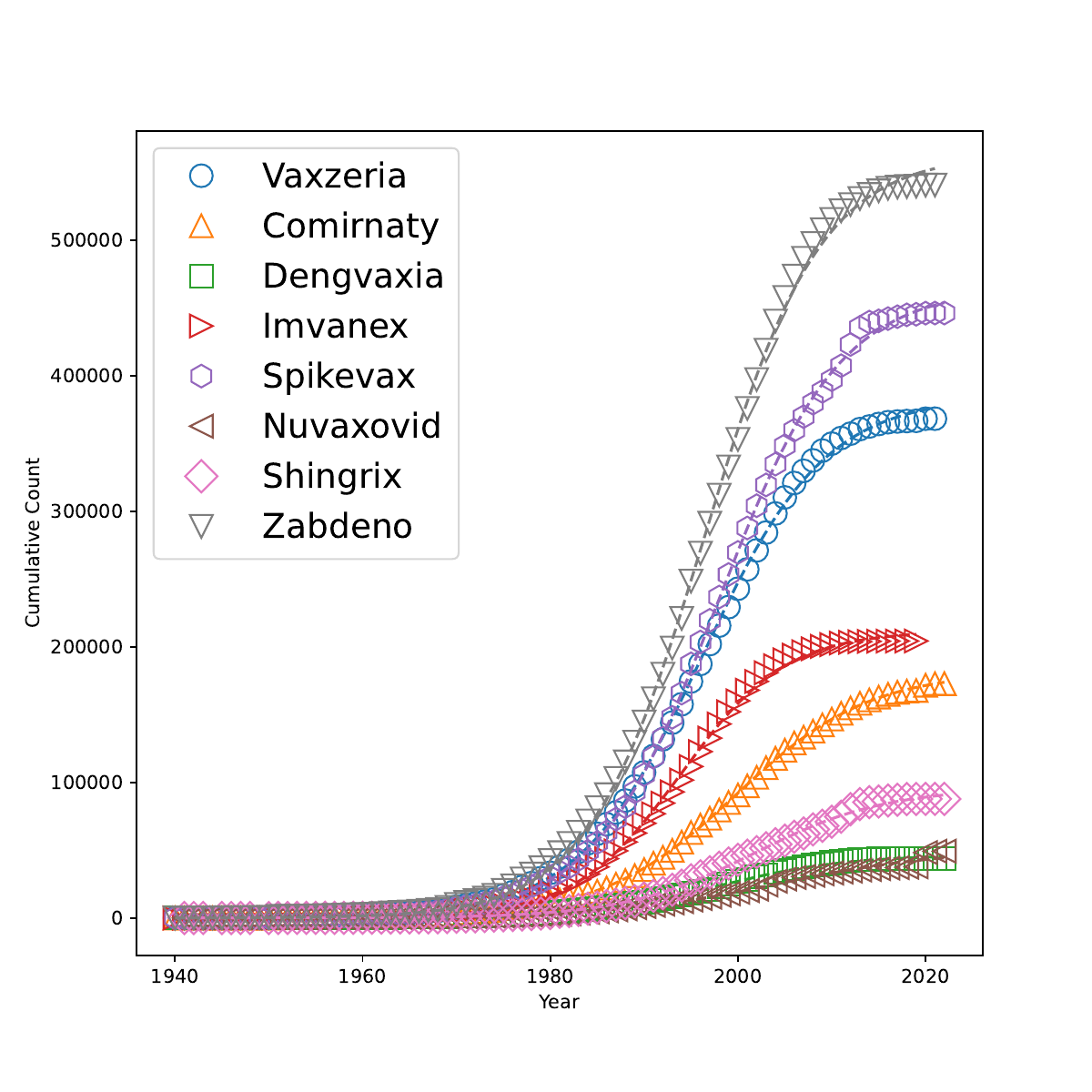} % produced by TSE using transdiss.py
		\caption{The cumulative number of dissemination events for each vaccine as a function of time.  }
		\label{fig:intrafit}
	\end{subfigure}
	\hfill
	\begin{subfigure}[t]{0.48\textwidth}
	\centering
	\includegraphics[width=\textwidth]{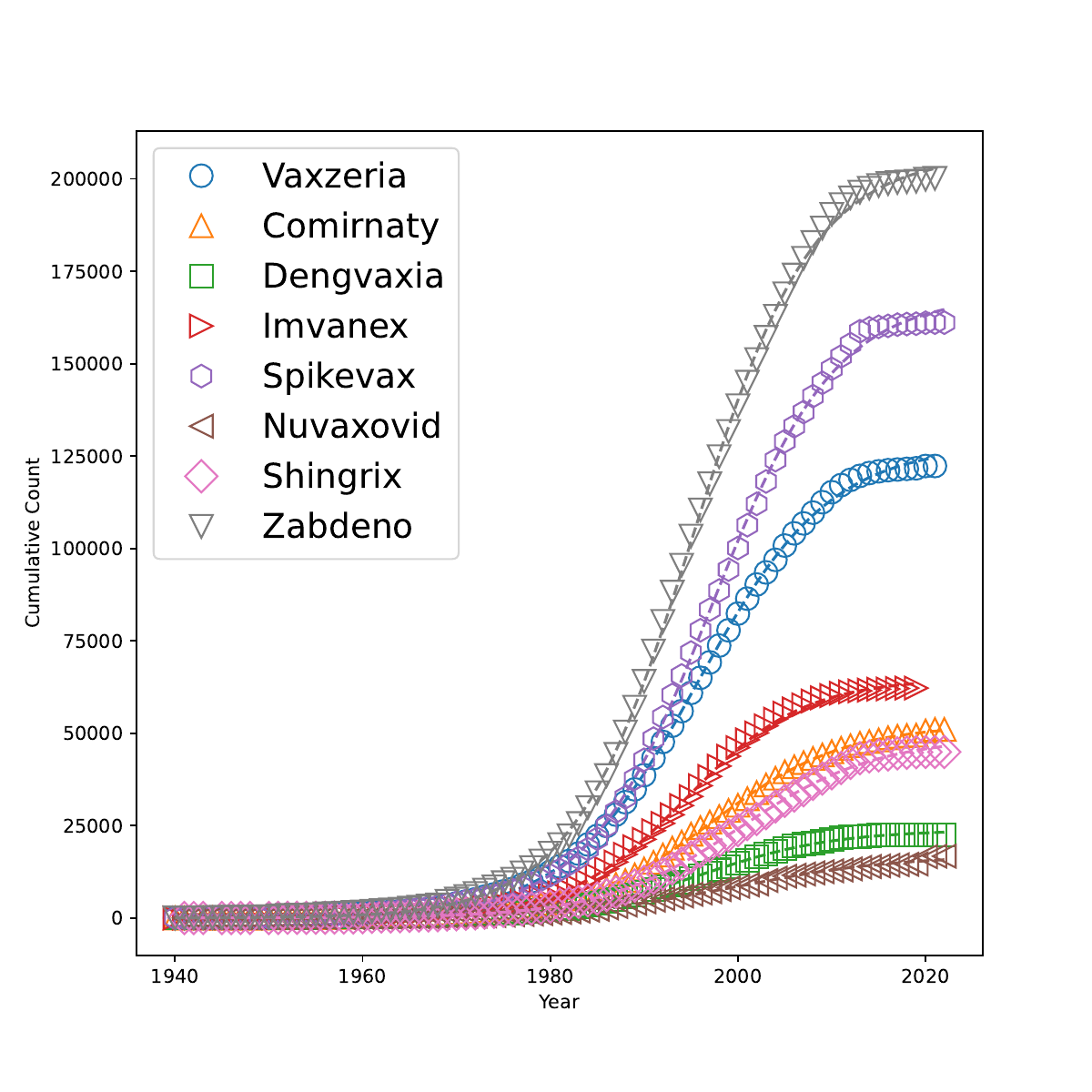} % produced by TSE using transdiss.py
	\caption{The cumulative number of translation events for each vaccine as a function of time.  }
	\label{fig:interfit}
	\end{subfigure}
	\\
	\begin{subfigure}[t]{0.48\textwidth}
	\centering
	\includegraphics[width=\textwidth]{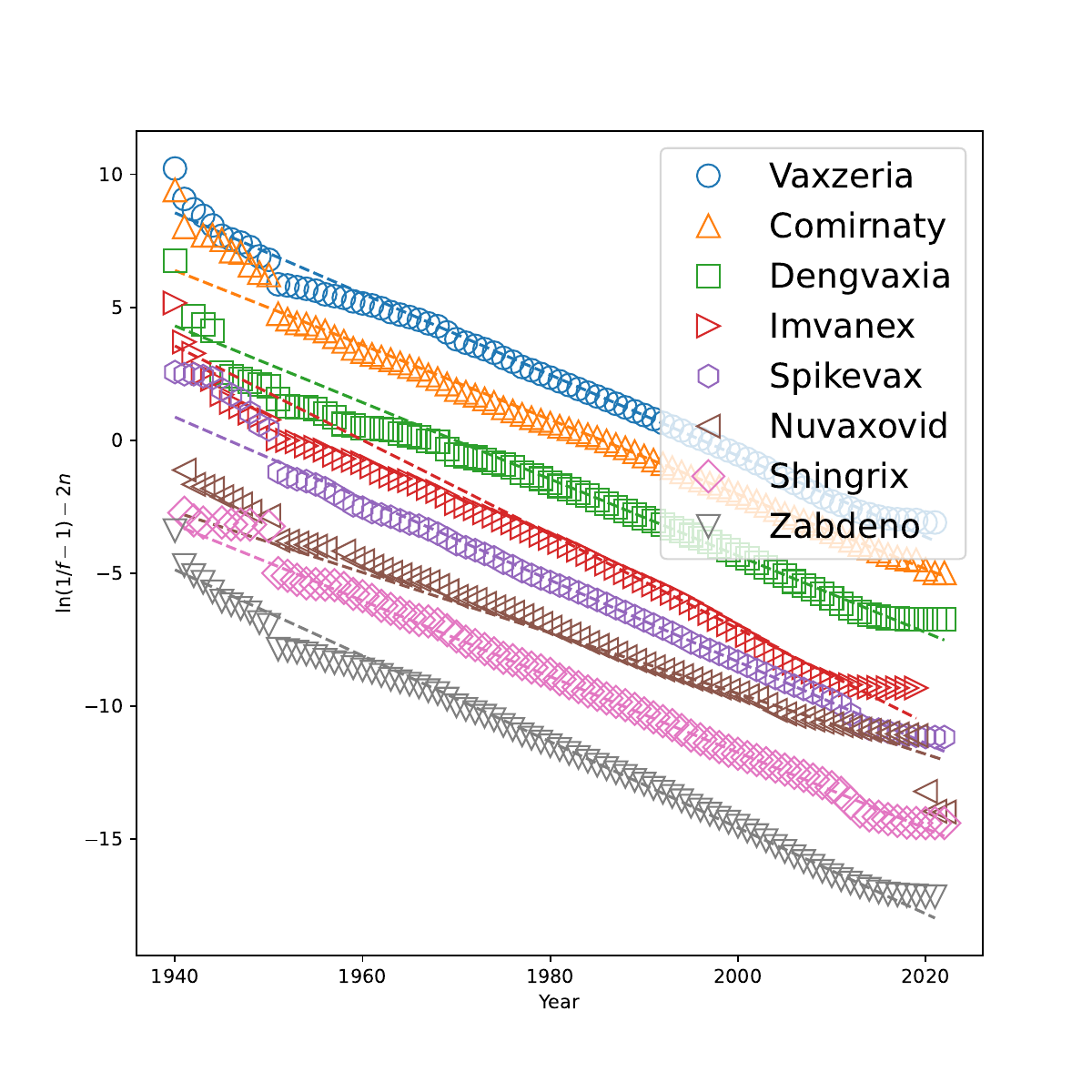} % produced by TSE using transdiss.py
	\caption{The cumulative fraction of dissemination events $f(t)$ for each vaccine plotted as $\ln( [f(t)]^{-1} -1 ) -2n$ where $n$ is the position in the sequence of vaccines shown, so $n=0$ for astra and $n=7$ for Zabdeno. }
	\label{fig:intratrans}
	\end{subfigure}
	\hfill
	\begin{subfigure}[t]{0.48\textwidth}
		\centering
		\includegraphics[width=\textwidth]{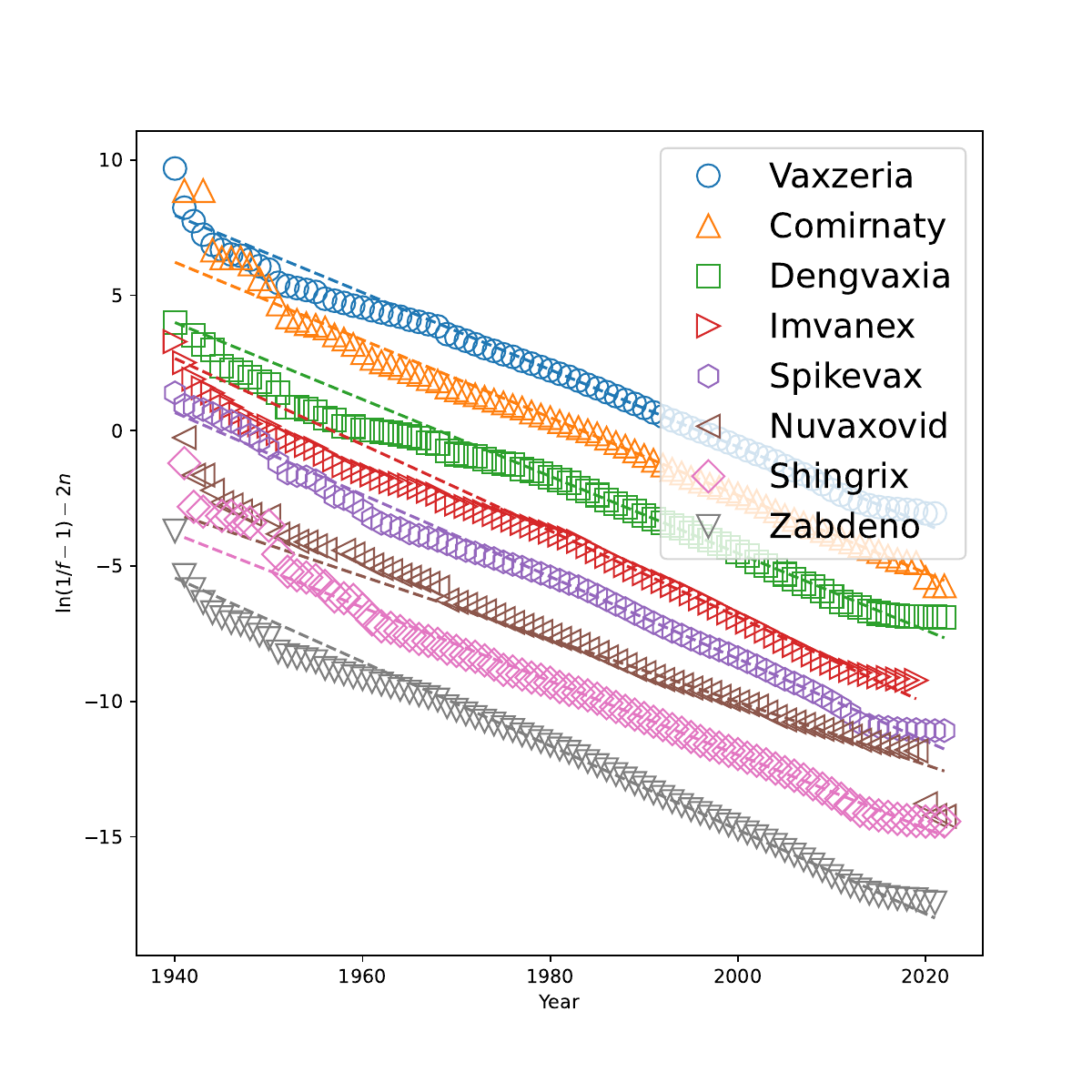} % produced by TSE using transdiss.py
		\caption{The cumulative fraction of translation events $f(t)$ for each vaccine plotted as $\ln( [f(t)]^{-1} -1 ) -2n$ where $n$ is the position in the sequence of vaccines shown, so $n=0$ for astro and $n=7$ for Zabdeno. }
		\label{fig:intertrans}
	\end{subfigure}
	\caption{Analysis of cumulative number of translation and dissemination events for each vaccine. The data is binned by year and represented by the points. The dashed line is from the best least-squares fit of the data to a logistic function \eqref{e:logisticfit}. 
	}
	\label{fig:scurve}
\end{figure}

Although the concept of an ``S-curve" is often used to describe technological progress metaphorically \cite{RN1916}, except the famous Moore's Law \cite{RN480}, the S-curve is rarely measured \cite{RN1356}. Here, we look at how the number of translation and dissemination events (inter- and intra-layer edges, respectively) evolve over time for each of our eight vaccines. In particular we look to see if the data is well described by a logistic function by fitting the cumulative number of events $n_\leq(t)$ for each vaccine as a function of time $t$ to the function
\beq 
 n_\leq(t) = \frac{\hat{A}}{1+ \exp(-(t-\hat{t}_0)/\hat{\tau}) }
  \, .
  \label{e:logisticfit}
\eeq
We do use by taking the data binned by year and minimising least-squares difference between data and logistic function \eqref{e:logisticfit} without assigning any error to the data points. The results for the fit are tabulated in \tabref{tab:scurve} and are shown visually in \figref{fig:scurve}. The fitted forms are a reasonable description of the data on the basis of a visual inspection. This view is supported by the similarity between the fitted value for the normalisation $\hat{A}$ and the total number of events for each case as shown in \tabref{tab:scurve}. While estimating uncertainties in the data points and using a more sophisticated fitting procedures might produce better results, the visual quality of this simple fit suggests that the results for the fitted parameters are unlikely to change much. We also do not claim that the logistic form itself is the best form possible but again, we feel that other functions, why they may give a better fit by some quantitative measure, are not going to be so much better in terms of the description they give.  Our data is after all is imperfect by amounts that are hard to quantity.  Overall, we feel this fit of the logistic function to our data gives a reasonable description of the data. 

One note about the quartile dates given in \tabref{tab:scurve}. Strictly the quartiles $T_q$ is the earliest year $t$ (integer valued) for which the cumulative total of papers $n_\leq(t)$ exceeds one quarter, one half and three-quarters of the total number of events. So strictly the figures given for quartiles in the table should be rounded up. However, our events occur spread over the whole year but our data does not always carry accurate the month information. So to give a slightly better estimate for quartile times we take the strict definition to give a $T_q$ integer value and then we linearly interpolate between $n_\leq(T_q-1)$ and $n_\leq(T_q)$ to find a better estimate $t_q$ (not an integer) which satisfies $T_q \leq t_q \leq T_q+1$.  That is\footnote{Remember that $n_\leq(t)$ represents total accumulation so includes contributions over the whole year $t$ right up to the start of year $t+1$. As a result the interpolation starts from time $T_{q}$ which, as a decimal number, represents the time at the start of the year included in the $n_\leq(T_q)$ value.}
\beq
 t_q = T_{q} + \left(\frac{n_\leq(t)- n_\leq(T_q-1)}{n_\leq(T_q)- n_\leq(T_q-1)} \right) 
 \label{e:tqestimate}
 \, .
\eeq

The results in  \tabref{tab:scurve} show that dissemination events are generally leading the translation events as shown by the values for the centre for the s-curve as indicated by the values for the fitted parameter centre $\hat{t}_0$ with the exception of the Imvanex vaccine. While we might expect dissemination to precede translation as we see in most of the data, we have to remember that these are averages over the development of many different aspects of each vaccine so this need not always come through in the data.  Also, no fit to data is a perfect representation of the data so the Imvanex result probably indicates that uncertainties of around a year in $\hat{t}_0$ are not unexpected. Given that most of the differences in the centre of the translation and dissemination curves appears to be negligible, perhaps reflecting that this indicator is an average over many parts of the process. The one case that stands out is the four year lag of the Nuvaxoid vaccine translation events behind the dissemination events. 
%%% \tsecomment{Can we add estimates of errors to these fits?  Maybe use Poisson uncertainty on counts.} 

We also note that Nuvaxoid vaccine stands out in other ways. The various measure of the time when work on this vaccine was active is noticeably later than the other seven vaccines. This might in part be because this vaccine has also been developed noticeably more slowly.  The width $\hat{\tau}$ of the fitted logistic function has the largest value, for dissemination $1.7$ years larger than second biggest value for dissemination events, 1.3 years bigger for translation events. The difference between quartile values also reflects this slower development of the Nuvaxoid vaccine.  In terms of years the development of Nuvaxoid is not much slower than the others but as a percentage, the development timescales of the Nuvaxoid vaccine is about 50\% slower than the timescales for Imvanex, the vaccines developed fastest, and around 25\% slower than the development times for the next slowest vaccine, Shingrix.

Five of the remaining seven vaccines show a hierarchy in terms of speed of development as measured by either inter-quartile times  or the values of the fitted $\hat{t}_0$ values, for both dissemination and translation. These five vaccines can be ordered from fastest to slowest as follows:
Imvanex,
Spikevax, %moderna,
Dengvaxia, %dengue,
Comirnaty, %biontech,
Shingrix.
The Zabdeno vaccine is also slower than dengue by any measure and is comparable to the speeds seen for Spikevax. However, the difference is not great, and interquartile range differs by only three years (from $13.0$ for Zabdeno dissemination events to $15.1$ for Shingrix translation events). This leaves the last vaccine, Vaxzeria, which shows one of the faster translation events development (between Comirnaty and 
Shingrix) but one of the slower dissemination speeds (between Imvanex and Spikevax). However, the differences in the timescales of seven vaccines, between  Imvanex and Shingrix at the two extremes, are small, around 25\%, three years difference in an interquartile range of around fourteen years.

\begin{table}[htb]
	\footnotesize
	\centering
	% this comes from allresultsedit.dat produced by scurve.py from TSE
	% Used https://www.tablesgenerator.com/latex_tables to create this from csv file
	\begin{tabular}{rcc || c | c | c | c || c || c | c | c}
Name      & Year & Type     &\multicolumn{4}{c||}{$t_q$/[year]}         & Maximum   & \multicolumn{3}{c}{Fitted Values}        \\ 
          &      &          & Q1     & Q2        & Q3     & Q3-Q1       &$n_\leq(t)$& norm. $\hat{A}$    &centre $\hat{t}_0$& width $\hat{\tau}$ \\ \hline \hline
Vaxzeria  & 2020 & trans.   & 1988.8 & 1996.1    & 2003.5 & 14.7        & 122322    & 128064             & 1995.8           & 7.0        \\
Comirnaty & 2020 & trans.   & 1990.7 & 1995.7    & 2005.3 & 14.6        & 51041     & 52257              & 1997.3           & 7.0        \\
Dengvaxia & 2019 & trans.   & 1989.0 & 1996.9    & 2003.4 & 14.4        & 22568     & 23827              & 1996.3           & 7.0        \\
Imvanex   & 2013 & trans.   & 1988.2 & 1995.3    & 2001.3 & 13.1        & 62235     & 64715              & 1994.4           & 6.3        \\
Spikevax  & 2020 & trans.   & 1990.5 & 1995.5    & 2004.5 & 14.0        & 161118    & 168499             & 1997.2           & 6.6        \\
Nuvaxovid & 2022 & trans.   & 1990.5 & 2000.5    & 2010.1 & 19.6        & 16741     & 16980              & 1999.8           & 8.6        \\
Shingrix  & 2017 & trans.   & 1992.3 & 2000.1    & 2007.4 & 15.1        & 44937     & 48850              & 2000.2           & 7.3        \\
Zabdeno   & 2020 & trans.   & 1989.0 & 1995.6    & 2002.8 & 13.9        & 200150    & 206520             & 1995.2           & 6.4        \\ \hline
Vaxzeria  & 2020 & diss.    & 1989.5 & 1995.7    & 2003.4 & 13.9        & 368267    & 384976             & 1996.1           & 6.6        \\
Comirnaty & 2020 & diss.    & 1992.6 & 2000.2    & 2007.1 & 14.5        & 172862    & 181583             & 1999.8           & 7.1        \\
Dengvaxia & 2019 & diss.    & 1990.2 & 1998.3    & 2004.5 & 14.4        & 43774     & 46621              & 1997.6           & 6.9        \\
Imvanex   & 2013 & diss.    & 1988.2 & 1995.0    & 2000.1 & 11.9        & 204288    & 211693             & 1993.9           & 5.6        \\
Spikevax  & 2020 & diss.    & 1991.4 & 1998.2    & 2005.0 & 13.6        & 446169    & 465005             & 1997.8           & 6.5        \\
Nuvaxovid & 2022 & diss.    & 1995.2 & 2004.1    & 2017.5 & 22.3        & 49304     & 50220              & 2004.1           & 8.8        \\
Shingrix  & 2017 & diss.    & 1994.4 & 2001.3    & 2009.2 & 14.8        & 87861     & 95819              & 2001.8           & 7.0        \\
Zabdeno   & 2020 & diss.    & 1990.4 & 1997.1    & 2003.4 & 13.0        & 540561    & 563196             & 1996.4           & 6.2              
    \end{tabular}      
	\caption{Values obtained by fitting a logistic function to the cumulative number $n_\leq(t)$ of translation and dissemination events for each vaccine. The data was binned by year with the third, fourth and fifth columns giving the quartile values for the times of events in the data as estimated using \eqref{e:tqestimate}. 
		The maximum $n_\leq(t)$ value is the total number of events.
		The last three columns on the right are the parameter values obtained by a least-squares fit of the data (no errors assigned to data points) to the function $n_\leq(t) = \hat{A}[1+ \exp(-(t-\hat{t}_0)/\hat{\tau}) ]^{-1}$.  Results are visualised in \figref{fig:scurvefit}.
}
\label{tab:scurve}
\end{table}

\begin{figure}[htb]
	\centering
	\includegraphics[width=0.48\textwidth]{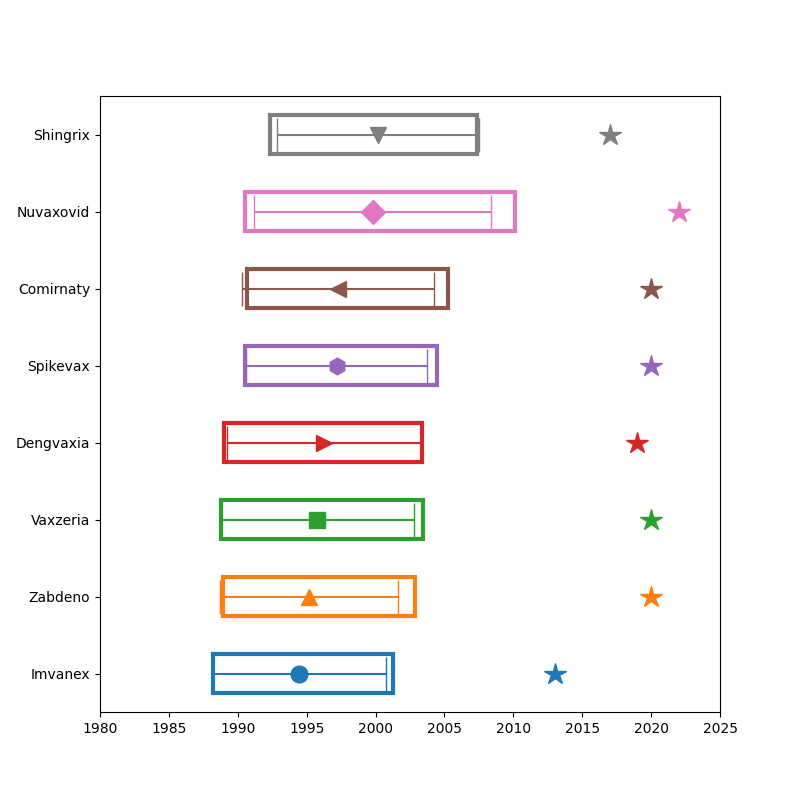} 
	\includegraphics[width=0.48\textwidth]{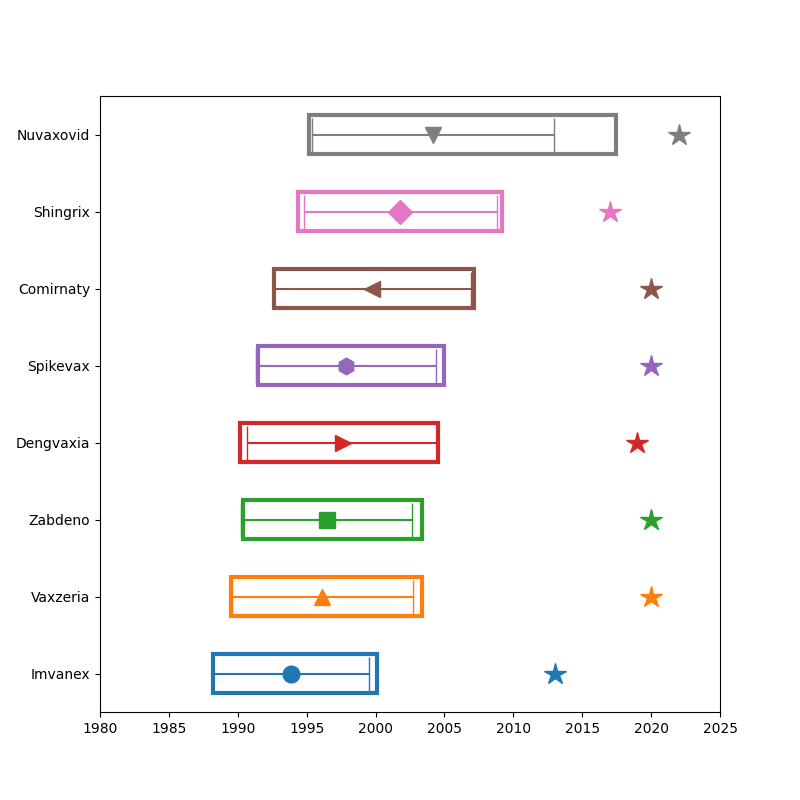} 
\caption{The S-curve properties of translation (left) and dissemination (right) events for each vaccine, as given in \tabref{tab:scurve}.  The boxes mark the years from the first to third quartile in terms of number of events, $n_\leq(t)=0.25$ to $0.75$ while the stars on the right mark the vaccine approval date. The data point is the fitted value for the centre parameter  $\hat{t}_0$ of the logistic function with the errors bars marking the width parameter $\pm\hat{\tau}$. 
}
\label{fig:scurvefit}
\end{figure}

\begin{figure}[htb]
	\centering
	\begin{subfigure}[t]{0.3\textwidth}
		\centering
		\includegraphics[width=\textwidth]{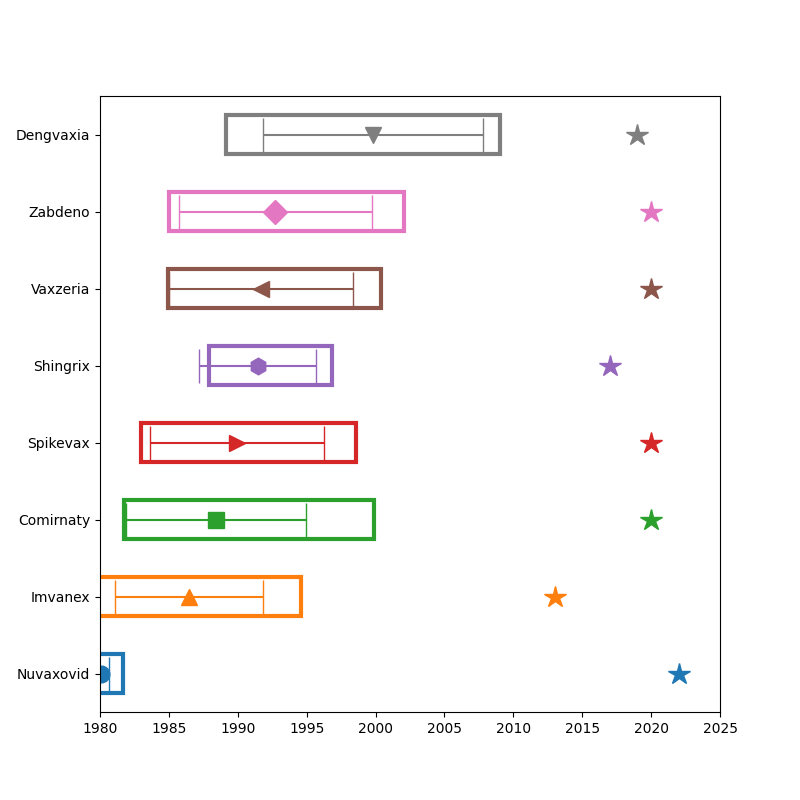} % produced by TSE using transdiss.py
		\caption{PAT$_{process}$→PAT$_{process}$}
	\end{subfigure}
	\hfill 
 	\begin{subfigure}[t]{0.3\textwidth}
		\centering
		\includegraphics[width=\textwidth]{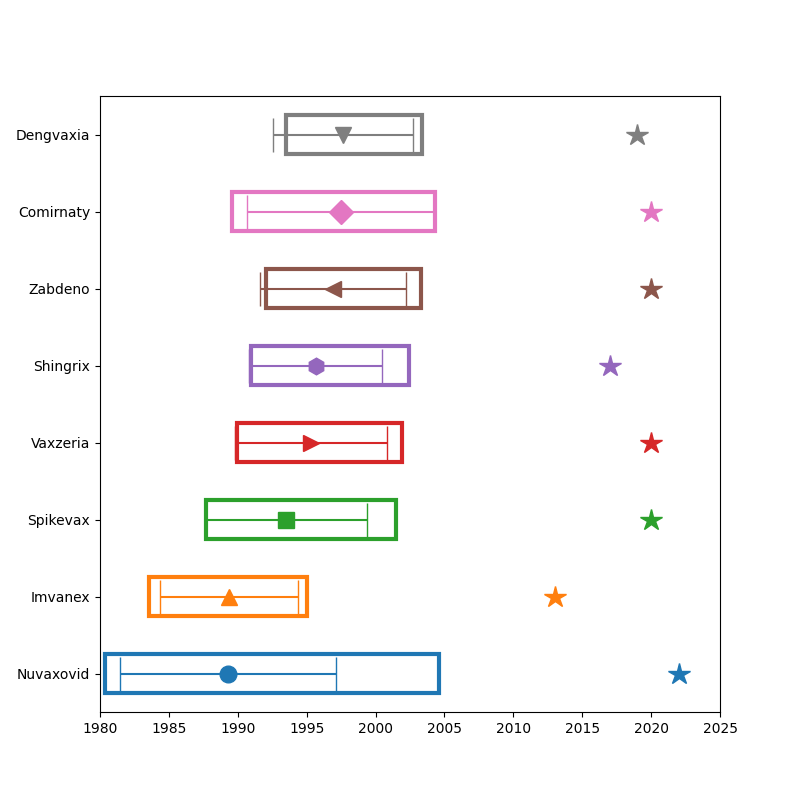} 
		\caption{PAT$_{process}$→PAT$_{product}$}
	\end{subfigure}
	\hfill 
    \begin{subfigure}[t]{0.3\textwidth}
		\centering
		\includegraphics[width=\textwidth]{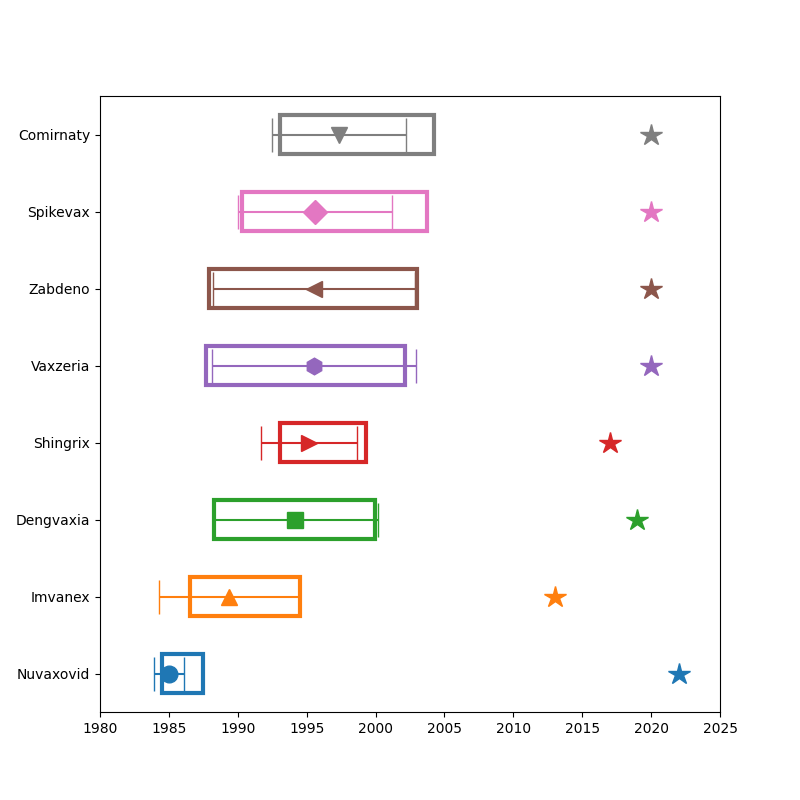} 
		\caption{PAT$_{process}$→PUB$_{applied}$}
	\end{subfigure}
	\hfill 
     \begin{subfigure}[t]{0.3\textwidth}
		\centering
		\includegraphics[width=\textwidth]{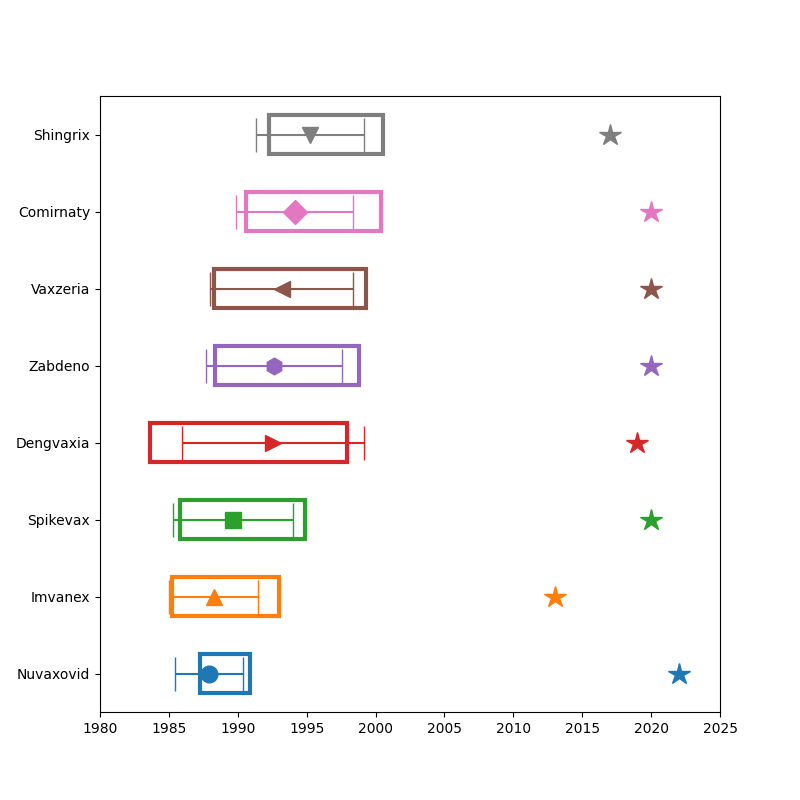} 
		\caption{PAT$_{process}$→PUB$_{basic}$}
	\end{subfigure}
	\hfill 
    \begin{subfigure}[t]{0.3\textwidth}
		\centering
		\includegraphics[width=\textwidth]{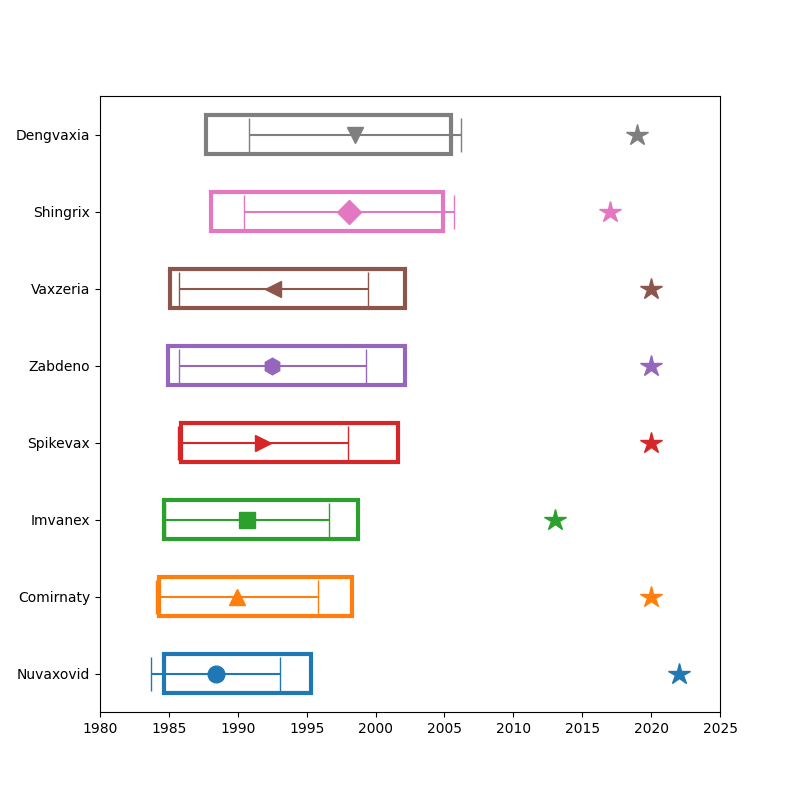} 
		\caption{PAT$_{product}$→PAT$_{process}$}
	\end{subfigure}
	\hfill
    \begin{subfigure}[t]{0.3\textwidth}
		\centering
		\includegraphics[width=\textwidth]{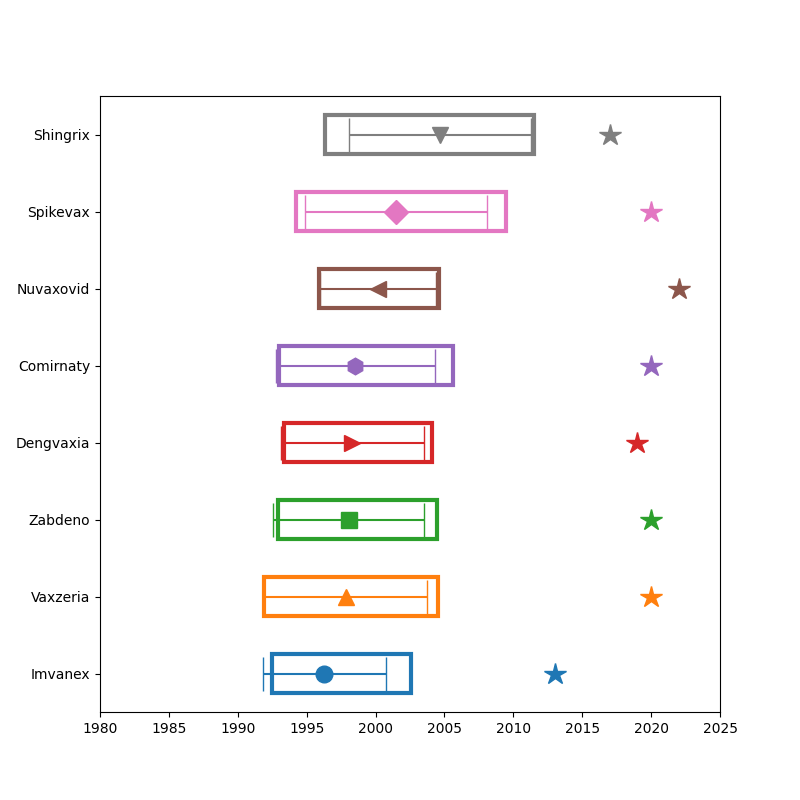} 
		\caption{PAT$_{product}$→PAT$_{product}$}
	\end{subfigure}
	\hfill
    \begin{subfigure}[t]{0.3\textwidth}
		\centering
		\includegraphics[width=\textwidth]{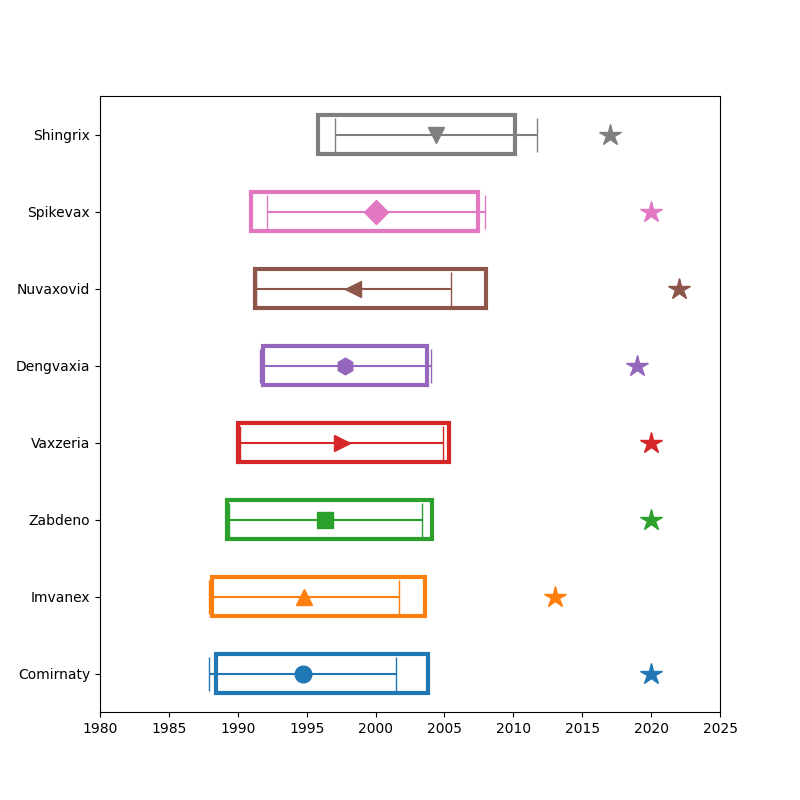} 
		\caption{PAT$_{product}$→PUB$_{applied}$}
	\end{subfigure}
	\hfill
    \begin{subfigure}[t]{0.3\textwidth}
		\centering
		\includegraphics[width=\textwidth]{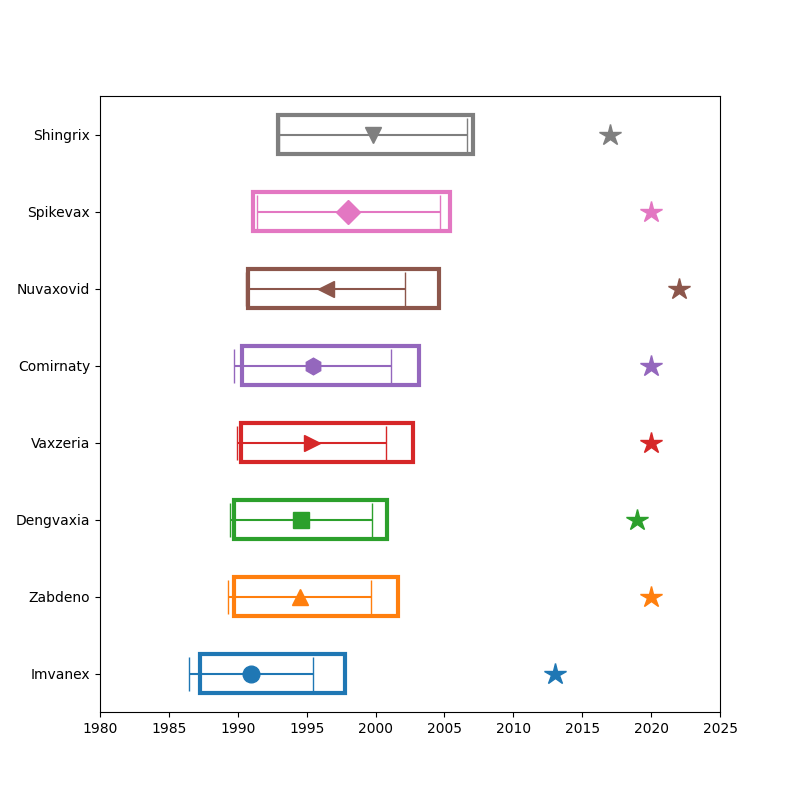} 
		\caption{PAT$_{product}$→PUB$_{basic}$}
	\end{subfigure}
	\hfill
    \begin{subfigure}[t]{0.3\textwidth}
		\centering
		\includegraphics[width=\textwidth]{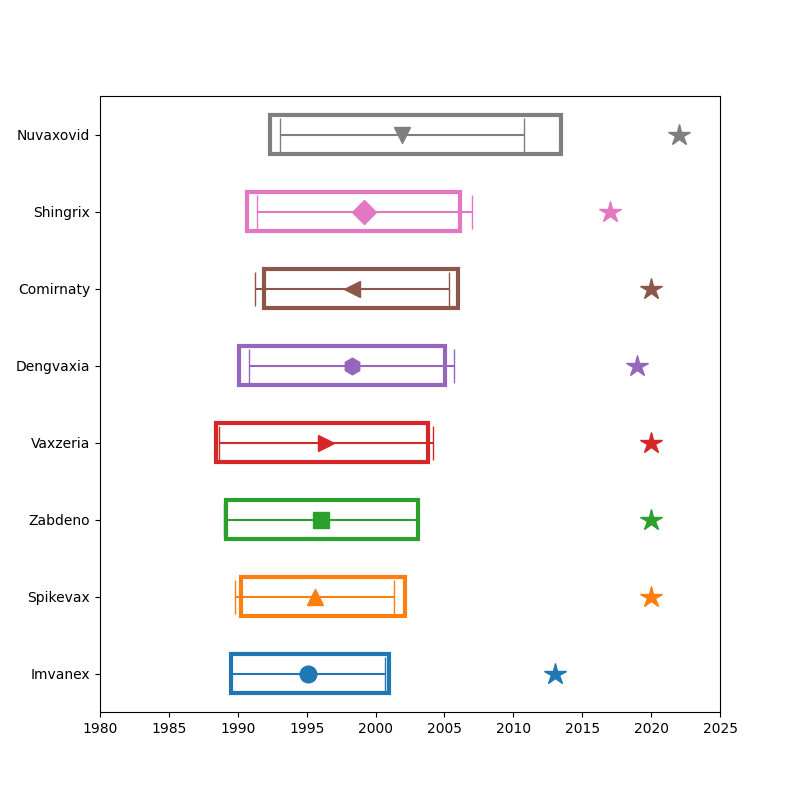} 
		\caption{PUB$_{applied}$→PUB$_{basic}$}
	\end{subfigure}
	\hfill
    \begin{subfigure}[t]{0.3\textwidth}
		\centering
		\includegraphics[width=\textwidth]{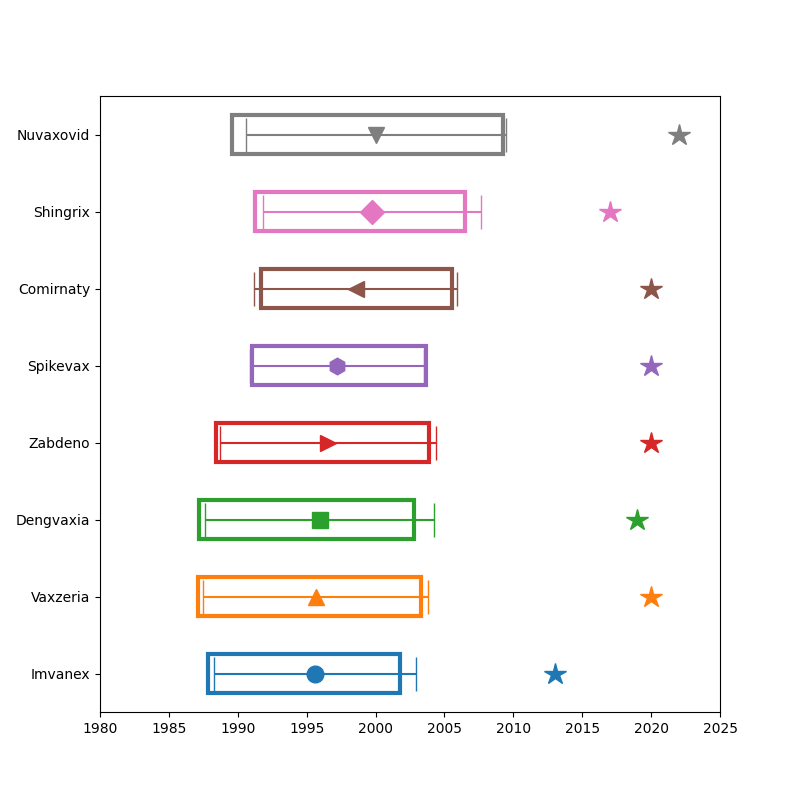} 
		\caption{PUB$_{basic}$→PUB$_{applied}$}
	\end{subfigure}
	\hfill
    \begin{subfigure}[t]{0.3\textwidth}
		\centering
		\includegraphics[width=\textwidth]{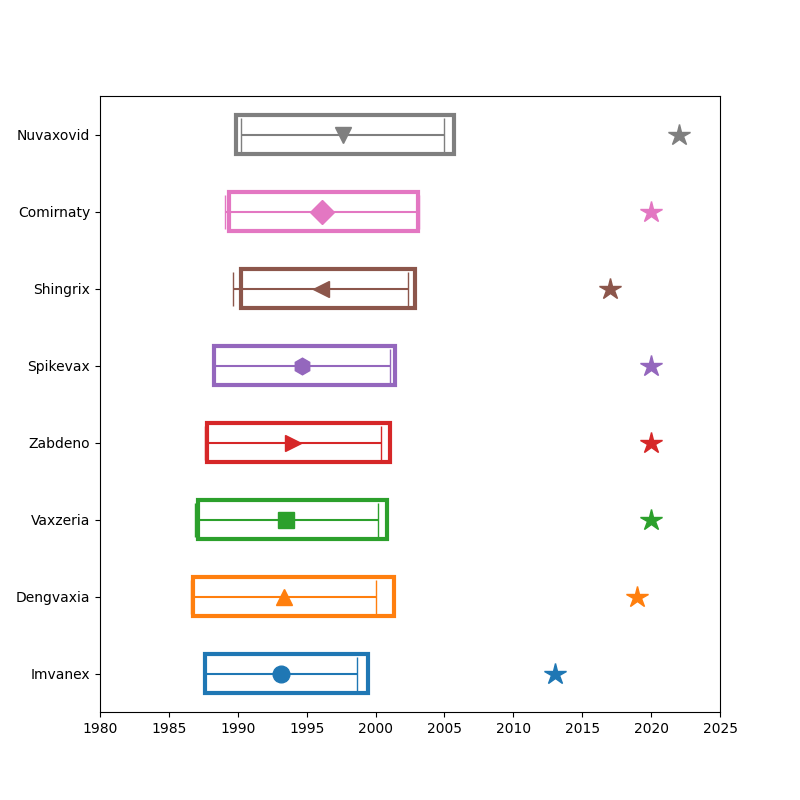} 
		\caption{PUB$_{basic}$→PUB$_{basic}$}
	\end{subfigure}
	\hfill
    \begin{subfigure}[t]{0.3\textwidth}
		\centering
		\includegraphics[width=\textwidth]{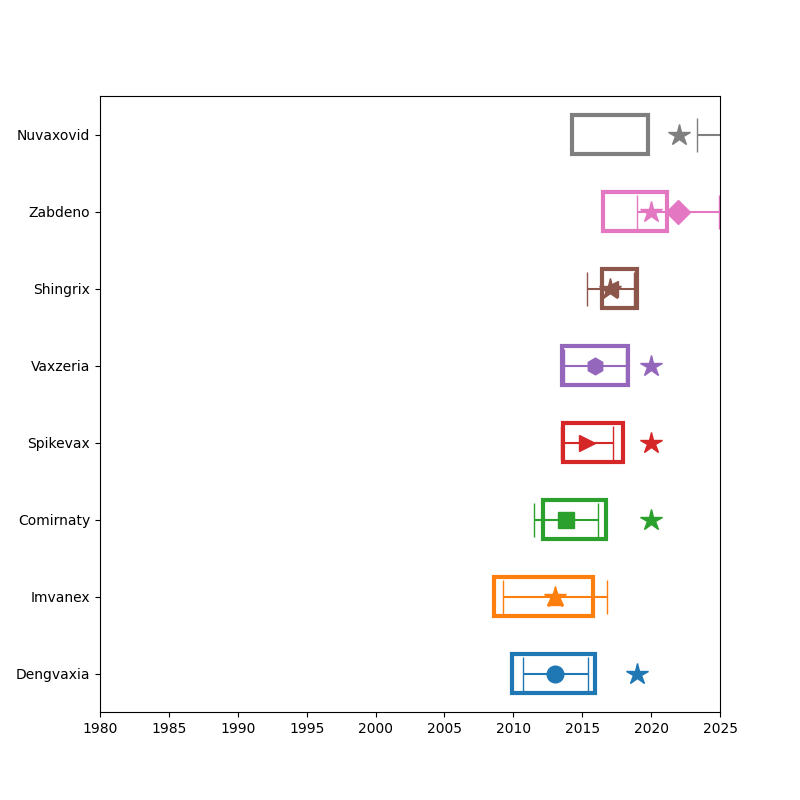} 
		\caption{CT→PAT$_{product}$}
	\end{subfigure}
	\hfill
\caption{The S-curve properties of translation and dissemination sub-types for each vaccine.  The boxes mark the years from the first to third quartile in terms of number of events, $n_\leq(t)=0.25$ to $0.75$ while the stars on the right mark the vaccine approval date. The data point is the fitted value for the centre parameter  $\hat{t}_0$ of the logistic function with the errors bars marking the width parameter $\pm\hat{\tau}$. Comparing the subplot provides an idea of which types of translation come and go first. Raw data available at \url{https://10.6084/m9.figshare.24422317}.
}
\label{fig:scurvefit_subtypes}
\end{figure}

\clearpage
% **************************************************************
\section{Funder Similarity and Clusters}
\label{app:fundersim}

We find the similarity between funders and cluster the funders as on a matrix $M_{ft}$. Here the label $f$ runs over the different funders 

The label $t$ runs over the twenty two different subtypes and the eight different vaccines. So for each vaccine we have the following fractional edge sub-type counts:-
\begin{quote}
clin\_clin,
clin-pat\_process,
clin-pat\_product,
clin-pubs\_basic,
clin-pubs\_applied,
pat\_process-pat\_process,
pat\_process-pat\_product,
pat\_process-pubs\_basic,
pat\_process-pubs\_applied,
pat\_product-pat\_process,
pat\_product-pat\_product,
pat\_product-pubs\_basic,
pat\_product-pubs\_applied,
pubs\_basic-pubs\_basic,
pubs\_basic-pubs\_applied,
pubs\_applied-pubs\_basic,
pubs\_applied-pubs\_applied,
\end{quote}

The notation gives each edge type as \texttt{source}-\texttt{target} where \texttt{source} (\texttt{target}) gives the document type acting as the source (target) node for that type of edge. The labels used for the type of document are one of five types:
\texttt{pubs\_basic} (publication basic science),
\texttt{pubs\_applied} (publication applied science), 
\texttt{pat\_process} (patent process), 
\texttt{pat\_product} (patent product), 
\texttt{clin} (clinical trial).
Note that three of the possible twenty five edge types are not included as the numbers of these edges are so small. 

The entries $M_{ft}$ are fractions for each vaccine individually, that is 
\beq 
 \sum_{t \in \mathcal{T}_v } M_{ft} = c_{fv}
\eeq 
is close to but a bit less than $1.0$ where the sum runs over edge subtypes for just one of the eight vaccines $v$.  
Put another way $\sum_{t} M_{ft}  = \sum_v c_{fv} = C_v$ is between $7.2$ and $7.7$ for all eight vaccines. 
This normalisation falls a little short as the type of edge and the date of documents can not always be identified reliably and such cases are not included in the counts. 
From this matrix we construct the distance matrix $d(f,g)$ between two funders $f$ and $g$ using cosine similarity where 
\beq
 d(f,g) = 1 - \sum_t \frac{M_{ft} }{Z_f} \frac{M_{gt}}{Z_g} \, , \quad
 (Z_f)^2 = \sum_t {M_{ft} }{M_{ft} } \, .
 \label{e:cosinesim}
 \eeq
 
\begin{figure}
	\centering
	\begin{subfigure}[t]{0.48\textwidth}
		\centering
		\includegraphics[width=\textwidth]{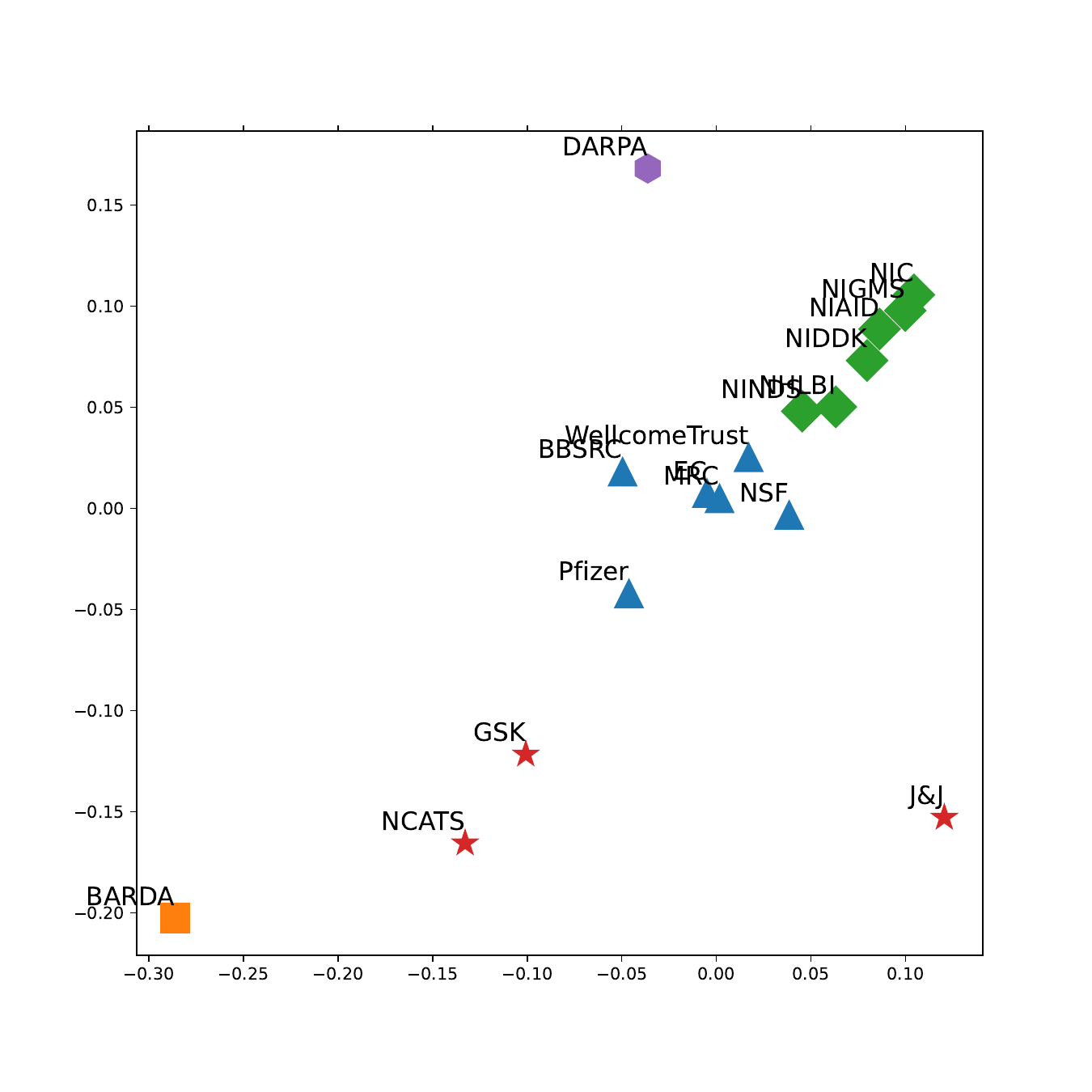} % produced by TSE using transdiss.py
		\caption{Multidimensional scaling analysis gives coordinates of funders using the two largest eigenvalues. The colours and shapes of points correspond to distinct clusters produced using k-means clustering methods with 5 clusters requested. }
		\label{fig:mdskm}
	\end{subfigure}
	\hfill
	\begin{subfigure}[t]{0.48\textwidth}
		\centering
		\includegraphics[width=\textwidth]{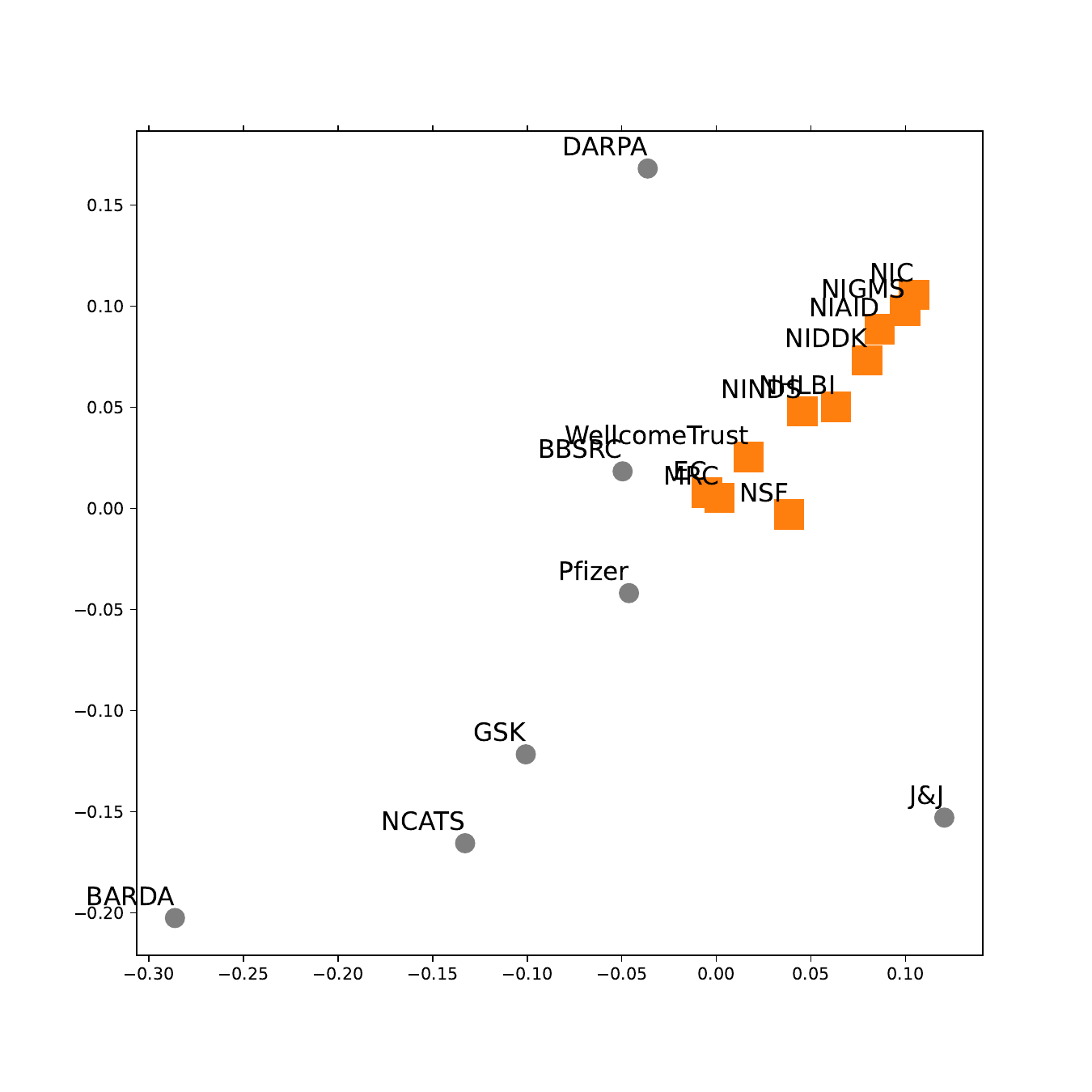} % produced by TSE using transdiss.py
		\caption{Multidimensional scaling analysis gives coordinates of funders using the two largest eigenvalues.  The coloured squares are the one cluster found using DBSCAN with a distance cutoff of $0.04$ and a minimum number of samples of $2$, grey circles represent outliers.}
		\label{fig:mdsdbs}
    \end{subfigure}
	\\
	\begin{subfigure}[t]{0.48\textwidth}
	\centering
	\includegraphics[width=\textwidth]{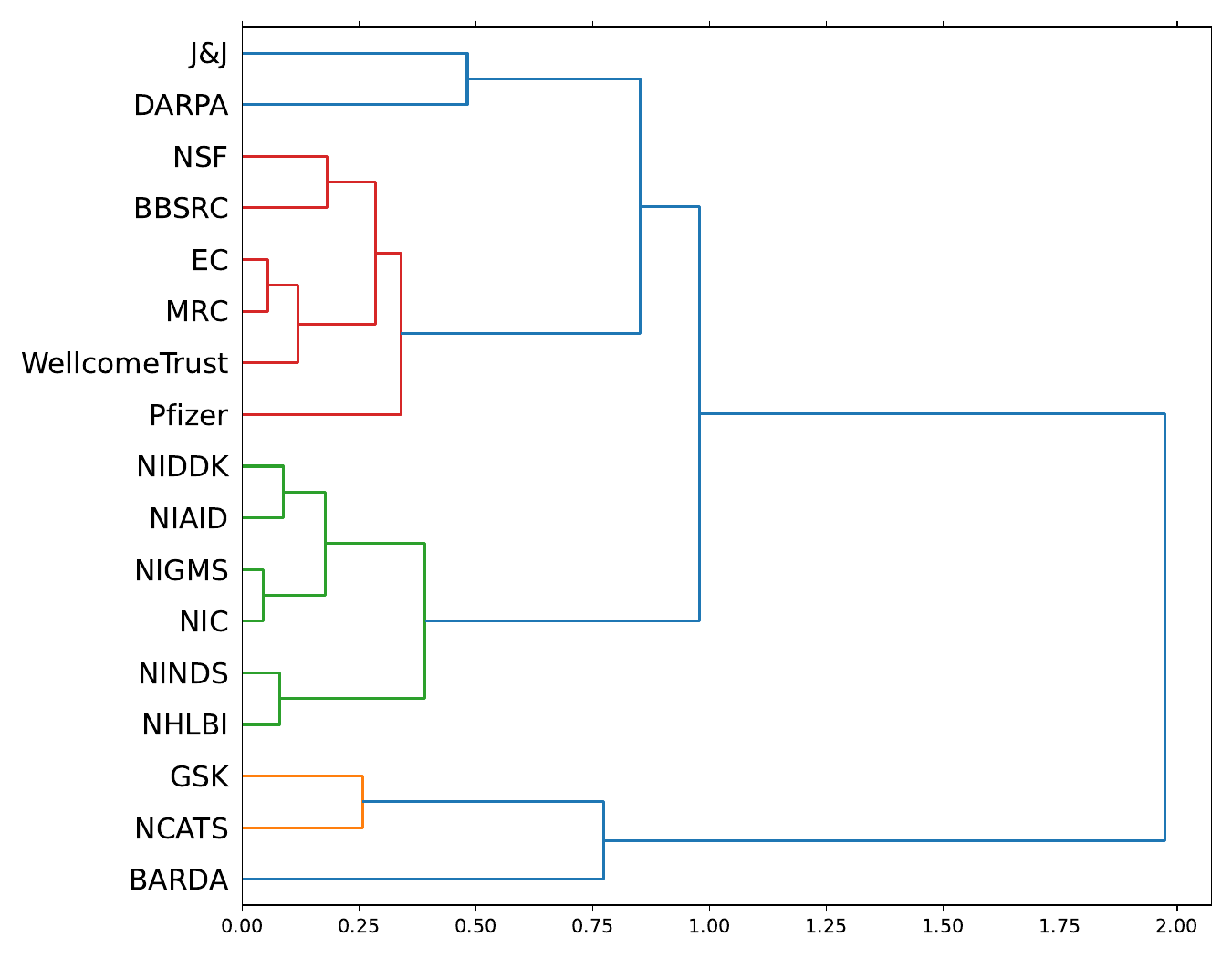} % produced by TSE using transdiss.py
		\caption{Hierarchical agglomerative clustering using the Ward method. Colours correspond to the clusters found with a cutoff value of $0.4$ }
	\label{fig:hac}
	\end{subfigure}
	\hfill
	\begin{subfigure}[t]{0.48\textwidth}
	\centering
	\includegraphics[width=\textwidth]{hello2_PCA_KM5.pdf} % produced by TSE using transdiss.py
		\caption{PCA scatter plot where coordinates come from the two most significant eigenvalues. The colours and shapes of points correspond to distinct clusters produced using k-means clustering methods with 5 clusters requested. }
	\label{fig:pca2}
	\end{subfigure}
	\caption{Analysis of similarity between funders based on a cosine distance matrix derived from the fraction of translation and dissemination sub-types per vaccine. For each funder, the fraction of edge subtypes for each vaccine is  compared to other funders using cosine similarity to produce a distance matrix.  This is then used for Multidimensional scaling in \figref{fig:mdskm} and \figref{fig:mdsdbs}, Hierarchical agglomerative clustering in \figref{fig:hac} and PCA in \figref{fig:pca2}.  
		For this distance matrix, all these methods show consistent results. 
     Edge data used for clustering available at \url{https://20231023_funder_translation_subtypes_figF1}
    }
	
	\label{fig:fundersim}
\end{figure}

This $d(f,g)$ is the distance matrix used in the PCA plot shown in the main text.  For completeness, we show a similar scatter plot based on the coordinates obtained from the same cosine distance matrix using the alternative Multidimensional scaling method. We also use the same distance matrix to cluster the funders using two different methods: k-means and DBSCAN. The results for a variety of parameter values is consistent with what is seen visually from both the PCA and the MDS methods in \figref{fig:mdskm}.  
The main observation is that the funders that are all part of the US \href{https://www.nih.gov/}{National Institutes of Health} have very similar edge sub-type patterns which presumably reflects that all members of the \href{https://www.nih.gov/}{National Institutes of Health} must follow common goals set by the parent US federal authority.  We also have a further group of funders that are close to the members of the NIH: the  MRC (a UK government funding agency similar in scope to NIH), the EC and the Welcome trust. Most of the other funders appear to be less similar to each other and to this main group. Of the three pharmaceutical companies two standout J\&J and GSK while the third, Pfizer is closer to the main group but still distinct. The UK funding agency the BBSRC is also set back a little from the second cluster, reflecting the fact that it has a wider set biological goals rather than a focus on purely medical research. Again that reflects clear UK policy in that its MRC and BBSRC agencies have a different focus.  Also two further US government agencies, DARPA and BARDA, are extreme outliers which seems to reflect their very different goals, ones complementary to the NIH goals, set for these agencies by the US federal government.

\clearpage
% **************************************************************

% **************************************************************
\section{Vaccine translation sub-type longitudinal distribution}
\label{app:figures}

These additional diagrams show that the cumulative fraction of translation is usually ahead of that of dissemination.

\begin{figure}[hbt]
	\centering
	\begin{subfigure}[b]{0.24\textwidth}
		\centering
		\includegraphics[width=\textwidth]{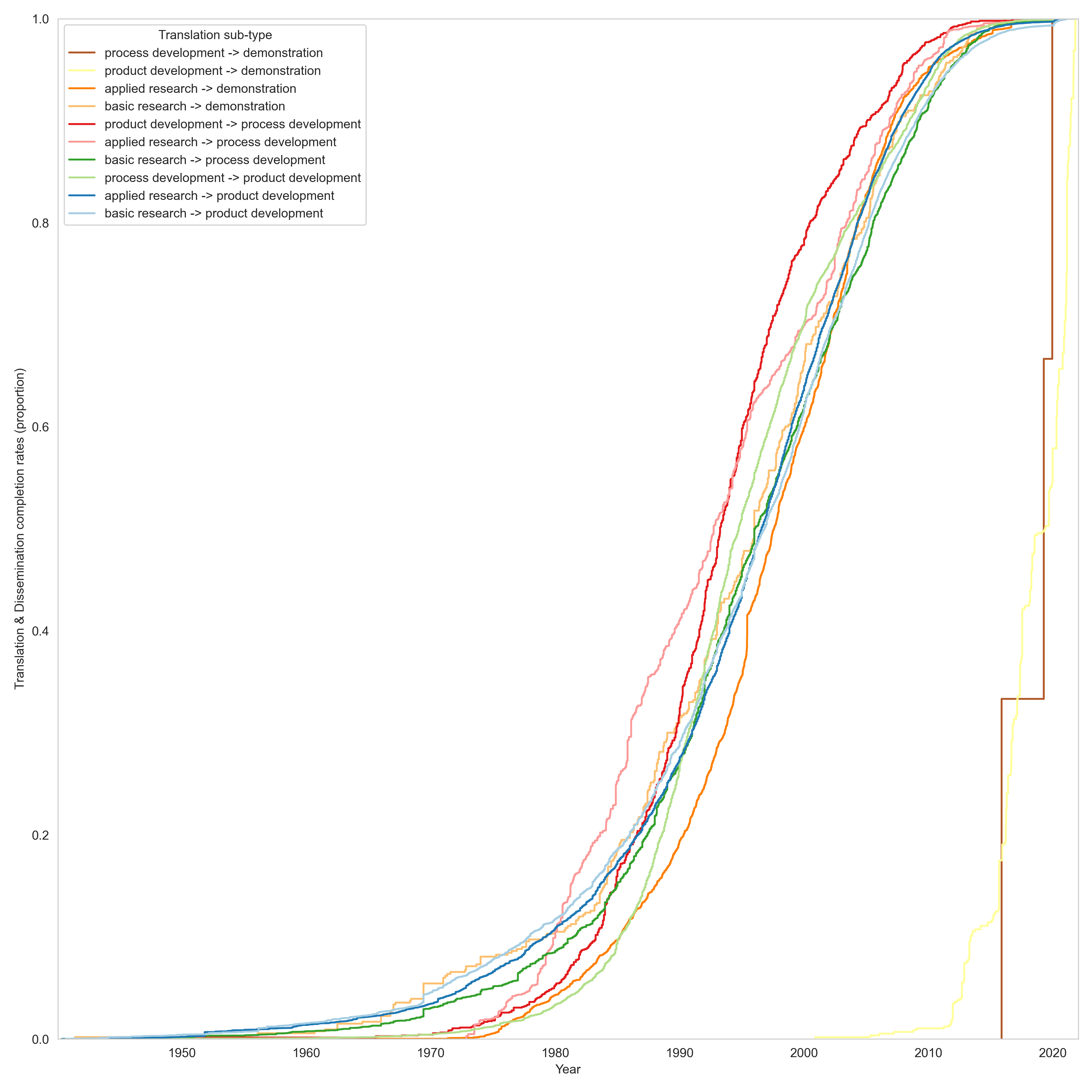}
		\caption{Zabdeno, Ebola, Janssen, 2020,\\AVV}
		%  \label{fig:h_v_d_zabdeno}
	\end{subfigure}
	\hfill
	\begin{subfigure}[b]{0.24\textwidth}
		\centering
		\includegraphics[width=\textwidth]{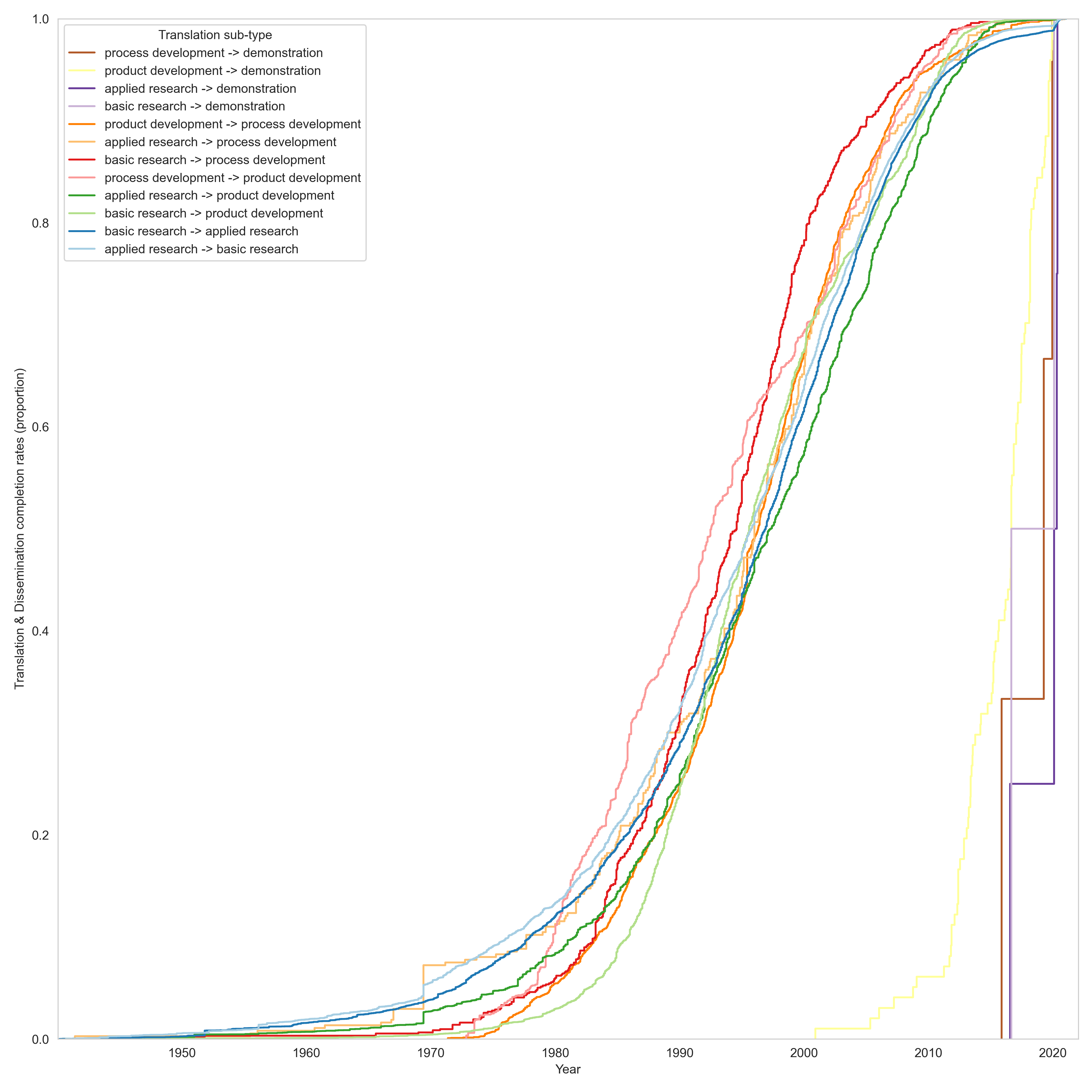}
		\caption{Vaxzevria, COVID-19, AstraZeneca, 2020, AVV}
		%  \label{fig:h_v_d_vaxzevria}
	\end{subfigure}
	\hfill
	\begin{subfigure}[b]{0.24\textwidth}
		\centering
		\includegraphics[width=\textwidth]{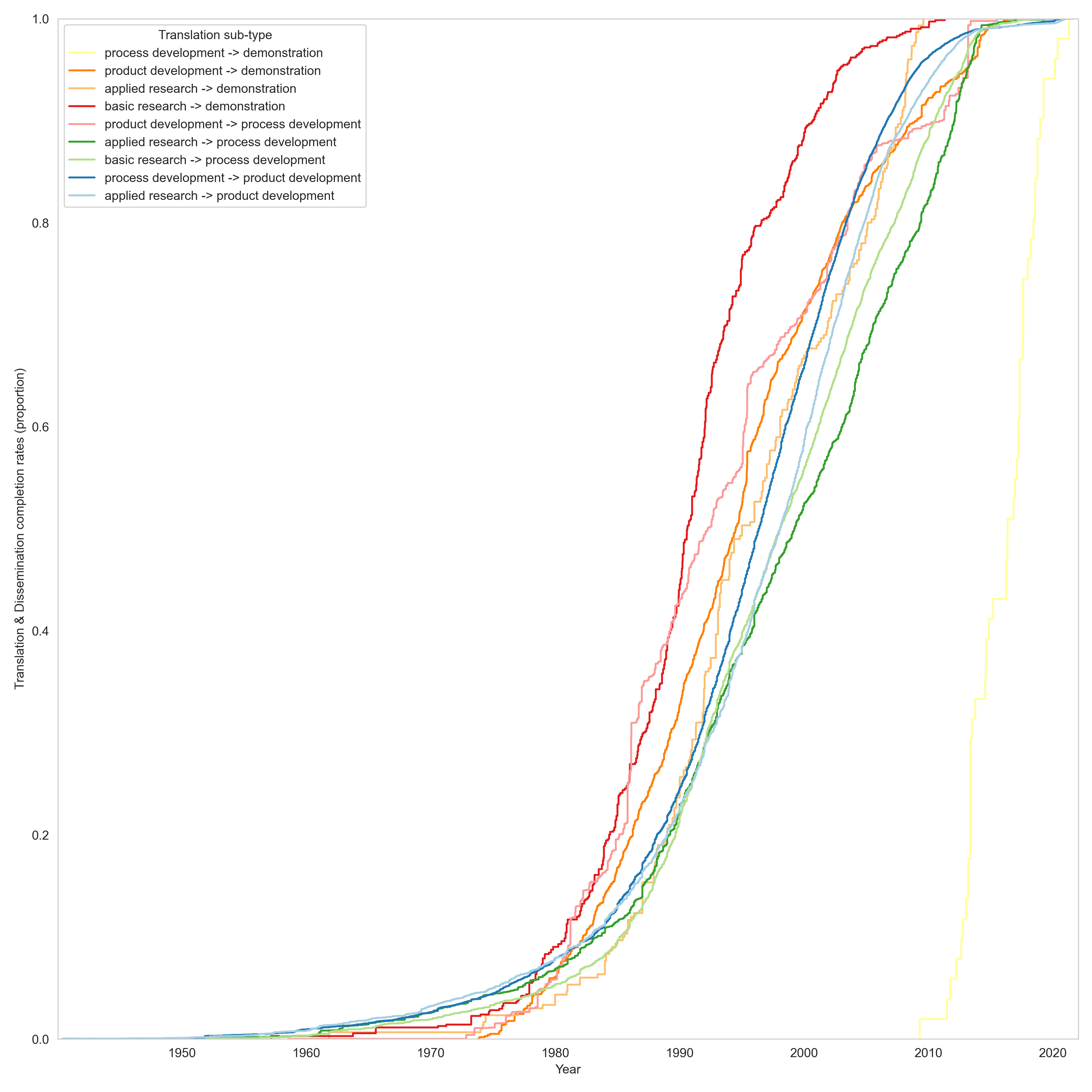}
		\caption{Spikevax, COVID-19, Moderna, 2020, mRNA}
		%  \label{fig:h_v_d_spikevax}
	\end{subfigure}
	\hfill
	\begin{subfigure}[b]{0.24\textwidth}
		\centering
		\includegraphics[width=\textwidth]{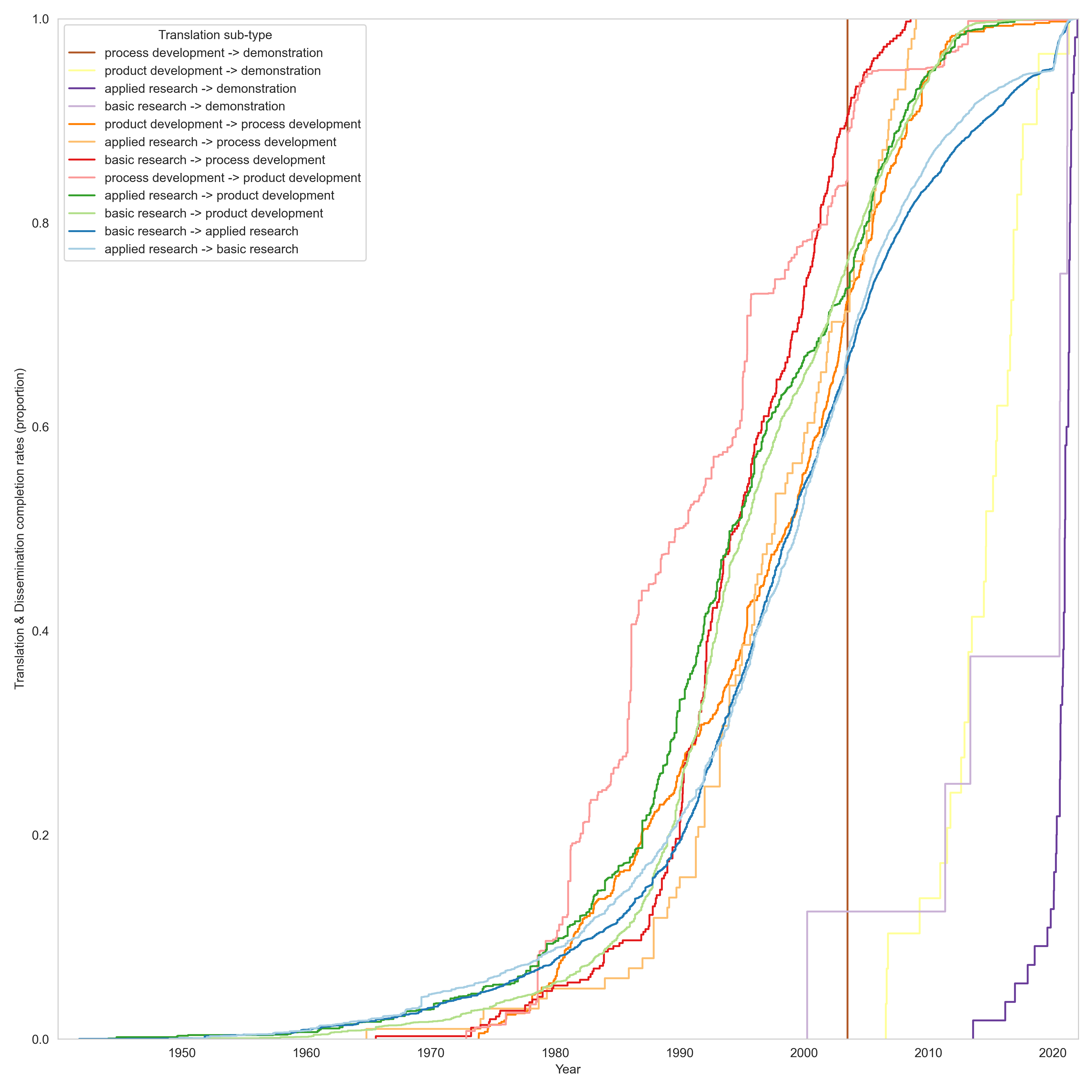}
		\caption{Comirnaty, COVID-19, BioNTech/Pfizer, 2020, mRNA}
		%  \label{fig:h_v_d_comirnaty}
	\end{subfigure}
	\hfill
	\begin{subfigure}[b]{0.24\textwidth}
		\centering
		\includegraphics[width=\textwidth]{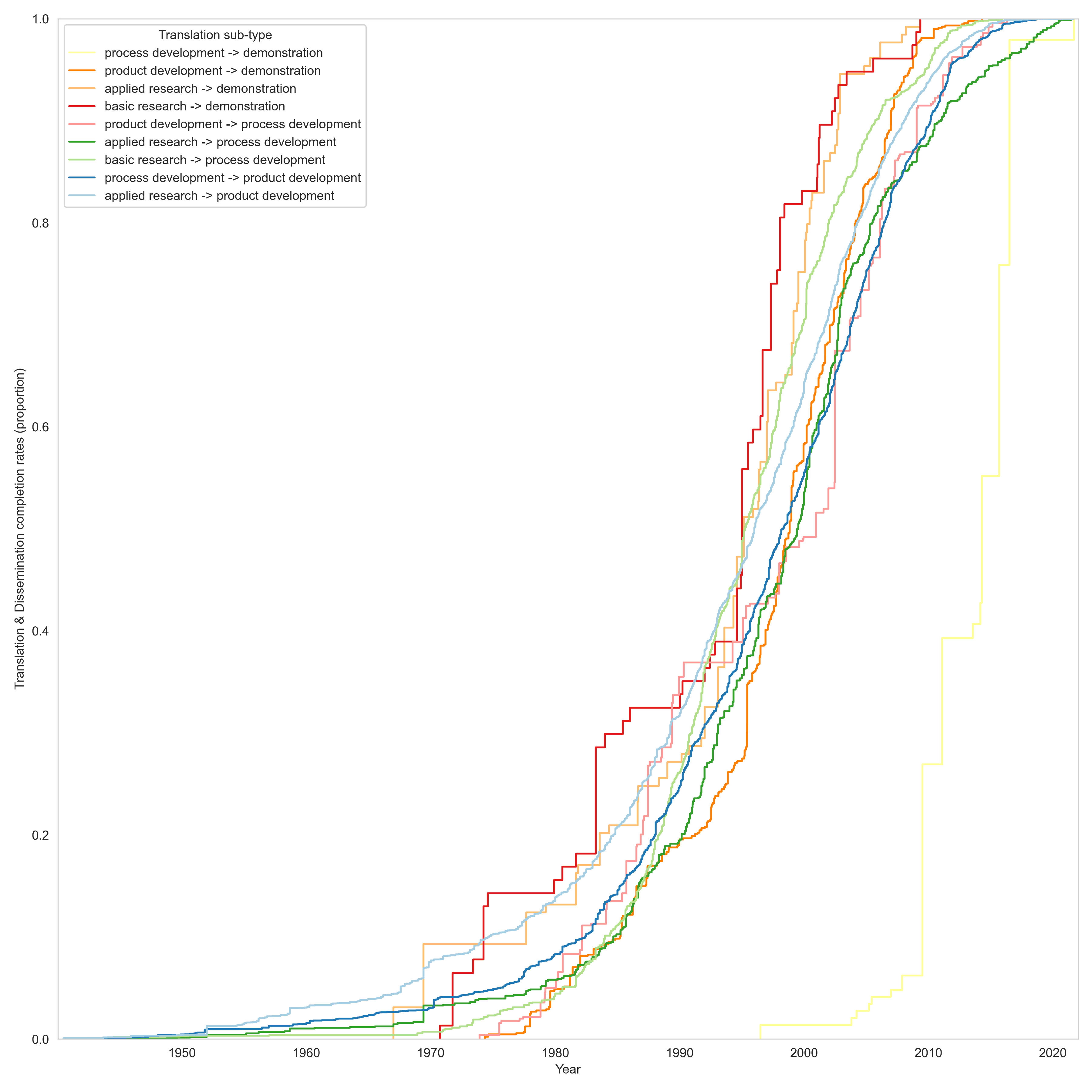}
		\caption{Dengvaxia, Dengue, Sanofi, 2019,\\ WPV}
		%  \label{fig:h_v_d_dengvaxia}
	\end{subfigure}
	\hfill
	\begin{subfigure}[b]{0.24\textwidth}
		\centering
		\includegraphics[width=\textwidth]{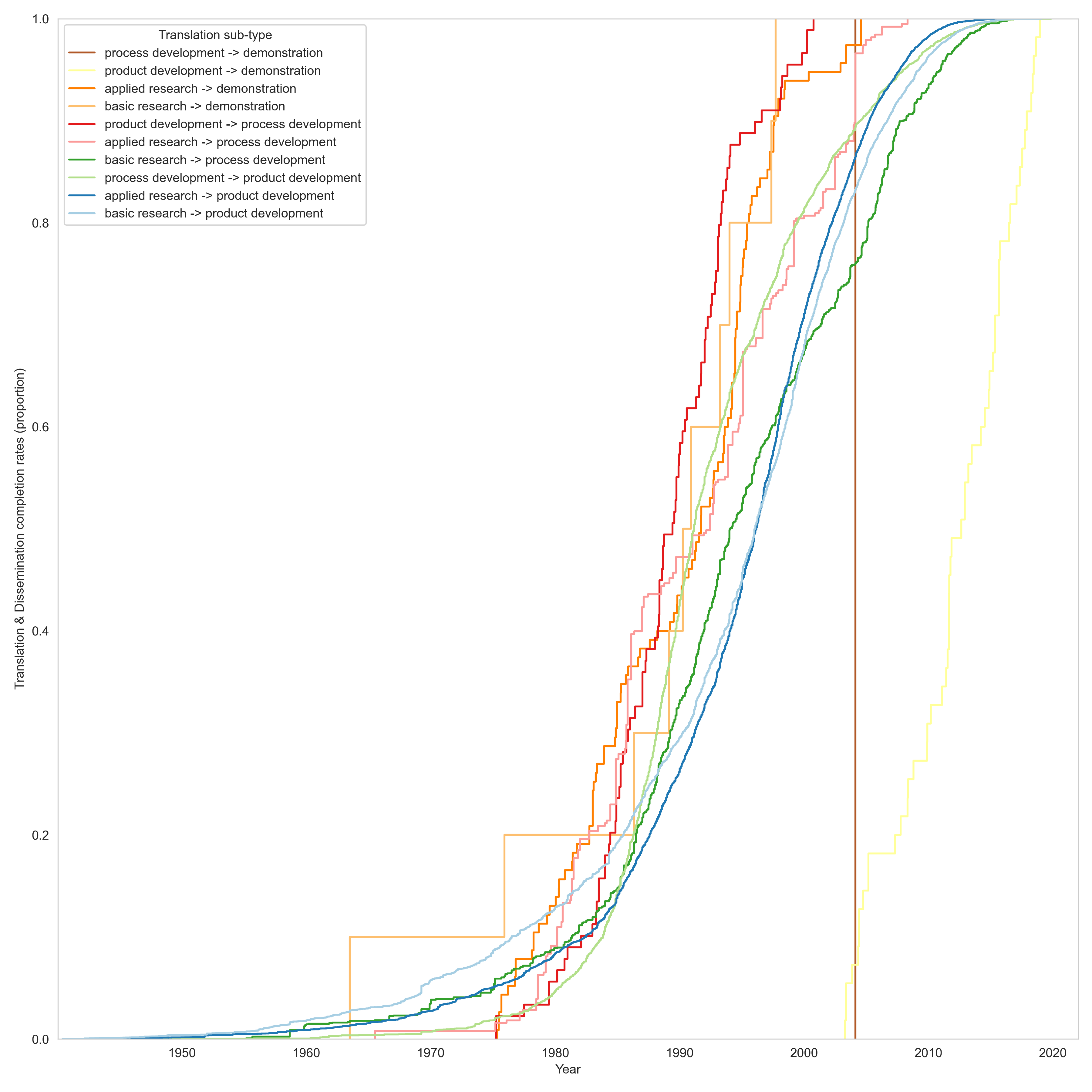}
		\caption{Imvanex, Smallpox, Bavarian Nordic, 2013, WPV}
		%  \label{fig:h_v_d_imvanex}
	\end{subfigure}
	\hfill
	\begin{subfigure}[b]{0.24\textwidth}
		\centering
		\includegraphics[width=\textwidth]{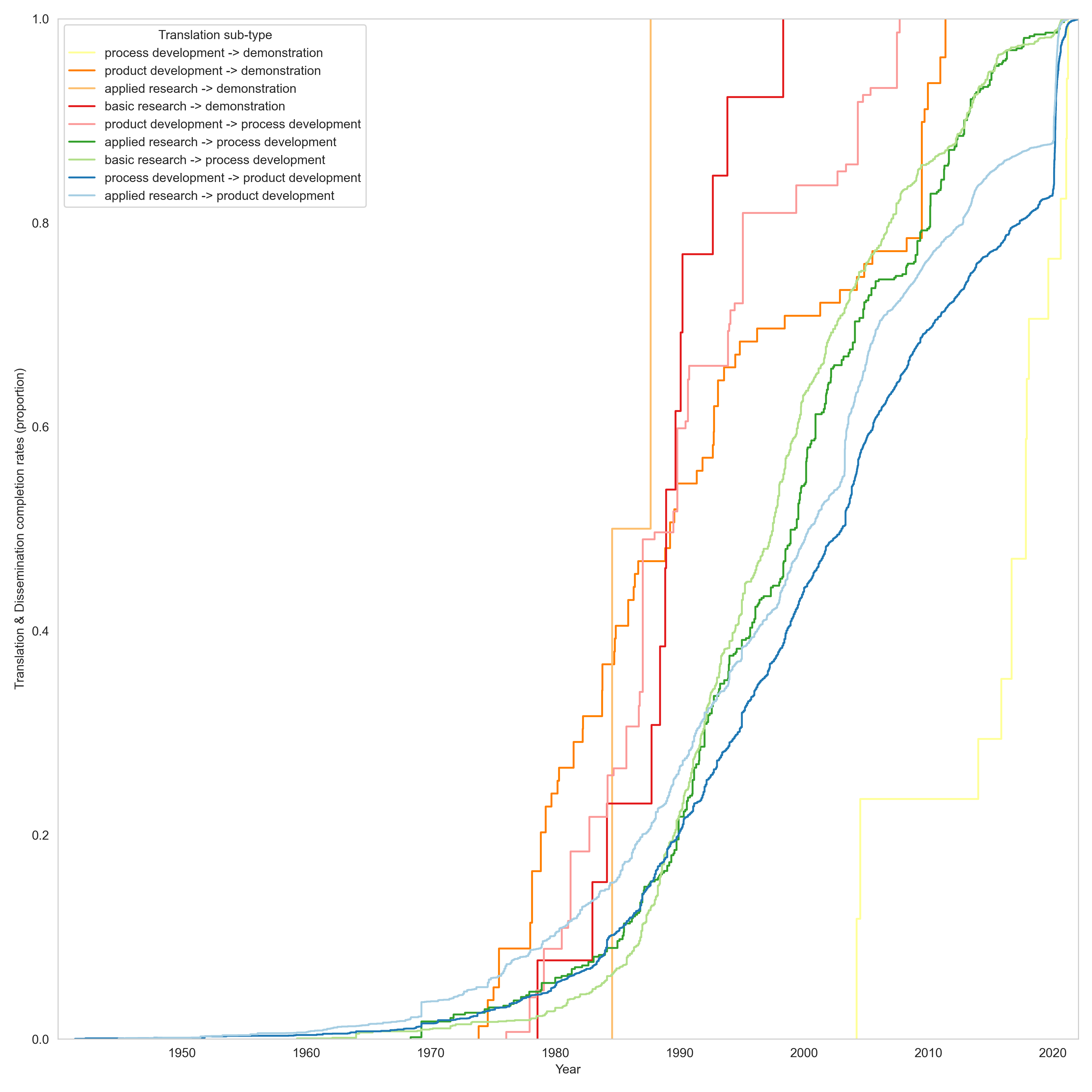}
		\caption{Nuvaxovid, COVID-19, Novavax, 2022, subunits}
		%  \label{fig:h_v_d_nuvaxovid}
	\end{subfigure}
	\hfill
	\begin{subfigure}[b]{0.24\textwidth}
		\centering
		\includegraphics[width=\textwidth]{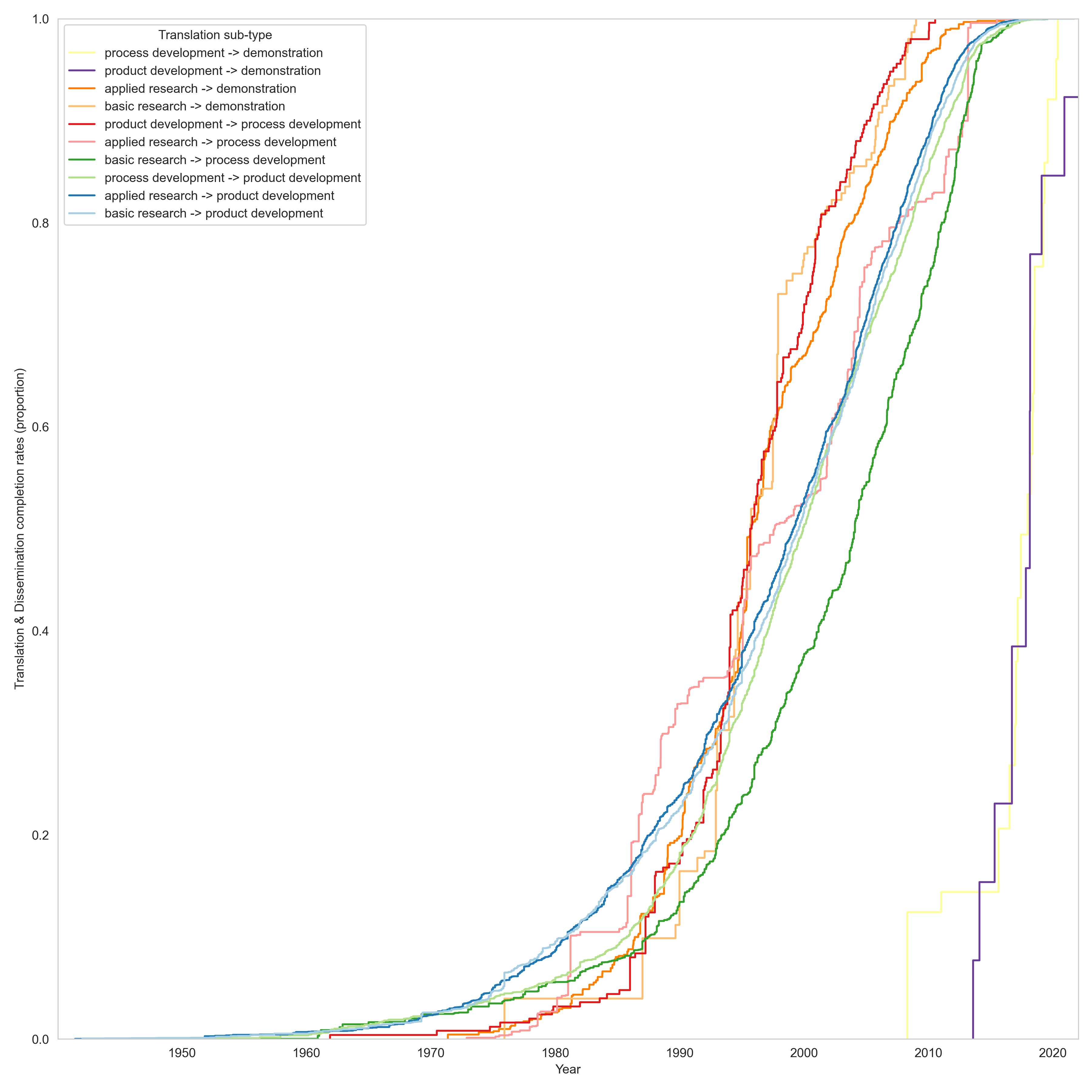}
		\caption{Shringrix, Shingles, GSK, 2017,\\ subunits}
		%  \label{fig:h_v_d_shringrix}
	\end{subfigure}
	\hfill
	\caption{Translation and dissemination sub-types completion rates as a function of time. Complete dataset for \figref{fig:inter_cdf}.} 
	% \label{fig:all_critical_paths}
\end{figure}

\begin{figure}[hbt]
	\centering
	\begin{subfigure}[b]{0.24\textwidth}
		\centering
		\includegraphics[width=\textwidth]{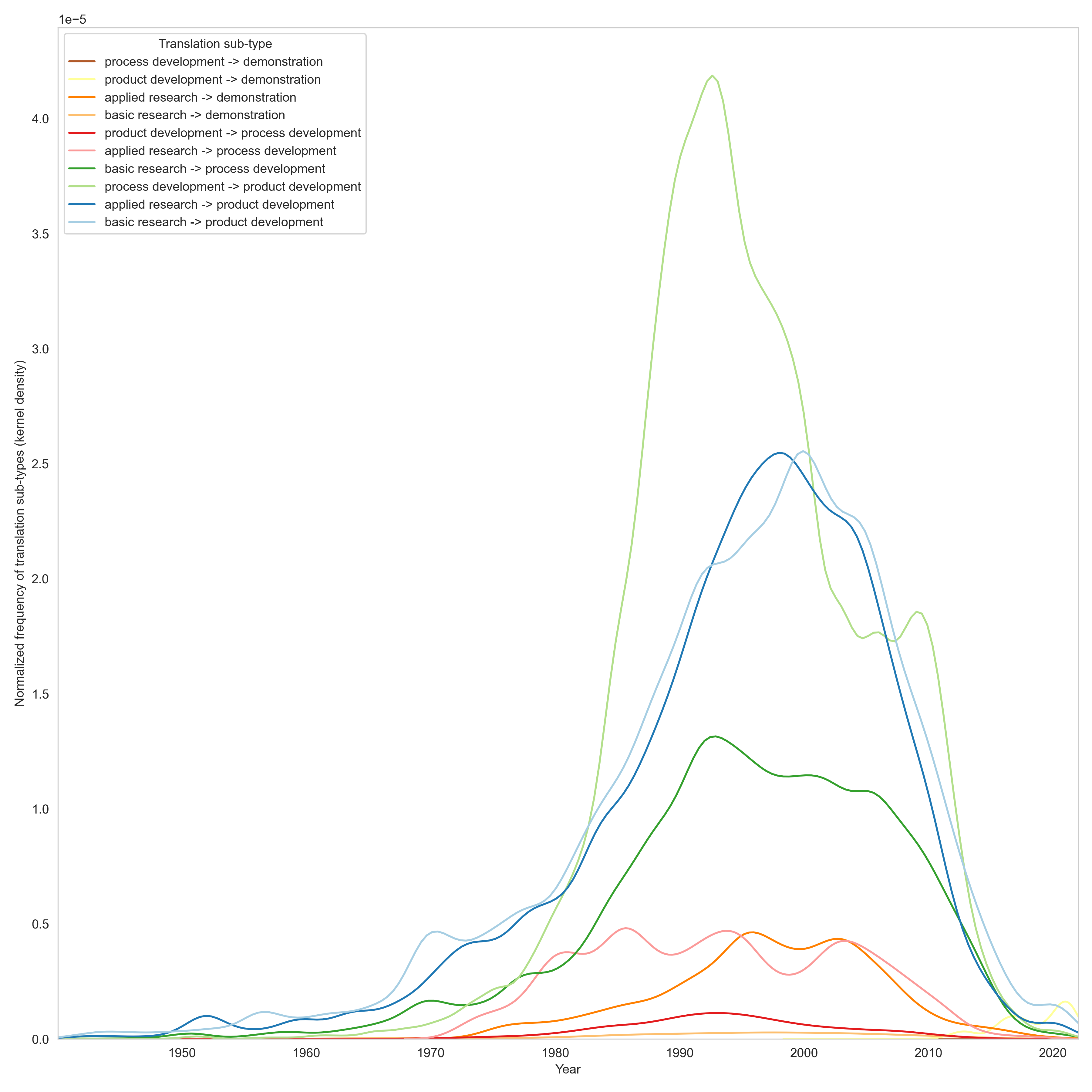}
		\caption{Zabdeno, Ebola, Janssen, 2020,\\AVV}
		%  \label{fig:h_v_d_zabdeno}
	\end{subfigure}
	\hfill
	\begin{subfigure}[b]{0.24\textwidth}
		\centering
		\includegraphics[width=\textwidth]{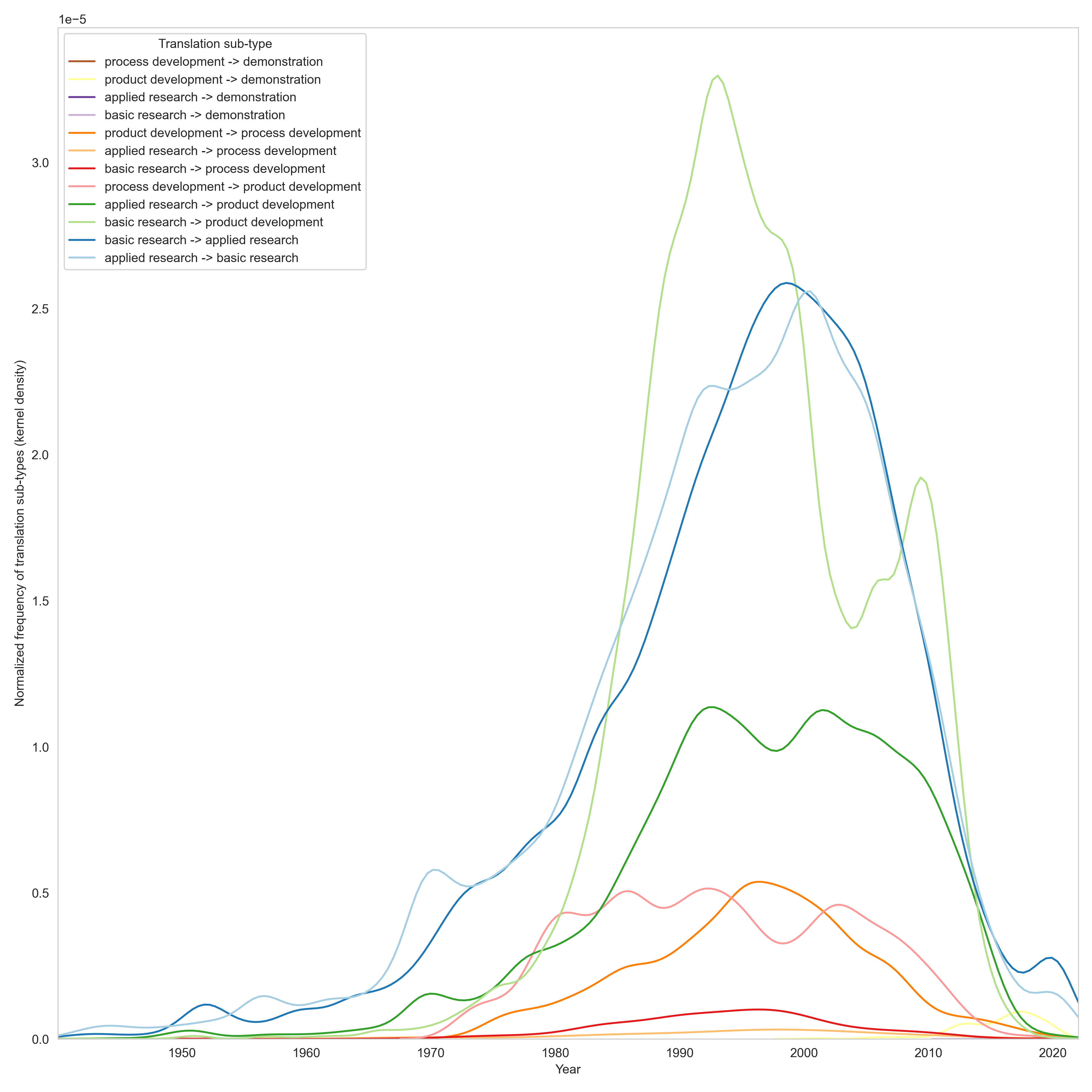}
		\caption{Vaxzevria, COVID-19, AstraZeneca, 2020, AVV}
		%  \label{fig:h_v_d_vaxzevria}
	\end{subfigure}
	\hfill
	\begin{subfigure}[b]{0.24\textwidth}
		\centering
		\includegraphics[width=\textwidth]{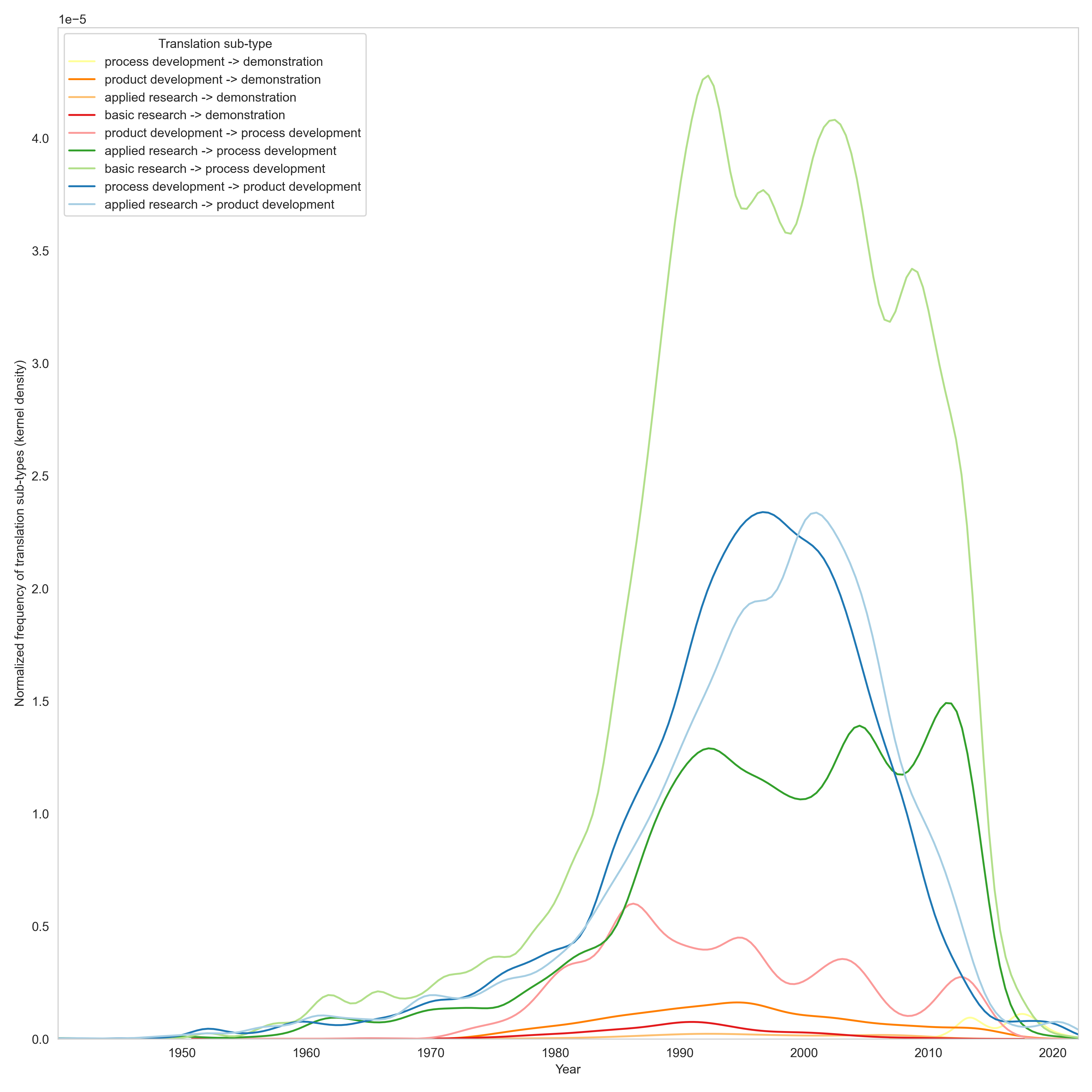}
		\caption{Spikevax, COVID-19, Moderna, 2020, mRNA}
		%  \label{fig:h_v_d_spikevax}
	\end{subfigure}
	\hfill
	\begin{subfigure}[b]{0.24\textwidth}
		\centering
		\includegraphics[width=\textwidth]{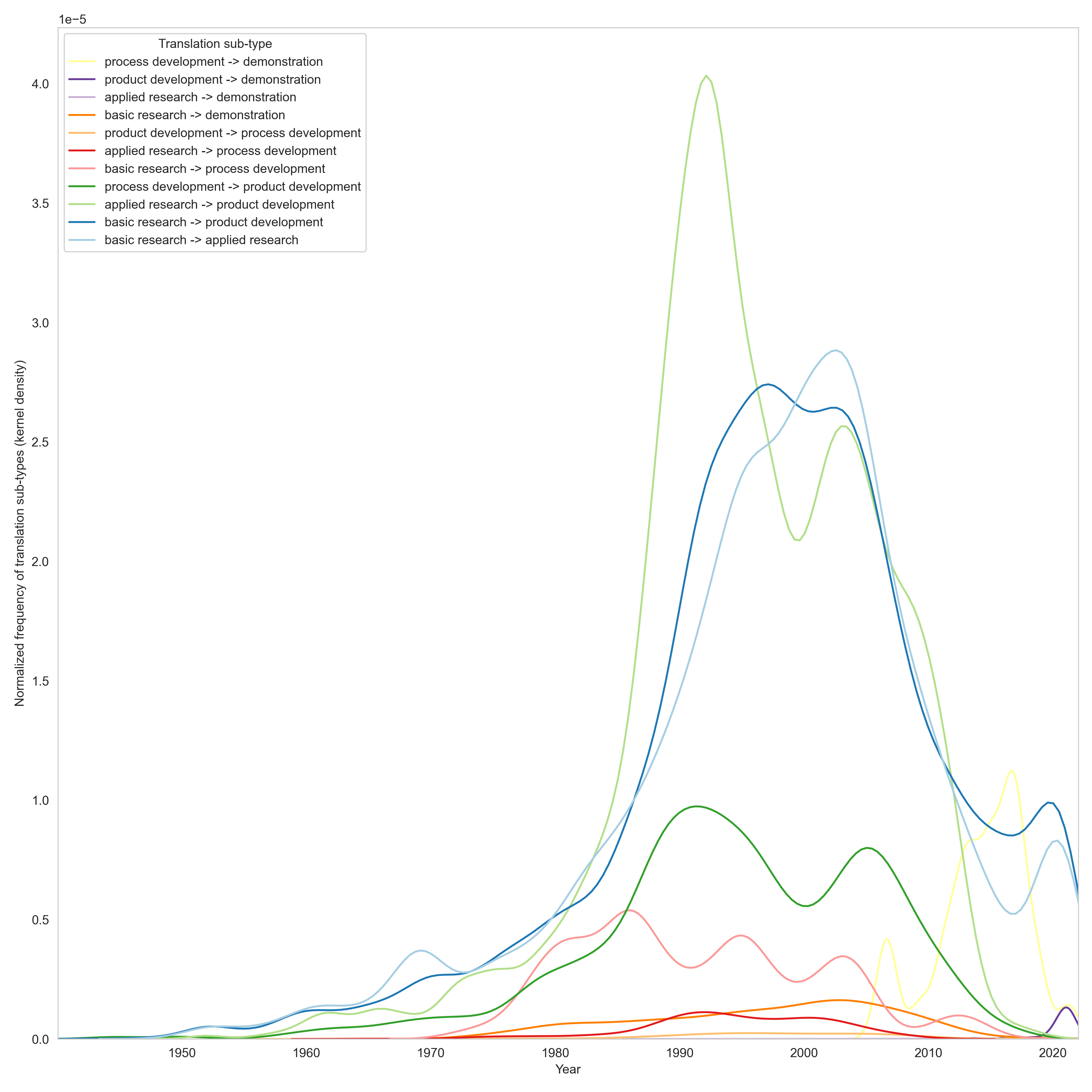}
		\caption{Comirnaty, COVID-19, BioNTech/Pfizer, 2020, mRNA}
		%  \label{fig:h_v_d_comirnaty}
	\end{subfigure}
	\hfill
	\begin{subfigure}[b]{0.24\textwidth}
		\centering
		\includegraphics[width=\textwidth]{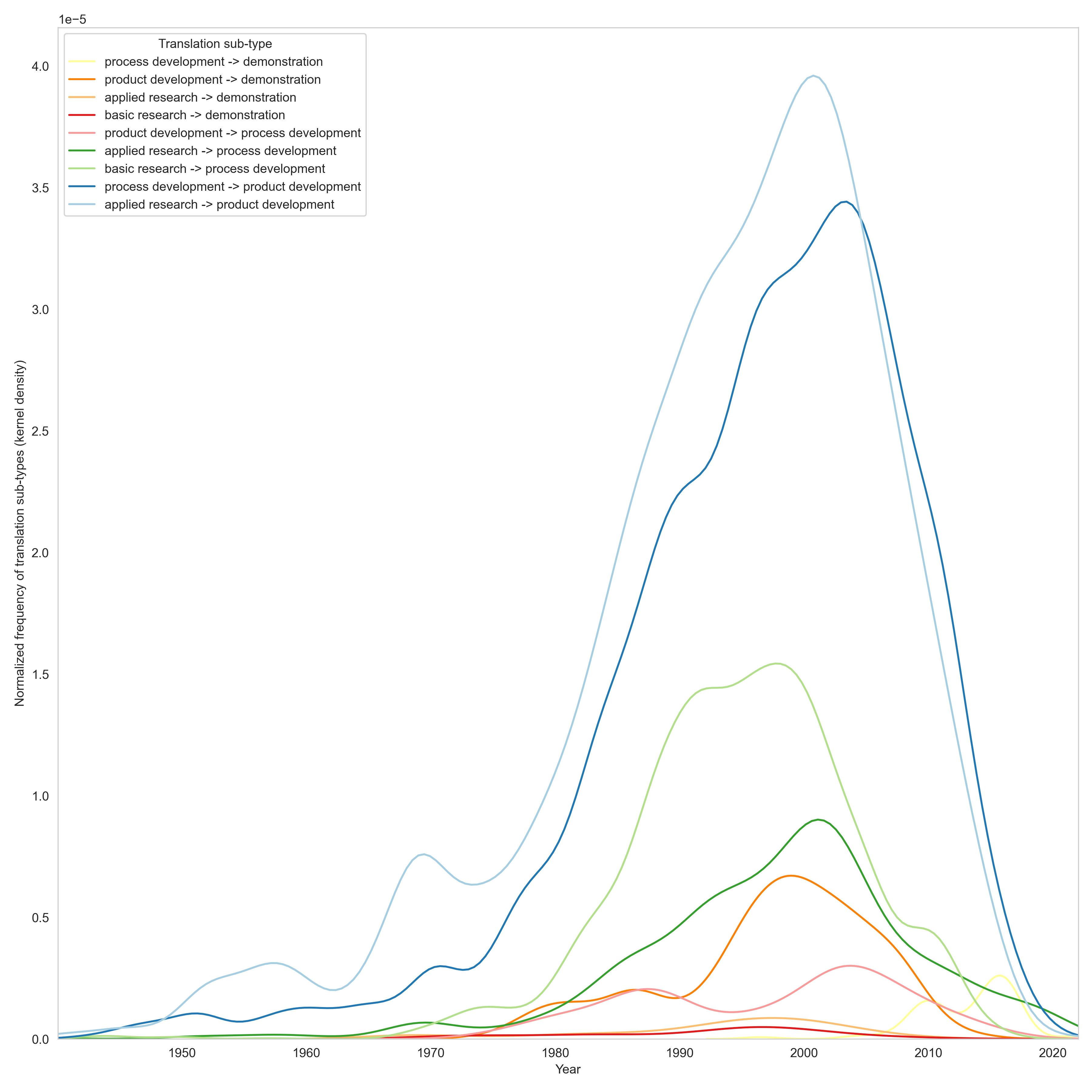}
		\caption{Dengvaxia, Dengue, Sanofi, 2019,\\ WPV}
		%  \label{fig:h_v_d_dengvaxia}
	\end{subfigure}
	\hfill
	\begin{subfigure}[b]{0.24\textwidth}
		\centering
		\includegraphics[width=\textwidth]{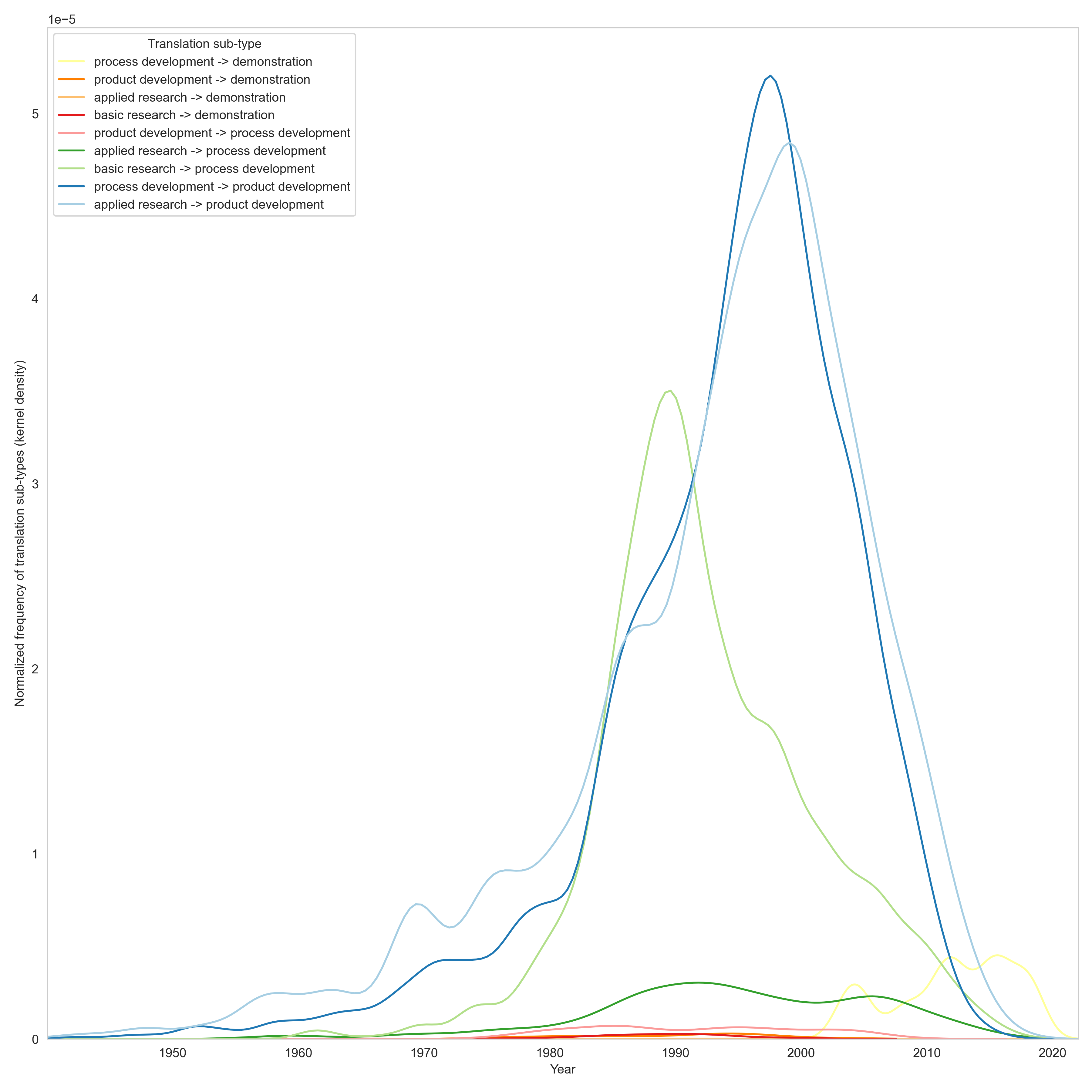}
		\caption{Imvanex, Smallpox, Bavarian Nordic, 2013, WPV}
		%  \label{fig:h_v_d_imvanex}
	\end{subfigure}
	\hfill
	\begin{subfigure}[b]{0.24\textwidth}
		\centering
		\includegraphics[width=\textwidth]{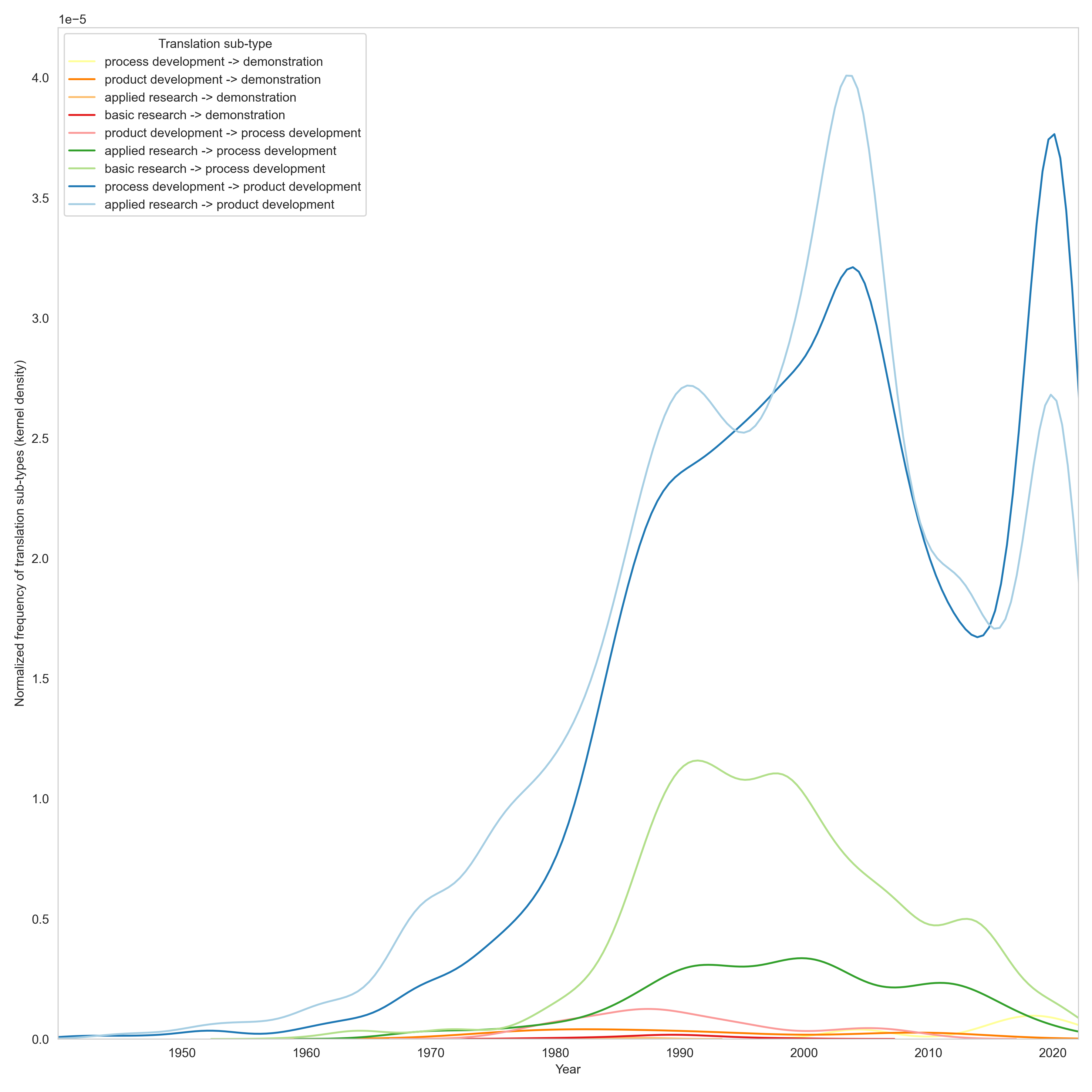}
		\caption{Nuvaxovid, COVID-19, Novavax, 2022, subunits}
		%  \label{fig:h_v_d_nuvaxovid}
	\end{subfigure}
	\hfill
	\begin{subfigure}[b]{0.24\textwidth}
		\centering
		\includegraphics[width=\textwidth]{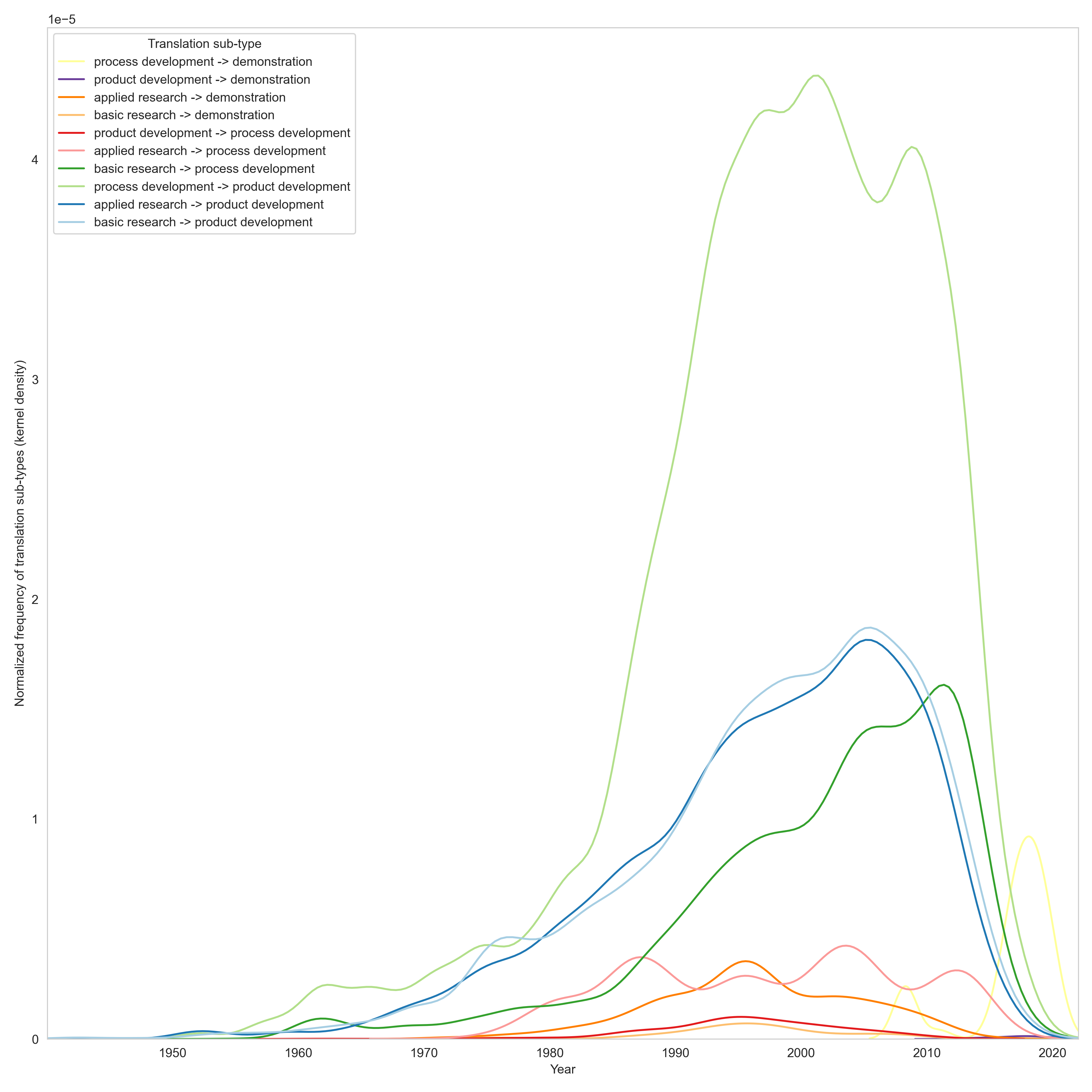}
		\caption{Shringrix, Shingles, GSK, 2017,\\ subunits}
		%  \label{fig:h_v_d_shringrix}
	\end{subfigure}
	\hfill
	\caption{Translation sub-types frequencies as a function of time. Complete dataset for \figref{fig:inter_kde}.} 
	% \label{fig:all_critical_paths}
\end{figure}

%%% The alternative is to comment out the line above and to compile 
%%% file innovationTranslationAppendix.tex to get the appendices 
%%% as a separate file, a supplementary material document. 

\end{document}